\documentclass[aps,prd,twocolumn,superscriptaddress,preprintnumbers,floatfix,longbibliography,nofootinbib]{revtex4-1}

\usepackage{graphicx}
\usepackage{amsmath}
\usepackage[caption=false]{subfig}
\usepackage{siunitx}
\usepackage{placeins}
\usepackage{color}
\usepackage{standalone}
\usepackage{dcolumn}
\usepackage{bm}
\usepackage{microtype}
\usepackage{etoolbox}
\usepackage{amssymb}
\usepackage{mathrsfs}
\usepackage{accents}
\usepackage[normalem]{ulem}
\usepackage[dvipsnames]{xcolor}
\usepackage[colorlinks,urlcolor=NavyBlue,citecolor=NavyBlue,linkcolor=NavyBlue,pdfusetitle]{hyperref}
\usepackage[all]{hypcap}
\usepackage{enumerate}
\usepackage{gensymb}

\AtBeginDocument{\usepackage{booktabs}}

\newtoggle{commentsoff}
\toggletrue{commentsoff}

\ifdefined\nocomments
    \toggletrue{commentsoff}
\fi

\iftoggle{commentsoff}{
  \newcommand{\mi}[1]{}
  \newcommand{\ls}[1]{}
  \newcommand{\rb}[1]{}
  \newcommand*{\comment}[1]{}
  \newcommand*{\warn}[1]{}
  \newcommand*{\todo}[1]{}
  \newcommand*{\red}[1]{#1}

}{
  \newcommand{\mi}[1]{{\color{cyan}~\textsf{[{\bf MI}: #1]}}}
  \newcommand{\ls}[1]{{\color{RedOrange}~\textsf{[{\bf LS}: #1]}}}
  \newcommand{\rb}[1]{{\color{green}~\textsf{[{\bf RB}: #1]}}}
  \newcommand*{\comment}[1]{{\color{blue} [{\bf NOTE}: #1]}}
  \newcommand*{\warn}[1]{{\color{red} [{\bf WARNING}: #1]}}
  \newcommand*{\todo}[1]{{\color{purple} \textsf{[{\bf TODO}: #1]}}}
  \newcommand*{\red}[1]{{\color{purple} #1}}

}

\graphicspath{{./fig/}}

\newcommand{\ts}{\textsuperscript}

\newcommand{\beq}{\begin{equation}}
\newcommand{\eeq}{\end{equation}}

\newcommand*{\eq}[1]{Eq.\ \eqref{eq:#1}}
\newcommand*{\fig}[1]{Fig.\ \ref{fig:#1}}
\newcommand*{\sect}[1]{Sec.\ \ref{sec:#1}}

\newcommand{\appropto}{\mathrel{\vcenter{
  \offinterlineskip\halign{\hfil$##$\cr
    \propto\cr\noalign{\kern2pt}\sim\cr\noalign{\kern-2pt}}}}}

\newcommand{\mbh}{M}
\newcommand{\obh}{\Omega_{\rm BH}}
\newcommand{\obhn}{\overline{\Omega}_{\rm BH}}

\newcommand{\rg}{r_g}

\newcommand{\lc}{\lambda_\mu}
\newcommand{\lbc}{\lambdabar_\mu}

\newcommand{\ob}{\omega}
\newcommand{\jb}{j}
\newcommand{\lb}{l}
\newcommand{\mb}{m}
\newcommand{\np}{\bar{n}}
\newcommand{\nr}{n}

\newcommand{\mc}{M_c}

\newcommand{\ogw}{\tilde{\omega}}
\newcommand{\lgw}{\tilde{l}}
\newcommand{\mgw}{\tilde{m}}
\newcommand{\fgw}{f}

\newcommand{\ra}{{\alpha_\star}}
\newcommand{\dec}{{\delta_\star}}

\newcommand{\ampfactor}{{\cal A}}
\newcommand{\fs}{\alpha}

\newcommand{\msun}{M_{\odot}}

\newcommand{\tinst}{\tau_{\rm inst}}
\newcommand{\tgw}{\tau_{\rm GW}}

\newcommand{\dcc}{LIGO--P1800270}

\begin{document}

\preprint{\dcc}

\title{Directed searches for gravitational waves from ultralight bosons}

\author{Maximiliano Isi}
\email[]{maxisi@mit.edu}
\thanks{NHFP Einstein fellow}
\affiliation{
LIGO Laboratory, Massachusetts Institute of Technology, Cambridge, Massachusetts 02139, USA
}%
\affiliation{LIGO Laboratory, California Institute of Technology, Pasadena,
California 91125, USA}

\author{Ling Sun}
\email[]{lssun@caltech.edu}
\affiliation{LIGO Laboratory, California Institute of Technology, Pasadena, California 91125, USA}
\affiliation{School of Physics, University of Melbourne, Parkville, Victoria 3010, Australia}
\affiliation{Australian Research Council Centre of Excellence for Gravitational Wave Discovery (OzGrav), University of Melbourne node, Parkville, Victoria 3010, Australia}

\author{Richard Brito}
\email[]{richard.brito@roma1.infn.it}
\affiliation{Max Planck Institute for Gravitational Physics (Albert Einstein Institute), Am M\"{u}hlenberg 1, Potsdam-Golm, 14476, Germany}
\affiliation{Dipartimento di Fisica, ``Sapienza'' Universit\`a di Roma \& Sezione INFN Roma1,\\Piazzale Aldo Moro 5, 00185, Roma, Italy}

\author{Andrew Melatos}
\email[]{amelatos@unimelb.edu.au}
\affiliation{School of Physics, University of Melbourne, Parkville, Victoria 3010, Australia}
\affiliation{Australian Research Council Centre of Excellence for Gravitational Wave Discovery (OzGrav), University of Melbourne node, Parkville, Victoria 3010, Australia}

\hypersetup{pdfauthor={Isi, Sun, Brito, Melatos}}

\date{\today}

\begin{abstract}
Gravitational-wave detectors can be used to search for yet-undiscovered ultralight bosons, including those conjectured to solve problems in particle physics, high-energy theory and cosmology.
In particular, ground-based instruments could probe boson masses between $10^{-15}$ eV to $10^{-11}$ eV, which are largely inaccessible to other experiments.
In this paper, we explore the prospect of searching for the continuous gravitational waves generated by boson clouds around known black holes.
We carefully study the predicted waveforms and use the latest-available numerical results to model signals for different black-hole and boson parameters.
We then demonstrate the suitability of a specific method (hidden Markov model tracking) to efficiently search for such signals, even when the source parameters are not perfectly known and allowing for some uncertainty in theoretical predictions.
We empirically study this method's sensitivity and computational cost in the context of boson signals, finding that it will be possible to target remnants from compact-binary mergers localized with at least three instruments.
For signals from scalar clouds, we also compute detection horizons for future detectors (Advanced LIGO, LIGO Voyager, Cosmic Explorer and the Einstein Telescope).
Among other results, we find that, after one year of observation, an Advanced LIGO detector at design sensitivity could detect these sources up to over 100 Mpc, while Cosmic Explorer could reach over $10^4$ Mpc.
These projections offer a more complete picture than previous estimates based on analytic approximations to the signal power or idealized search strategies.
Finally, we discuss specific implications for the followup of compact-binary coalescences and black holes in x-ray binaries.
Along the way, we review the basic physics of bosons around black holes, in the hope of providing a bridge between the theory and data-analysis literatures.
\end{abstract}

\maketitle

\section{Introduction}

After decades of dedicated effort, the Advanced Laser Interferometer Gravitational-wave Observatory (aLIGO) \cite{aLIGO} and Advanced Virgo \cite{Virgo} detectors have inaugurated gravitational-wave (GW) astronomy with the observation of several compact-binary coalescences (CBCs) \cite{gw150914,gw151226,o1bbh,gw170104,gw170608,gw170814,gw170817}.
Arguably, one of the most exciting prospects in this new era of astronomy is to use gravitational waves to learn about fundamental physics.
Common examples of this are attempts to probe the nature of gravity by testing general relativity \cite{gw150914_tgr,Yunes:2016jcc}, or to probe the nature of nuclear matter through the neutron-star equation of state \cite{gw170817,gw170817_pe,gw170817_eos}.
Another exciting possibility is that of using gravitational waves to learn about particle physics.
In particular, it may be possible to search for new ultralight bosons with gravitational-wave detectors, a prospect that has recently garnered much attention \cite{Arvanitaki2011,Yoshino2014,Yoshino2015,Arvanitaki2015,Arvanitaki2017,Brito2017-letter,Brito2017,Baryakhtar2017}.
In this paper, we explore the potential of achieving this using directed searches for continuous gravitational waves with ground-based detectors.

There are strong theoretical reasons to believe in the existence of new weakly-interacting, ultralight scalar (spin 0) or vector (spin 1) particles.
The prime example of this is the axion, a (pseudo-)scalar particle originally proposed to explain the strong constraints on the existence of charge-parity (CP) violating terms in the strong nuclear force sector~\cite{Peccei1977,Peccei1977PhRvD,Weinberg1978}.
This quantum-chromodynamics (QCD) axion is among the best-motivated extensions of the standard model, but there are others.
For example, string theory predicts the existence of a variety of axion-like particles (potentially including the QCD axion) with masses populating each decade between $10^{-33}$ eV and $10^{-10}$ eV as a result of the compactification of extra spatial dimensions \cite{Arvanitaki2010}.
Similarly, a hidden sector of light vector particles also naturally arises in compactifications of string theory~\cite{Goodsell2009}.
Besides motivations from particle and high-energy physics, these bosons are also popular dark matter candidates (see e.g.~\cite{Jaeckel:2010ni,Essig:2013lka,Hui:2016ltb}).

Because of their weak couplings to the standard model and their vanishingly-small mass, all these proposed new particles would be extremely hard to detect by conventional means.
In particular, all existing constraints on the existence of the QCD axion rely on its expected coupling to Standard Model particles, a property that is heavily model-dependent~\cite{Kim:2008hd}.
Such observations loosely constrain the mass of the QCD axion to be $\lesssim 10^{-3}$ eV \cite{Raffelt2006}, but values of order $10^{-10}$ eV or lower are favored by theory \cite{Arvanitaki2011}.
For other kinds of conjectured ultralight bosons, whose potential interactions with the standard model are very weak or inexistent, constraints and detection can only provided through their gravitational coupling.

Given the substantial challenge in detecting ultralight bosons, there has been considerable excitement about the proposal to look for these particles by taking advantage of the universal character of gravitational couplings.
The idea hinges on the phenomenon of superradiance \cite{Penrose1969,Press1972,Bekenstein1973,Brito2014,Brito2015}, by which macroscopic clouds of these bosons should form around rapidly spinning black holes (BHs) and, in turn, produce a varied set of observational signatures~\cite{Arvanitaki2011,Yoshino2014,Yoshino2015,Arvanitaki2015,Arvanitaki2017,Brito2017-letter,Brito2017,Baryakhtar2017,Plascencia:2017kca,Cardoso2018,Baumann:2018vus,Hannuksela:2018izj,Zhang:2018kib}. Indeed, constraints based on BH spin measurements of x-ray binaries have already been put, and exclude roughly the mass interval $[10^{-12},10^{-11}]$ eV for non-interacting massive scalar fields \cite{Arvanitaki2017,Cardoso2018} and $[10^{-13},10^{-11}]$ eV for non-interacting massive vector fields \cite{Baryakhtar2017,Cardoso2018}.  Constraints derived from observations of x-ray binaries should however be interpreted with caution, since there is large uncertainty about the age and history of these systems, as well as caveats about the systematics affecting their spin measurements \cite{Reynolds:2013qqa,McClintock:2013vwa}.

A much more clean observational signature is the emission of potentially-detectable gravitational waves at a frequency of roughly twice the boson mass~\cite{Arvanitaki2015,Arvanitaki2017,Brito2017-letter,Brito2017,Baryakhtar2017}.
As it turns out, this means that clouds formed around stellar mass BHs should emit signals within the most sensitive band of ground-based detectors, probing boson masses in the theoretically interesting mass-range of the order of $10^{-15}$ eV through $10^{-11}$ eV~\cite{Arvanitaki2015,Brito2017}.

In this paper, we explore the prospect for the direct detection of continuous gravitational waves emitted by boson clouds.
In particular, we focus on searches in data from present and future ground-based detectors directed at known BHs.
As the main observational scenario, we consider the followup of remnants from compact-binary coalescences detected through gravitational waves, but also examine BH candidates known from electromagnetic observations.
Besides treating the data analysis, we study in detail the morphology of boson signals and use numerical calculations, combined with the latest analytic results, to estimate their amplitude and other relevant features.
This allows us to more accurately predict the potential signals that may be expected from clouds around a given BH.
In the case of scalar clouds, we use those estimates to obtain detection horizons for second-generation and proposed third-generation instruments.
Along the way, we review the theoretical basics of BH--boson superradiance in a language that we hope facilitates future work by gravitational-wave analysts interested in the topic.

In \sect{theory}, we review the theory of boson clouds around BHs and the gravitational-wave emission mechanism.
In \sect{signal} we discuss at length the specific morphology of the expected signals, as seen by ground-based detectors.
In \sect{searches}, we introduce hidden Markov model tracking as an ideal strategy to search for these signals, validate the method and estimate the sensitivity through Monte-Carlo simulations; we also discuss implications for the followup of compact-binary mergers and x-ray binaries as potential sources. 
Finally, we provide a summary and conclusions in \sect{conclusion}.

\section{The boson cloud}
\label{sec:theory}

The physics of boson fields around BHs has been extensively studied in different limits using both analytic and numerical methods \cite{Ternov:1978gq,Zouros1979,Detweiler1980,Dolan2007,Arvanitaki2010,Arvanitaki2011,Brito2014,Brito2015,Brito2017,Witek2013,East2017}.
We summarize those results below, in a way that is best suited for the data-analysis framework introduced later in the paper.
We explain how a macroscopic boson cloud spontaneously arises around fast-spinning BHs and proceeds to emit gravitational radiation, providing some essential mathematical detail.

Consider first a Kerr BH of mass $\mbh$ and angular momentum $J$.
The characteristic length associated with the BH mass will be $\rg\equiv G\mbh/c^2$, or half the Schwarzschild radius.
The other characteristic length, given by the BH spin, is the usual Kerr parameter, $a\equiv J/(\mbh c)$, from which we can in turn define the dimensionless spin, $\chi \equiv a c^2 /(G\mbh)$.
In terms of these quantities, the radius of the hole's outer horizon, in Boyer-Lindquist coordinates, is
\beq \label{eq:rplus}
r_+ = \rg \left(1 + \sqrt{1-\chi^2}\right) \equiv \rg\, \bar{r}_+ \, ,
\eeq
where $\bar{r}_+$ is defined here to be dimensionless.
At this location, the BH will then have a frame-dragging angular speed (with respect to infinity) of
\beq \label{eq:obh}
\obh = \frac{1}{2} \frac{c}{\rg} \frac{\chi}{1 + \sqrt{1-\chi^2}} \equiv \frac{c}{\rg}\, \obhn \, ,
\eeq
where $\obhn$ is defined here to be dimensionless.
(See, e.g., \cite{Teukolsky2015} for a recent review of the Kerr metric.)

Now, imagine that, beyond the usual particles in the standard model, there exists an ultralight boson of mass 
\beq
m_b \equiv \mu / c^2\, ,
\eeq
where $\mu$ is the boson's rest energy.
The corresponding length and time scales are given by the Compton wavelength, $\lc \equiv 2\pi \lbc \equiv h / (m_b c)$, and angular frequency, $\omega_\mu = c/\lbc = \mu/\hbar$.
As soon as the BH is born,%
\footnote{Meaning, as soon as it is sufficiently close to the ideal Kerr metric.}
(perhaps, as a result of stellar collapse or a binary coalescence) the usual quantum fluctuations in the boson field will cause pairs of particles to spontaneously appear in the hole's vicinity, causing a number of them to unavoidably fall in.
What happens next will depend on the properties of the BH and the infalling excitations: under most circumstances, the particle will simply disappear behind the horizon never to return; however, for the right sets of parameters, the excitation in the boson field will scatter off the BH with a boost in amplitude, effectively increasing the number of particles (occupation number) around the BH \cite{Penrose1969,zeldovich1,zeldovich2,Press1972,Bekenstein1973,Bekenstein:1998nt,Brito2015}.

From the second law of BH thermodynamics \cite{Bekenstein1973} (or more generic kinematic arguments \cite{Bekenstein:1998nt,Arvanitaki2010,Brito2015}), we may expect the boson-wave amplification to occur when the following {\em superradiance condition} is satisfied:
\beq \label{eq:sr}
\ob_\mu / \mb < \obh\, ,
\eeq
where $\mb$ is the (magnetic) quantum number corresponding to the projection of the particle's {\em total} angular momentum along the BH spin direction.
This amplification extracts energy from the BH just as in the classical Penrose process \cite{Penrose1969}, which is itself another manifestation of BH superradiance (see \cite{Brito2015} for a review).
Because of the field's nonzero mass, a scattered boson will generally tend to be bound to the BH, attracted by its gravitational pull.
Consequently, scattered particles may remain confined in that region, facilitating successive scatterings and the associated compounded amplification of the field.
This process is similar to the ``BH bomb'' devised by Press and Teukolsky \cite{Press1972}, with the mirror replaced by the boson mass \cite{Damour:1976kh,Zouros1979,Detweiler1980,Furuhashi:2004jk,Strafuss:2004qc,Dolan2007}.

We may anticipate that the boson amplification will be maximized when the field and BH have comparable characteristic lengthscales, i.e.~$\lc \sim \rg$.
If this is the case, then the field amplitude will grow at an exponential rate in a (quasi-)bound state around the BH (\sect{theory_spectrum}).
As the field grows, it draws energy and angular momentum from the BH until the condition of \eq{sr} is no longer satisfied.
Because the field is bosonic, there are \emph{a priori} no limits to the occupation number of any given energy level: the number of particles in a superradiant state will grow exponentially to form a macroscopic ``cloud''. However, if formed solely due to the superradiant instability, the cloud will extract at most ${\sim}10\%$ of the BH's mass~\cite{East:2017ovw,Herdeiro:2017phl}.
This cloud will slowly fade away, as its energy is radiated away in the form of gravitational waves over very long timescales compared to the superradiant rate (\sect{theory_gws}).

\subsection{Black-hole and boson interactions}
\label{sec:theory_spectrum}

\subsubsection{Energy levels}

The qualitative picture laid out above is backed up by analytic and semi-analytic calculations of boson fields over a Kerr background \cite{Ternov:1978gq,Zouros1979,Detweiler1980,Dolan2007,Arvanitaki2010,Arvanitaki2011,Brito2014,Brito2015,Brito2017}.
In the nonrelativistic ($\fs/\jb\ll 1$) regime implied by \eq{sr}, the influence of the BH is effectively reduced to a simple inverse-radius gravitational potential.
This potential causes the bosons to present quasi-bound energy eigenstates essentially identical to those in the hydrogen atom, but with gravity replacing electromagnetism as the relevant interaction.
More carefully solving the Schr\"odinger equation over a Kerr background, one indeed finds that, in this regime, the system has hydrogenic energy levels \cite{Ternov:1978gq,Detweiler1980},
\beq \label{eq:energy_levels}
E_{\np} \approx \mu \left(1 - \frac{1}{2}\frac{\alpha^2}{\np^2} + \dots \right) ,
\eeq
for $\np= \nr + \lb + 1$ the principal quantum number, $\nr$ the radial quantum number, and $\lb$ the {\em orbital} azimuthal quantum number.%
\footnote{Although the spin parameter does not appear at leading order in $\fs$ \cite{Dolan2007}, the fact that the BH is spinning does affect the angular part of the boson eigenfunctions: these have to be described using spin-weighted {\em spheroidal}, rather than spherical, harmonics \cite{Brito2017} (see \sect{signal}).
Higher-order corrections to the energy eigenvalues, including corrections due to the BH spin, can be found in~\cite{Baumann:2018vus}.}
As usual, we have $|\jb-s| \leq \lb \leq \jb+s$ and  $-\jb \leq \mb \leq \jb$, where $\jb$ and $s$ are respectively the total and spin angular-momentum quantum numbers.
All superradiant levels are hydrogenic with a spectrum well-described by \eq{energy_levels} \cite{Dolan2007}.
The quantity $\alpha$ in \eq{energy_levels} plays exactly the same role as the fine-structure constant in the hydrogen atom, and takes the value of the ratio of the two relevant lengthscales (or, equivalently, timescales):
\beq \label{eq:alpha}
\fs \equiv \frac{\rg}{\lbc} = \frac{G \mbh}{c} \frac{m_b}{\hbar} = \frac{G \mbh}{c^3} \ob_\mu\, ,
\eeq
where $\lbc \equiv \lc / (2\pi)$.
Importantly, \eq{sr} implicitly constrains $\fs$ as a function of the BH spin.
If we want superradiance to take place, then \eq{sr} and \eq{obh} demand:
\beq \label{eq:sr2}
\fs < \frac{1}{2} \mb \chi  \left(1 + \sqrt{1-\chi^2}\right)^{-1} < \frac{m}{2}\, ,
\eeq
where the second inequality is obtained by noting $0\leq \chi < 1$.
Because $\mb\leq\jb$, this condition justifies working in the nonrelativistic, $\fs < \jb$, limit in the first place \cite{Arvanitaki2011}.

\subsubsection{Cloud growth}

Unlike the hydrogen atom, however, the BH-boson system is non-Hermitian due to the ingoing boundary condition for waves at the horizon.
This means that the occupation number of the different energy eigenstates need not be constant---in fact, they will most certainly {\em not} be so for the superradiant states we are interested in.
For small $\fs/\mb$, the \red{occupation number}\footnote{The rate of change of the occupation number is \emph{twice} that of the field amplitude, which is itself given by the imaginary part of the wavefunction frequency: $\Gamma = 2\, \mathrm{Im}(\omega)$.} of a given quantum state will grow exponentially at a rate that may be analytically approximated as \cite{Ternov:1978gq,Detweiler1980,Baryakhtar2017}:
\beq \label{eq:growth}
\Gamma_{\jb\lb\mb\nr} \approx \red{2} \fs^{2\jb+2\lb + 5} \bar{r}_+ \left(\mb \obh - \omega_\mu\right) C_{\jb\lb\mb\nr} \, ,
\eeq
with $C_{\jb\lb\mb\nr}$ a dimensionless factor, and $\bar{r}_+$ the dimensionless radius defined in \eq{rplus}.
In the case of a scalar boson, the orbital angular momentum is necessarily the total angular momentum ($\jb=\lb$), and it can be shown that \cite{Ternov:1978gq,Detweiler1980}:
\begin{align}
C_{\jb\lb\mb\nr}^{\rm (scalar)} &= \frac{2^{4\lb+2}(2\lb+\nr+1)!}{(\lb+\nr+1)^{2\lb+4} \nr!} \left[\frac{\lb!}{(2\lb)!(2\lb+1)!}\right]^2 \\
&\times \prod_{k=1}^\lb \left[k^2 \left(1 - \chi^2 \right) + \frac{4 r_+^2}{c^2} \left(\mb \obh - \omega_\mu\right)^2 \right] . \nonumber
\end{align}
Expressions for vector $C_{\jb\lb\mb\nr}$ up to leading order in $\fs$ can be found in Appendix A of \cite{Baryakhtar2017}.
The validity of this approximation in the regime of interest has been confirmed numerically for scalars \cite{Dolan2007} and, more recently, vectors \cite{Pani:2012bp,East2017,Cardoso2018,Frolov:2018ezx,Dolan:2018dqv}.

Irrespective of boson spin, there are three key features of the occupation growth rate, $\Gamma$ in \eq{growth}, that can be distilled from the above results:
\begin{enumerate}[(i)]
\item the sign of $\Gamma$ depends solely on $\left(\mb \obh - \omega_\mu\right)$, implying that indeed energy levels satisfying \eq{sr} will grow exponentially, while others will be depleted;
\item $\Gamma$ is a high power of $\fs$, growing with the sum $\jb+\lb$;
\item for a fixed angular momentum ($\jb$, $\lb$) and $\fs$, $\Gamma$ decreases mildly with $n$.
\end{enumerate}
Because we are working in the small-$\fs$ limit, these three facts mean that, for a given system (i.e.~a given $\fs$ and $\chi$), \emph{the fastest growth will occur for the fundamental radial harmonic of the level with the {\em smallest} possible total angular-momentum, $\jb$, that still supports a magnetic number, $|\mb|\leq \jb$, sufficiently large to satisfy \eq{sr}.}
In other words, if the boson has spin-weight $s=0,1$, the level with the fastest superradiant growth in a given system will have angular quantum numbers $\{\jb,\, \lb,\, \mb\}$ given by
\beq
\jb = \lb + s = \mb = {\rm ceil}(\fs/\obhn)\, ,
\eeq
where ``ceil'' stands for the operation of rounding up to the closest integer.
In particular, the fastest-possible level over {\em all} values of $\fs$ and $\chi$ will then be
\beq \label{eq:best_level}
\jb= \lb + s = \mb = 1~,~\nr=0\, .
\eeq
Given this, it follows from \eq{growth} that vector clouds will tend to grow significantly faster than scalar ones ($\Gamma^{\rm (v)}/\Gamma^{\rm (s)} \sim \fs^{-2}$).

\subsubsection{Final state}

As the particle number grows, the energy and angular momentum required to populate the boson energy levels are extracted from the BH.\footnote{As this happens, \eq{sr} guarantees that the BH area increases, satisfying the second law of thermodynamics~\cite{Bekenstein1973,Bekenstein:1998nt}.}
Consequently, the BH quickly loses mass and spin until \eq{sr} is asymptotically saturated and the growth rate, \eq{growth}, vanishes.
As implied by \eq{sr2}, the spin of the BH at the end of this process will then be
\beq \label{eq:final_spin}
\chi_f = \frac{4 \fs_f \mb}{4\fs^2_f + \mb^2}\, ,
\eeq
where $\fs_f$ is given by \eq{alpha} for the {\em final} BH mass.
If no other processes (like accretion) take place in the relevant timescale, the final mass of the cloud ($\mc$) will simply be given by the difference between the initial (${\mbh}_i$) and final (${\mbh}_f$) BH masses.
If only one level is populated, then it may be shown that this will be approximately \cite{Brito2017}
\beq \label{eq:mc}
\mc = \mbh_i - \mbh_f \approx M_i \frac{\fs_i \red{\chi_i}}{\mb}\,,
\eeq
with the last equality being valid for $\alpha\lesssim 0.1$.

A more exact value for this quantity may be obtained by numerically solving a set of difference equations, e.g.~Eqs.~(17)--(21) in \cite{Brito2014}, assuming a quasi-adiabatic evolution.
If superradiant growth is the dominant factor, so that we can ignore other processes like GW emission and accretion, these are just
\begin{subequations} \label{eq:difeqs}
  \begin{align}
    &\dot{\mbh} =  - \Gamma_{\jb\lb\mb\nr} \mc\, , \\
    &\dot{\mc} = -\dot{\mbh}\, , \\
    &\dot{J} = -m \Gamma_{\jb\lb\mb\nr} \omega_{\np}^{-1} c^2 \mc\, \\
    &\dot{J_c} = - \dot{J}\, ,
  \end{align}
\end{subequations}
where dots indicate time derivatives, $(\mbh,\, J)$ and $(\mc,\, J_c)$ are the instantaneous mass and angular momentum for the BH and boson cloud respectively.
In the following sections, we will use the exact value for $\mc$ computed this way to characterize the signal.
In any case, it may be shown that the boson cloud may reach a size of at most $\mc \approx 0.1\times M_i$~\cite{East:2017ovw,Herdeiro:2017phl}.

The time it takes a single-level cloud to grow to its full size is simply the time it takes the BH to reach the spin of \eq{final_spin}.
In the absence of significant interaction with the environment (e.g.~through accretion or strong gravitational-wave emission), this is inversely linked to the ``instability timescale'' implied by of \eq{growth}, namely
\beq \label{eq:tinst}
\tinst \equiv 1/\Gamma_{\jb\lb\mb\nr}\, .
\eeq
This corresponds to an $e$-folding in the occupation number of level $(\jb,\lb,\mb,\nr)$.
In the nonrelativistic limit ($\fs\ll 1$), $\tinst$ can be approximated for the dominant scalar level ($\lb=\mb=1$, $\nr=0$) by \cite{Brito2017}
\beq \label{eq:tinst_scalar}
\tinst^{\rm (s)} \approx 27\, {\rm days} \left(\frac{\mbh}{10\, \msun}\right) \left(\frac{0.1}{\fs}\right)^9 \frac{1}{\chi_i}\, ,
\eeq
which is generally slower than the timescale for the dominant vector level ($\jb=\mb=1$, $\nr=0$), approximated by \cite{Baryakhtar2017}
\beq \label{eq:tinst_vector}
\tinst^{\rm (v)} \approx 2\, {\rm minutes} \left(\frac{\mbh}{10\, \msun}\right) \left(\frac{0.1}{\fs}\right)^7 \frac{1}{\chi_i}\, ,
\eeq
in the same small-$\fs$ limit.

As it turns out, gravitational interactions between levels might prevent the simultaneous population of more than one state \cite{Arvanitaki2011}.
Whether this is true or not, the hierarchy of involved timescales also suggests that we need only consider single-level clouds.
For instance, for most parameters, the $\lb=\mb=2$ scalar level has an instability timescale larger or comparable to the depletion timescale of the fastest-growing level $\lb=\mb=1$, so we should expect the latter to be unoccupied by the time the former reaches any significant size.
The same holds for vectors, as discussed in~\cite{Baryakhtar2017}.
This means that we need only consider a single level at a time: as soon as the BH is born, the fastest-growing level will be quickly populated; %
because the field has integer spin, there is no limit to how many particles can occupy it, and the growth will continue until the BH reaches the spin of \eq{final_spin}; %
the next level, with second-largest $\Gamma$, will begin to grow only after the first one is depleted (through gravitational wave emission, as discussed below, or any other reason).

All of the results in this section were obtained from perturbative analyses that consider the nonrelativistic behavior of a boson field over a static Kerr background.
These do not take into account effects like back-reaction of the field onto the background metric, gravitational wave emission by the boson condensate, or interaction between energy levels.
However, the validity of Eqs.~\eqref{eq:energy_levels}--\eqref{eq:growth} is confirmed by full numerical relativity simulations for the case of vector fields \cite{Witek2013,East:2017ovw,East2017,East:2018glu}.

\subsection{Gravitational-wave emission}
\label{sec:theory_gws}

Once the boson cloud has reached a macroscopic size, it will emit a significant amount of gravitational radiation.
There are three main mechanisms by which this may happen:
({i}) emission due to annihilation of bosons into gravitons;
({ii}) boson transitions between energy levels, analogous to electron jumps in the hydrogen atom;
and ({iii}) abrupt collapse of the cloud due to particle self-interactions (``bosenova'').

Due to the high occupation numbers involved, the first two processes can be described purely classically, with GW emission stemming from a time-varying quadrupole (and higher-multipoles, to a lesser degree) in the cloud's stress energy.
Transitions only become important if more than one level is occupied with comparable numbers.
Therefore, it could take over thousands of years after the birth of the BH for such a signal to become detectable~\cite{Arvanitaki2015,Arvanitaki2017}, making transitions interesting for very old BHs only.
Unfortunately, the typical duration of transition signals would be of order years or shorter~\cite{Arvanitaki2015}, which makes their observation from old potential sources highly unlikely.
Meanwhile, bosenovae are only relevant in the presence of large boson self-interactions~\cite{Yoshino:2012kn,Yoshino:2015nsa}; in particular, they are not expected to occur for the QCD axion~\cite{Arvanitaki2015}.
In any case, the typical duration of bosenova signals would be of the order of milliseconds and the numerical simulations required to produce their waveforms are still in their infancy~\cite{Yoshino:2012kn,Yoshino:2015nsa}---this makes bosenovae possibly relevant for unmodeled-burst analyses (e.g., \cite{Klimenko:2008fu,Klimenko:2015ypf,Cornish2014}), rather than the continuous-wave searches we are concerned with.

Given the above considerations, we restrict ourselves to annihilation signals, which are the best understood and most relevant for ground-based detectors.
To understand this, consider a BH that has been maximally spun down due to the growth of the boson cloud surrounding it.
We should expect this cloud to be composed of a vast number of particles in a coherent state corresponding to the fastest-growing energy level, as determined by \eq{growth}.
Indeed, we may think of the cloud as a macroscopic object with particle density given by the norm of the boson wavefunction plus a time-varying component proportional to $\alpha^2$, and rotating with angular frequency $\ob_{\np} = \mb \obh$, as implied by saturation of \eq{sr}. Treating this object purely classically, we may then expect the cloud to radiate gravitational waves at twice its rotational frequency \cite{Thorne1987}. We note however that, unlike typical GW sources, the radiation wavelength is always smaller than the cloud's extent and therefore the assumptions behind  the quadrupole approximation do not apply~\cite{Arvanitaki2011}.

A detailed description of the cloud's gravitational-wave emission can instead be obtained by using the Teukolsky formalism to solve the linearized Einstein equations for the cloud's stress energy, given by the wavefunction of the relevant quantum state \cite{Yoshino2014,Brito2017}.
A purely monochromatic boson field has a stress-energy tensor proportional to $e^{-2 i\ob_{\np}t}$ therefore one indeed finds that the cloud emits gravitational waves with angular frequency
\beq \label{eq:fgw}
\ogw = 2 \ob_{\np}\, ,
\eeq
and that the emission pattern is described by a set of spin-weighted spheroidal harmonics corresponding to the spin of the final BH, with azimuthal numbers $\lgw \geq 2 \jb$, and magnetic quantum number {\em fixed} to $\mgw = 2 \mb$ (see \sect{signal_waveform}).
Both vectors and scalars emit GWs with the same angular pattern (for a given $\jb$ and $\mb$), with the fastest-growing level radiating mostly in the $\lgw=\mgw=2$ mode.

The gravitational power radiated in each angular mode $\lgw$ may be written as
\beq \label{eq:power}
\dot{E}_{\rm GW}(\lgw,\mgw,\ogw) = \frac{1}{4\pi} \frac{c^{5}}{G} \left(\frac{c}{\rg \ogw}\right)^{2} \left(\frac{\mc}{\mbh_f}\right)^2\, \ampfactor_{\lgw\mgw}^2(\fs,\chi_i) \, ,
\eeq
where $\mbh_f$ and $\rg=G\mbh_f/c^2$ are respectively the mass and lengthscale of the \emph{final} BH, and $\ampfactor_{\lgw\mgw}(\fs,\chi_i)$ is a \red{dimensionless} factor.
For scalars, we compute the $\ampfactor_{\lgw\mgw}$'s numerically using BH perturbation theory as in \cite{Brito2017},%
\footnote{Note that $\ampfactor=|Z|\times(\mbh^2/\mc)$, for $|Z|$ as defined in \cite{Brito2017}.}
but this is not currently feasible for vectors.
Regardless of boson spin, as long as \eq{sr2} is satisfied, the emitted power will be a steep function of $\alpha$ \cite{Arvanitaki2015,Yoshino2014,Brito2017}.
In fact, for small $\alpha$, the power emitted in the dominant angular mode ($\lgw=2$) by the fastest-growing scalar level ($\lb=\mb=1$, $\nr=0$) may be roughly approximated as \cite{Brito2014}
\beq \label{eq:power_scalar}
\dot{E}_{\rm GW}^{\rm (s)} \approx 7\times10^{41}\, {\rm erg/s} \left(\frac{\fs}{0.1}\right)^{16} \chi_i^2\, ,
\eeq
and for the fastest-growing vector level ($\jb=\mb=1$, $\lb=\nr=0$) as \cite{Baryakhtar2017}
\beq \label{eq:power_vector}
\dot{E}_{\rm GW}^{\rm (v)} \approx 2\times10^{49}\, {\rm erg/s} \left(\frac{\fs}{0.1}\right)^{12} \chi_i^2\, .
\eeq
The difference in the $\fs$ dependence in these two expressions arises from the fact that the fastest vector level has no orbital angular momentum ($\lb=0$), and so lies closer to the BH yielding a more compact cloud.

The energy in the gravitational radiation is drawn from the cloud itself, which slowly fades away as its component particles annihilate into gravitons \cite{Arvanitaki2011,Arvanitaki2015}.
As a result, the signal will be almost monochromatic, with a slowly decreasing amplitude and slowly increasing frequency.
The evolution of both these quantities is tied to the timescale implied by \eq{power},
\beq \label{eq:tgw}
\tgw \equiv \mc c^2 / \dot{E}_{\rm GW}\, .
\eeq
This ``gravitational-wave timescale'' is just the time it takes for half of the rest-energy of the cloud to be radiated away, and can be thought of as the typical duration of the signal. 
Using the approximations of \eq{mc} and \eq{power_scalar}, we get for the dominant scalar level in the nonrelativistic limit, a signal duration of
\beq \label{eq:tgw_scalar}
\tgw^{\rm (s)} \approx 6.5 \times 10^{4}\, {\rm yr} \left(\frac{\mbh}{10\, \msun}\right) \left(\frac{0.1}{\fs}\right)^{15}\hspace{-2pt} \frac{1}{\chi_i} .
\eeq
Similarly, using \eq{power_vector}, for the dominant vector level we get
\beq \label{eq:tgw_vector}
\tgw^{\rm (v)} \approx 1\, {\rm day} \left(\frac{\mbh}{10\, \msun}\right) \left(\frac{0.1}{\fs}\right)^{11}\hspace{-2pt} \frac{1}{\chi_i} .
\eeq
Clearly, the vector processes tend to take place at a much faster pace than scalar ones, as expected from the higher radiated power.
In both cases, however, the duration of the signal is significantly longer than the time it takes for the cloud to grow, as given by \eq{tinst}.
This an important, general feature that justifies the separate treatment of the early growth and late emission stages in the first place.

\section{The signal}
\label{sec:signal}

Having reviewed the physics of boson clouds around BHs in \sect{theory}, we will now focus on the properties of the gravitational signal produced by one of these systems, as seen by differential-armlength detectors on the ground.
Given that we only expect one quantum state to be significantly populated at any given time, we will restrict our discussion to signals from single-level clouds.
In any case, the signal from a multilevel cloud can be produced trivially by the addition of several single-level waveforms described below, assuming negligible interaction between levels.
We describe the signal morphology in \sect{signal_waveform} and elaborate on the most salient features in \sect{discuss}.

\subsection{Waveform}
\label{sec:signal_waveform}

Consider a cloud made up of bosons in a single quantum state that has just stopped growing, after drawing enough energy and angular momentum from its host BH to saturate \eq{sr}.
In that case, as anticipated in \sect{theory_gws}, we expect the cloud to emit a continuous gravitational signal with a small spin-up and amplitude depending on the properties of both BH and boson.
The strain signal, $h_I$, seen by a given differential-armlength detector, $I$, can be written in the usual form as a sum over polarizations,
\beq
h^I(t) = F^I_+(t)\, a_+ \cos \phi(t) + F^I_\times(t)\, a_\times \sin \phi(t) ,
\eeq
where the $F^I_p$'s are the antenna-response functions of detector $I$ to signals of plus ($+$) and cross ($\times$) polarizations (see, e.g., Appendix B in \cite{Anderson2001} for explicit expressions).
These depend implicitly on the relative location and orientation of the detector and the source, usually parametrized by its right ascension ($\ra$), declination ($\dec$), and polarization angle ($\psi$).
This last parameter determines how the frame in which the polarizations are defined is oriented in the plane of the sky; 
for our purposes, this will be the angle between the spin of the BH and the projection of the celestial north onto the plane normal to the line of sight.

The polarization amplitudes, $a_+$ and $a_\times$, are made up of contributions from several angular multipoles, indexed by the wave azimuthal number, \red{$\lgw\geq2\jb$}, and with fixed magnetic number, $\mgw = 2 \mb$,
\beq \label{eq:polamps}
a_{+/\times} = - \sum_{\lgw \geq 2 l}  h_{0}^{(\lgw)}  \left[ _{-2}S_{\lgw \mgw \ogw} \pm {}_{-2}S_{\lgw -\mgw -\ogw} \right] ,
\eeq
with the plus (minus) sign on the right-hand side corresponding to the $+$ ($\times$) polarization \cite{Brito2017}.
Because the boson cloud does not emit GWs isotropically, these amplitudes depend on the orientation of the source relative to the detector.

The angular dependence is encoded in \eq{polamps} via the {\em spin-weighted spheroidal harmonics} \cite{Teukolsky1973}, which are analogous to the usual spin-weighted spherical harmonics but account for the nonsphericity of the space around the Kerr BH.
As such, these are functions of BH spin, signal frequency and orientation with respect to the source:
\beq 
{}_{s}S_{\lgw \mgw \ogw} \equiv {}_{s}S_{\lgw \mgw} (a\ogw/c,\cos\iota) \, ,
\eeq
where for us the spin weight is always $s=-2$, as needed to describe GWs.
The inclination $\iota$ is defined as the angle between the BH spin and the line of sight.
We compute these eigenfunctions numerically using Leaver's method \cite{Leaver1985, Berti2006}.

The characteristic amplitude of each mode can be written as
\beq \label{eq:h0}
h_0^{(\lgw)} = \frac{c^4}{G} \frac{M_c}{M_{\rm BH}^2} \frac{1}{2\pi^2 \fgw^2 r} \ampfactor_{\lgw \mgw}(\fs, \chi_i), \, 
\eeq
where the dimensionless factor $\ampfactor_{\lgw \mgw}$ encodes the relative amount of energy that the source deposits in each mode, as in \eq{power}.
Assuming the BH has been fully spun down by the cloud, this is only a function of the initial BH spin and the fine-structure constant $\fs$, and can be computed numerically from BH perturbation theory following \cite{Brito2017}.
In the nonrelativistic limit ($\fs\ll 1$), this can be approximated by \cite{Yoshino2014,Brito2014,Arvanitaki2017} 
\beq \label{eq:h0_scalar_approx}
h_0^{\rm (s)} \approx 8 \times 10^{-28} \left(\frac{M}{10 \msun}\right)
\left(\frac{\alpha}{0.1}\right)^7 \left(\frac{\rm Mpc}{r}\right) 
\left(\frac{\chi - \chi_f}{0.1}\right)
\eeq
for the dominant scalar mode ($\lb=\mb=1$, $\nr=0$, $\lgw=\mgw=2$), and by \cite{Baryakhtar2017}
\beq \label{eq:h0_vector_approx}
h_0^{\rm (v)} \approx 4 \times 10^{-24} \left(\frac{M}{10 \msun}\right)
\left(\frac{\alpha}{0.1}\right)^5 \left(\frac{\rm Mpc}{r}\right) 
\left(\frac{\chi - \chi_f}{0.1}\right)
\eeq
for the dominant vector mode ($\jb=\mb=1$, $\nr=0$, $\lgw=\mgw=2$), corresponding to the approximations in \eq{power_scalar} and \eq{power_vector} respectively.
Equations \eqref{eq:h0_scalar_approx} and \eqref{eq:h0_vector_approx} include an explicit spin-dependent correction factor to account for the dependence of the mass cloud on $\alpha$, with $\chi_f$ itself a function of $\alpha$ defined in \eq{final_spin}.
Although not fully accurate, these expressions will be useful when studying the scalings of the expected signal amplitude---especially in the case of vectors, for which the $\ampfactor$ factors have yet to be computed numerically.

In both the scalar and vector cases, the amplitude of the signal will decrease as a function of time, starting from the peak values given by \eq{h0}.
This is due to the progressive dissipation of the boson cloud sourcing the GW signal.
This weakening occurs over a timescale of the order of the signal duration, \eq{tgw}.

In a frame inertial with respect to the source, the phase evolution of the signal corresponds to a simple monotone that may potentially evolve slowly in frequency.
In the frame of the detector, extra timing corrections are needed, so that, in terms of the time $t$ measured at Earth, the phase evolution can be written as
\beq \label{eq:phase}
\phi (t) = 2\pi \sum_{j=0}^N \frac{\partial_t^{(j)} \fgw}{(j+1)!} \left[t-t_0+\delta t(t) \right]^{(j+1)} + \phi_0\, ,
\eeq
where $\partial_t^{(j)} \fgw$ is the $j$\ts{th} time derivative of $\fgw$, the GW frequency measured at fiducial time $t_0$, and $\phi_0$ is a phase offset.
The timing corrections $\delta t (t)$ account for delays due to the relative motion of the source and detector, general- and special-relativistic effects, as well as potential corrections due to the presence of a companion if the source is part of a binary (e.g.~\cite{Blandford1976,Taylor1989, Riles2017}).

As implied by \eq{fgw}, the source-frame frequency will be given by $\fgw = \ogw/ (2\pi) = \ob_{\np} / \pi$.
Noting that $\ob_{\np}\approx \ob_\mu$ by \eq{energy_levels}, this may be approximated as a function of BH mass and fine-structure constant as
\beq \label{eq:fgw_approx}
\fgw \approx \frac{\fs}{\pi\rg} \approx 645\, {\rm Hz} \left(\frac{10 \msun}{\mbh}\right) \left(\frac{\alpha}{0.1}\right) .
\eeq
This means that stellar-mass BHs should support boson clouds that emit gravitational waves at frequencies within the sensitive band of existing and planned ground-based detectors.

Once the superradiance instability has shut down, the GW signal will expect a (slight) positive change in frequency (a ``spinup'').
This is expected on purely classical grounds, as is typical of any gravitationally-bound system (as in the characteristic ``chirp'' of compact-binary coalescences).
The value of $\partial_t^{(1)} f \equiv \dot f$ can be computed from the rate of change in the cloud's binding energy \cite{Baryakhtar2017}.
For scalars, \eq{power_scalar} implies a signal frequency derivative (see Appendix \ref{app:fdot})
\beq \label{eq:fdot_scalar}
\dot \fgw^{\rm (s)} \approx 3\times 10^{-14}\, {\rm Hz/ s} \left(\frac{10M_{\odot}}{M}\right)^2 \left(\frac{\alpha}{0.1}\right)^{19} \chi_i^2\, ,
\eeq
while, for vectors \eq{power_vector} implies \cite{Baryakhtar2017}
\beq \label{eq:fdot_vector}
\dot \fgw^{\rm (v)} \approx 1\times 10^{-6}\, {\rm Hz/ s} \left(\frac{10M_{\odot}}{M}\right)^2 \left(\frac{\alpha}{0.1}\right)^{15} \chi_i^2\, .
\eeq
The frequency drift is faster for vectors, as corresponds to quicker cloud dissipation.
Besides this spinup, we will also want to allow for higher-order derivatives of the frequency in \eq{phase} to incorporate potential perturbations caused by the astrophysical environment, presence of a companion, level interactions, or theoretical uncertainty.

\subsection{Projected properties}
\label{sec:discuss}

\newcommand{\mbhex}{\red{60~M_\odot}}
\newcommand{\chiex}{\red{0.70}}
\newcommand{\dex}{\red{5 ~{\rm Mpc}}}
\newcommand{\alphaex}{\red{0.179}}
\newcommand{\muex}{\red{4\times10^{-13}\, {\rm eV}}}
\newcommand{\fex}{\red{191\, {\rm Hz}}}
\newcommand{\hEx}{\red{5.2\times10^{-26}}}

We are interested in making statements about the presence of ultralight bosons based on searches for GW signals from known BHs.
This means that we need to know what strain frequencies and amplitudes we may expect from our target BH, without knowing the true mass of the boson (if it exists).
In principle, a particular BH could support clouds for a range of boson masses, which we will parametrize implicitly via the fine-structure constant, $\fs$ of \eq{alpha}.
Although a particular single-level cloud is expected to emit a GW quasimonotone, an unknown $\fs$ means that we could expect signals at a variety of frequencies.
This is clear from \eq{power_scalar}, in the small-$\fs$ limit, or more generally from the fact that the $\ampfactor_{\lgw\mgw}$ factors in \eq{h0} will be nonzero for a range of $\fs$'s.
In other words, a given BH could ``resonate'' with different bosons, allowing us to probe a (narrow) range of particle masses; hence there is a {\em band} of signal frequencies to be potentially expected from a given BH.

In this section, we study in detail the properties of continuous signals that can be expected from clouds around a given BH.
Although so far we have kept the discussion general, we now focus on scalars to provide concrete examples.
We use numerical techniques to compute the power emitted by different systems, obtaining estimates that should be more reliable than previously published projections.
In particular, for each value of $\fs$ and initial BH parameters, we numerically solve the differential equations governing the evolution of the cloud to obtain the final BH parameters, as in \cite{Brito2014}.
We then use the numerical results of \cite{Brito2017} for the $\ampfactor_{\lgw\mgw}$ factors to compute the radiated amplitude by means of \eq{h0}.
Unfortunately, at the moment, it is not possible to do this for vectors, since the corresponding perturbative calculations are significantly more difficult and have yet to be carried out.

\begin{figure}
\centering
\includegraphics[width=\columnwidth]{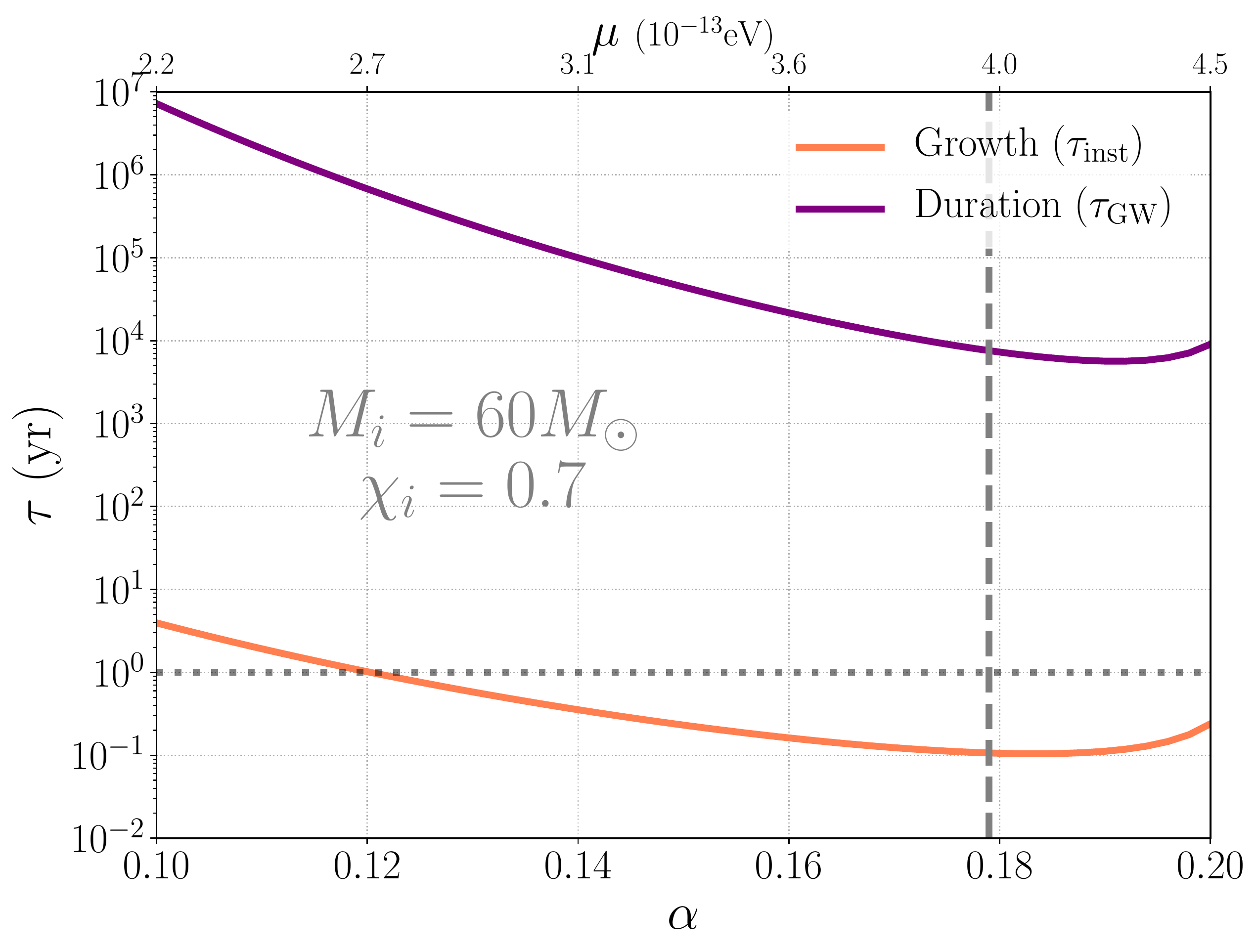}
\caption[Signal growth and duration timescales for example BH (scalar cloud)]{{\em Signal growth and duration timescales for example BH (scalar cloud)}.
Curves show the signal duration (purple, top), \eq{tinst}, and growth (orange, bottom), \eq{tgw}, timescales for a scalar cloud ($\lb=\mb=1,\, \nr=0$) as a function of the fine-structure constant $\fs$ from \eq{alpha}.
The BH is assumed to have an initial mass of $\mbh=\mbhex$ and spin of $\chi=\chiex$.
The vertical dashed line marks the value $\fs$ that yields peak emission for such BH, for which $\tinst = 39$ days and $\tgw = 7.5\times10^{3}$ yr.
Note that values of $\fs>0.2$ preclude superradiance given this spin, cf.~\eq{sr2}.}
\label{fig:tau_alpha}
\end{figure}

To summarize the key points from the discussion below: any given BH can allow us to probe a narrow range of boson masses set by its mass and, to a lesser extent, its spin (Figs.~\ref{fig:h0_fgw_alpha}--\ref{fig:fgw_chi_h0});
heavier BHs will ``resonate'' with lighter bosons and produce louder signals at lower frequencies (\fig{peak_mbh_chi}); and signals from heavier BHs (lighter bosons) will both grow and vanish more slowly, resulting in smaller frequency derivatives (Figs.~\ref{fig:tau_alpha}--\ref{fig:alpha_mbh_tgw}).
We expect similar conclusions to hold for vectors, except that the overall radiated power will be stronger and the timescales shorter, cf.~\eq{growth} and \eq{power_vector}.

\subsubsection{Amplitude and frequency}
\label{sec:discuss_amplitude}

\begin{figure}
\centering
\includegraphics[width=\columnwidth]{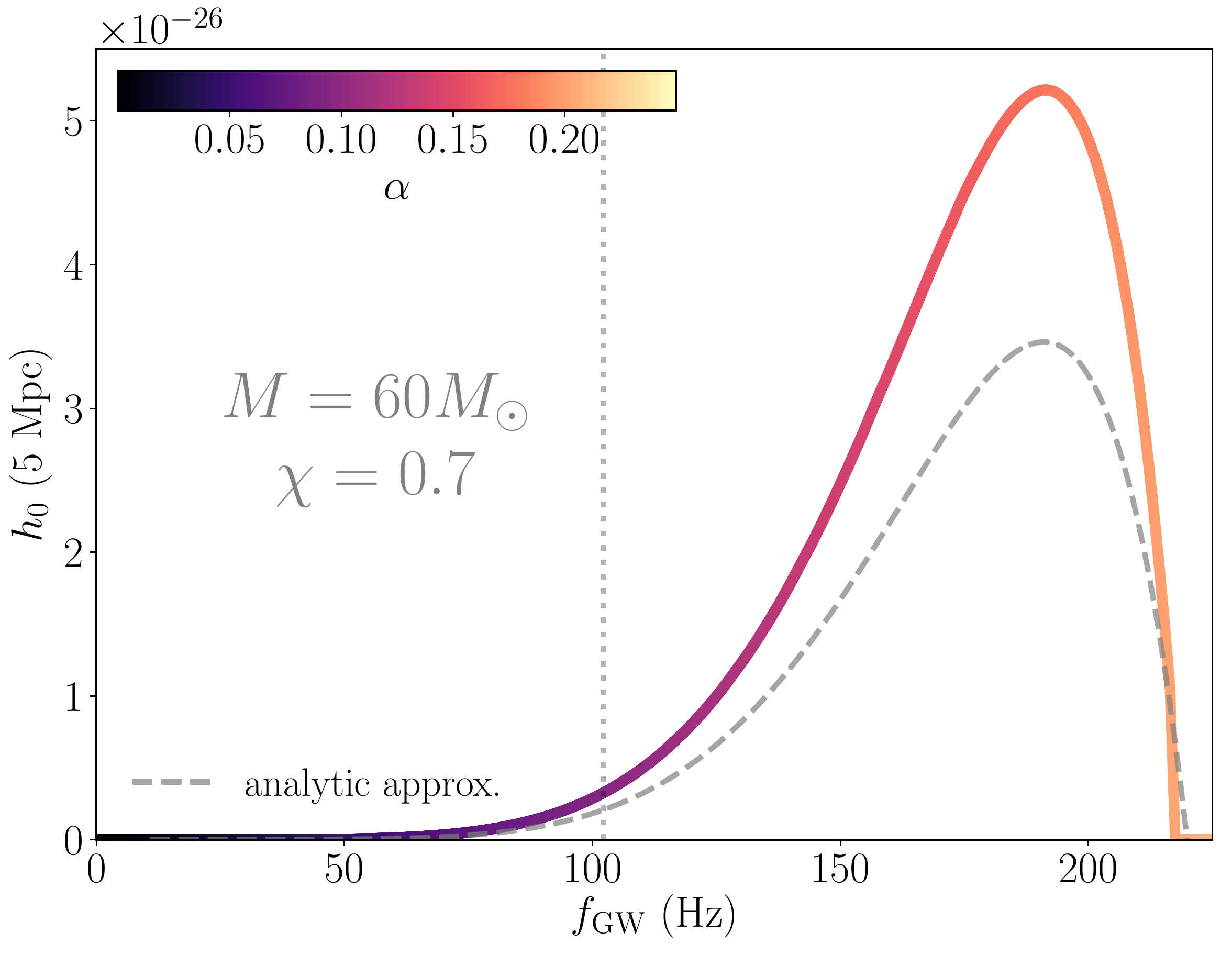}
\caption[Strain amplitude vs frequency for example black hole]{{\em Strain amplitude vs frequency for example black hole (scalar cloud)}.
The $x$-axis shows different frequencies at which we might expect a GW signal from a scalar ($\lb=\mb=1,\, \nr=0$) cloud around a BH with initial mass of $\mbh_i=\mbhex$ and initial spin $\chi_i=\chiex$.
The colored curve shows the corresponding characteristic amplitude, $h_0$ in \eq{h0} assuming $r=\dex$, parametrized by the fine-structure constant, $\fs$ in \eq{alpha}, as indicated by the colorbar.
For reference, the other curve shows the small-$\fs$ approximation of \eq{h0_scalar_approx}, including the spin correction responsible for the amplitude turnover.
Points to the left of the vertical dotted line have $\tinst>1\, {\rm yr}$.}
\label{fig:h0_fgw_alpha}
\end{figure}

Begin with the example of a BH with initial mass $\mbh = \mbhex$ and dimensionless spin $\chi=\chiex$, parameters consistent with the remnant from LIGO's first detection \cite{gw150914}.
Assume that the BH is then maximally spun down by the presence of a \emph{scalar} cloud, and consider the amplitude of GWs emitted immediately after. 
We will assume the cloud is dominated by the fastest-growing energy level ($\lb=\mb=1$, $\nr=0$), and restrict ourselves to the dominant GW mode ($\lgw=\mgw=2$).
For concreteness, place the source 5 Mpc away and consider the amplitude of the signal as seen from Earth for different values of $\fs$.
Our BH could potentially support clouds emitting GWs at different frequencies and with different amplitudes, depending on the true value of the boson mass.

The above fact is clearly illustrated in \fig{h0_fgw_alpha}, which shows the characteristic amplitude of waves produced by clouds around our example BH for different initial values of $\fs$. 
In the best case scenario for this BH, if there existed a boson with $\mu \approx \muex$ such that $\fs\approx \alphaex$, then we would observe a signal with characteristic strain amplitude $h_0 \approx \hEx (5\,{\rm Mpc}/r)$ at $\fgw\approx\fex$, corresponding to the peak in \fig{h0_fgw_alpha}.
For this value of the boson mass, at the end of the superradiant process the BH will have reached a final spin of $\chi=0.62$, having lost $1.7\%$ of its mass to the cloud.

\fig{h0_fgw_alpha} also implicitly defines the range of frequencies of interest for searches directed at this BH to be, say, within \red{150} Hz and \red{200} Hz.
This range could be broader or narrower, depending on the sensitivity of the search and how long one waits from the birth of the BH to make an observation.
For instance, points to the left of the dotted gray line correspond to boson masses for which the signal would take longer than 1 year to complete one $e$-folding in amplitude (more on timescales in \sect{discuss_timescales} below).
According to \eq{sr2}, the maximum value of $\fs$ for this source is ${\sim}0.2$, for which the amplitude vanishes.
Note that, because $\mu$ depends linearly on $\fs$, this means that any given BH will allow us to probe a very narrow range of boson masses.

\begin{figure}%
\centering
\includegraphics[width=\columnwidth]{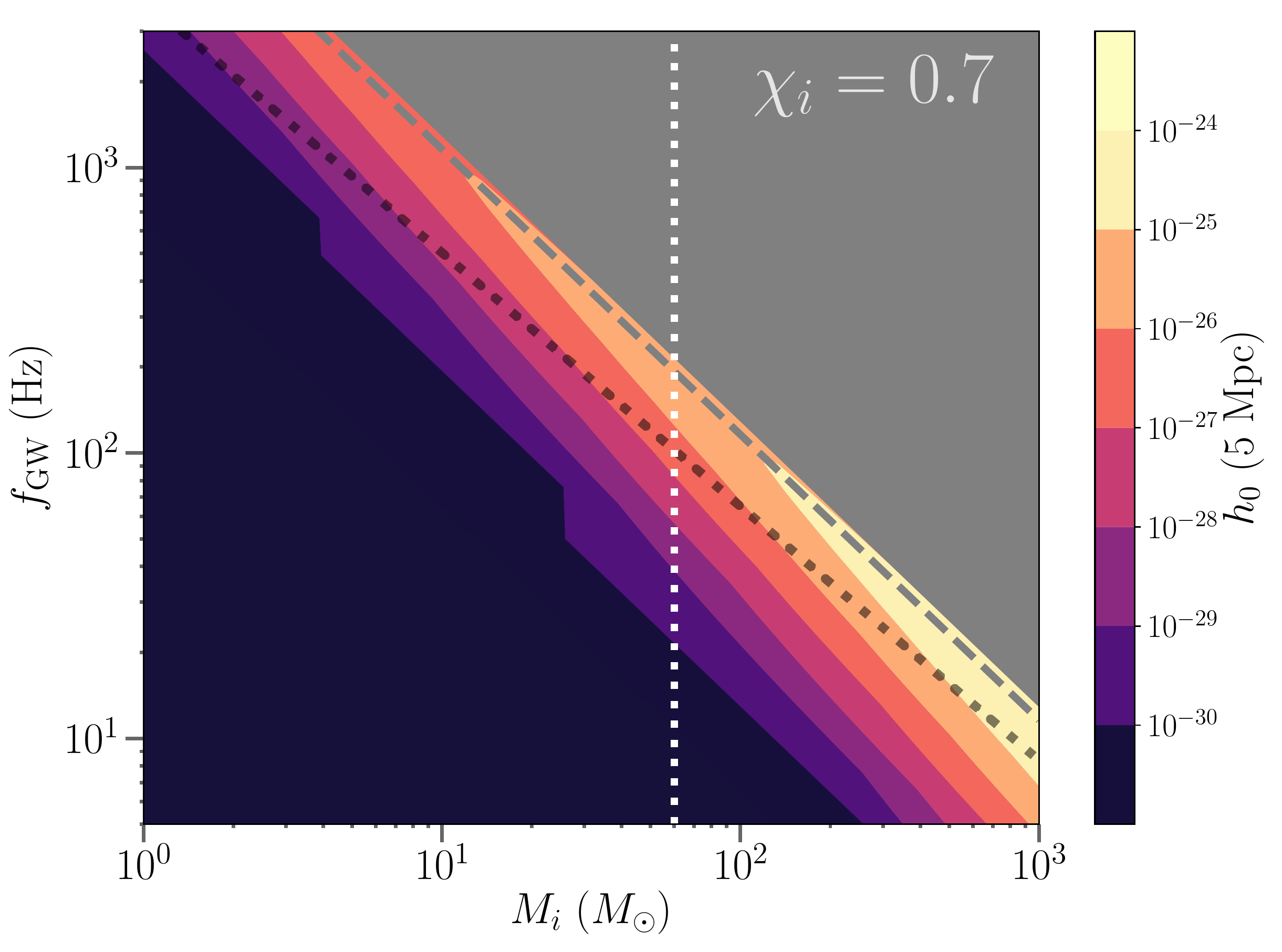}
\caption[Strain amplitude at different GW frequencies vs initial BH mass for fixed initial spin (scalar cloud)]{{\em Strain amplitude at different GW frequencies vs initial BH mass for fixed initial spin  (scalar cloud)}.
Color shows the characteristic strain amplitude, \eq{h0}, from a scalar cloud ($\lb=\mb=1,\, \nr=0$) that would be emitted at different frequencies ($y$-axis), i.e.~for different $\fs$'s (not shown), vs initial BH mass ($x$-axis).
The gray dashed line marks the peak amplitude.
The source is assumed to lie at $r=\dex$, with initial spin $\chi_i=\chiex$.
For ease of display, we set an arbitrary lower cutoff of $h_0\geq\red{10^{-30}}$: amplitudes for points in the bottom \red{purple} region vanish asymptotically for lower $\fgw$, while amplitudes for points in the top \red{gray} region vanish identically since superradiance cannot occur for those parameters.
Points below the dotted gray line have $\tinst > 1\,{\rm yr}$.
A vertical cross-section at $\mbh=\mbhex$ (vertical dotted line) yields \fig{h0_fgw_alpha}.}
\label{fig:fgw_mbh_h0}
\end{figure}

While the overall shape of the curve in \fig{h0_fgw_alpha} will be generally the same for all BHs, the location and width of the peak will be a strong function of the initial BH mass and spin.
This is represented in \fig{fgw_mbh_h0}, in which we have fixed the BH distance and spin to the values above, but allowed its mass to vary.
Color in this figure represents the strain amplitude of GWs emitted at a given frequency ($y$-axis) as a function of initial BH mass ($x$-axis), while the gray dashed line marks the peak amplitude for a given BH mass.
Although $\fs$ is not explicitly shown, it should be clear from \fig{h0_fgw_alpha} that moving vertically toward higher frequencies and amplitudes corresponds to increasing $\fs$.
In fact, a vertical cut of \fig{fgw_mbh_h0} at $\mbh_i=\mbhex$ would yield \fig{h0_fgw_alpha}.

\begin{figure}
\centering
\includegraphics[width=\columnwidth]{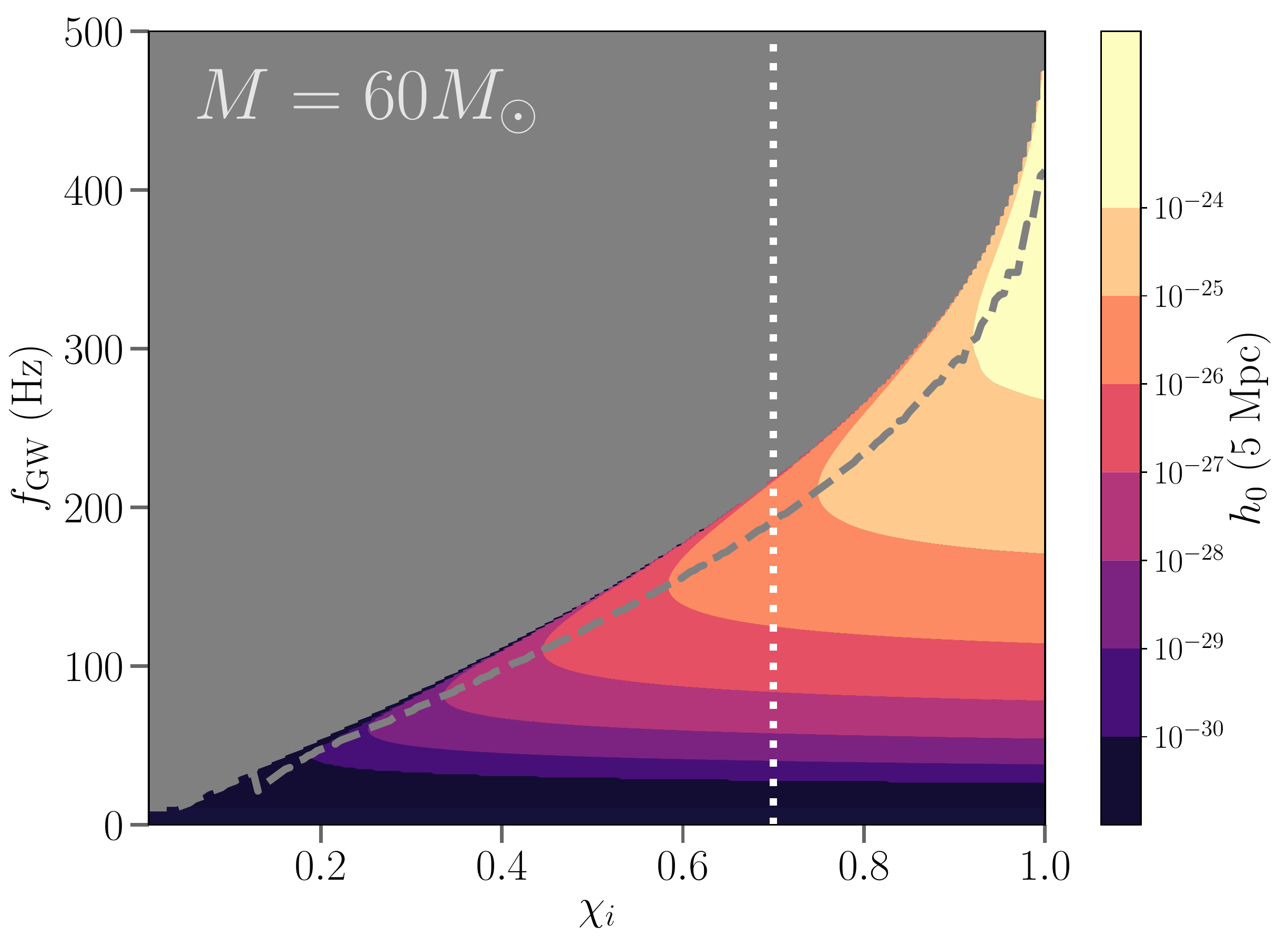}
\caption[Strain amplitude at different GW frequencies vs BH \red{initial} spin for fixed mass (scalar cloud)]{{\em Strain amplitude at different GW frequencies vs BH \red{initial} spin for fixed mass (scalar cloud)}.
Color shows the characteristic strain amplitude, \eq{h0}, from a scalar cloud ($\lb=\mb=1,\, \nr=0$) that would be emitted at different frequencies ($y$-axis), i.e.~for different $\fs$'s (not shown), vs BH \red{initial} spin ($x$-axis).
The gray dashed line marks the peak amplitude.
The source is assumed to lie at $r=\dex$, with mass $\mbh=\mbhex$.
For ease of display, we set an arbitrary lower cutoff of $h_0\geq\red{10^{-30}}$: as can be inferred from \fig{h0_fgw_alpha}, amplitudes for points in the bottom \red{purple} region vanish asymptotically for lower $\fgw$, while amplitudes for points in the top \red{gray} region vanish identically since superradiance cannot occur for those parameters.
A vertical cross-section at $\chi_i=\chiex$ (vertical dotted line) yields \fig{h0_fgw_alpha}.}
\label{fig:fgw_chi_h0}
\end{figure}

\fig{fgw_mbh_h0} shows that heavier BHs can support clouds that emit GWs at lower frequencies but greater amplitudes.
This was expected from the discussion in \sect{theory_gws}: (i) heavier BHs are also larger, and so must be the boson cloud surrounding it, thus yielding lower GW frequencies; and (ii) heavier BHs result in a heavier cloud, as dictated by \eq{mc}, which will in turn radiate more strongly, per \eq{power}.
Because the overall radiated power decreases with BH mass, this also means that the band of detectable frequencies is narrower for lighter BHs, which is also visible in \fig{fgw_mbh_h0}.
The fact that the peak frequency (dashed line) decreases linearly with BH mass was already anticipated in \eq{fgw_approx}.
This can be understood from the observations that (i) $\fgw \sim \ob/\pi \sim 2/\lbc$ and that (ii) $\lbc \sim \rg$ for the boson and BH sizes to match and maximize superradiance.
As one moves vertically up the plot (increasing $\fgw$ or, equivalently, $\fs$), the emitted power vanishes abruptly at a point defined by the saturation of \eq{sr2};
in this case, because $\chi_i=\chiex$ and $\mb=1$, this corresponds to $\fs=\red{0.2}$.
Finally, as in \fig{h0_fgw_alpha}, clouds corresponding to points below the dotted gray line would take longer than 1 year from the birth of the BH to complete one $e$-folding during growth.

The properties of the GW emission will also vary with the initial spin of the BH.
This is illustrated in \fig{fgw_chi_h0}, in which we have fixed the BH distance and mass ($\mbh_i = \mbhex$), but allowed its initial spin to vary.
As in \fig{fgw_mbh_h0}, color represents the characteristic strain amplitude emitted at a given frequency ($y$-axis) for different values of the initial spin ($x$-axis); again, the dashed line traces the peak amplitude.
It is no surprise to find that BHs with greater initial spins yield louder GWs: the faster the BH is spinning before the superradiant process kicks off, the longer the cloud may grow without saturating \eq{sr} and, consequently, the more mass it will extract before its growth stalls.
The fact that higher-spin BHs result in heavier clouds is reflected in \eq{mc}, and the corresponding dependence of $h_0$ on $\chi_i$ can be glimpsed from \eq{power_scalar}, which is valid for $\fs\ll 1$.
Similarly, for a fixed initial BH mass, a faster initial spin results in a lighter final BH [see \eq{mc} again], and so yields higher GW emission frequencies per \eq{fgw_approx}.
As in \fig{fgw_mbh_h0}, the upper frequency cutoff is given by \eq{final_spin}.

\begin{figure}
\centering
\includegraphics[width=\columnwidth]{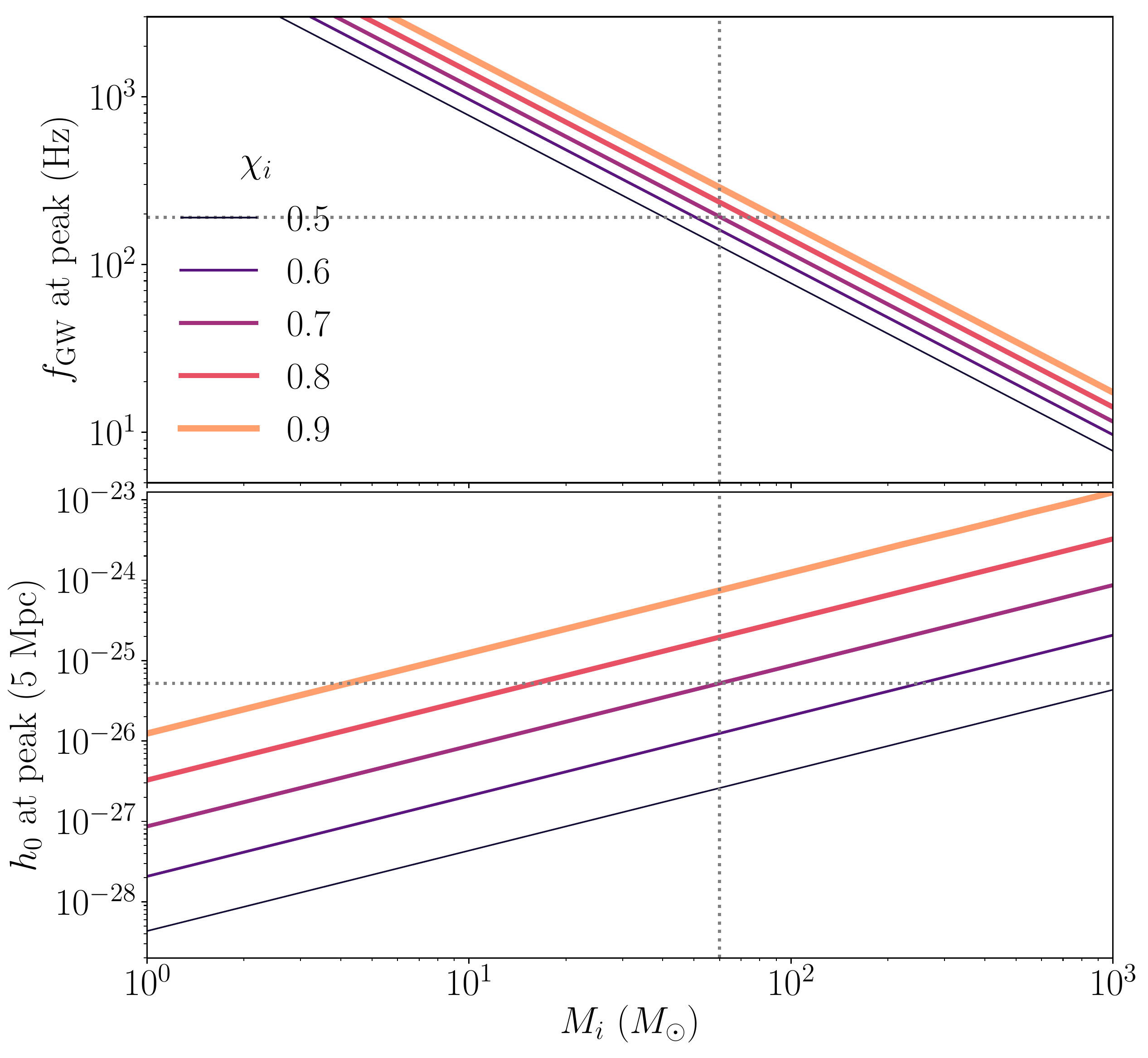}
\caption[Optimal strain frequency and amplitude vs initial BH mass for different initial spins (scalar cloud)]{{\em Optimal strain frequency and amplitude vs initial BH mass for different initial spins (scalar cloud)}.
Frequency (top) and characteristic amplitude (bottom) of the strain produced by the best-possible cloud (best-possible $\fs$) as a function of \red{initial} BH mass.
Different curves correspond to different initial spins, showing that higher spins result in stronger emission.
We assume that the source is situated at $r=\dex$, and that the scalar cloud is dominated by the fastest level ($\lb=\mb=1,\, \nr=0$).
The intersection of the $\chi_i=\chiex$ line with a vertical cut at $\mbh=\mbhex$ (dotted vertical line) give the amplitude and frequency of the peak in \fig{h0_fgw_alpha} (dotted horizontal lines).}
\label{fig:peak_mbh_chi}
\end{figure}

\begin{figure}
\centering
\includegraphics[width=\columnwidth]{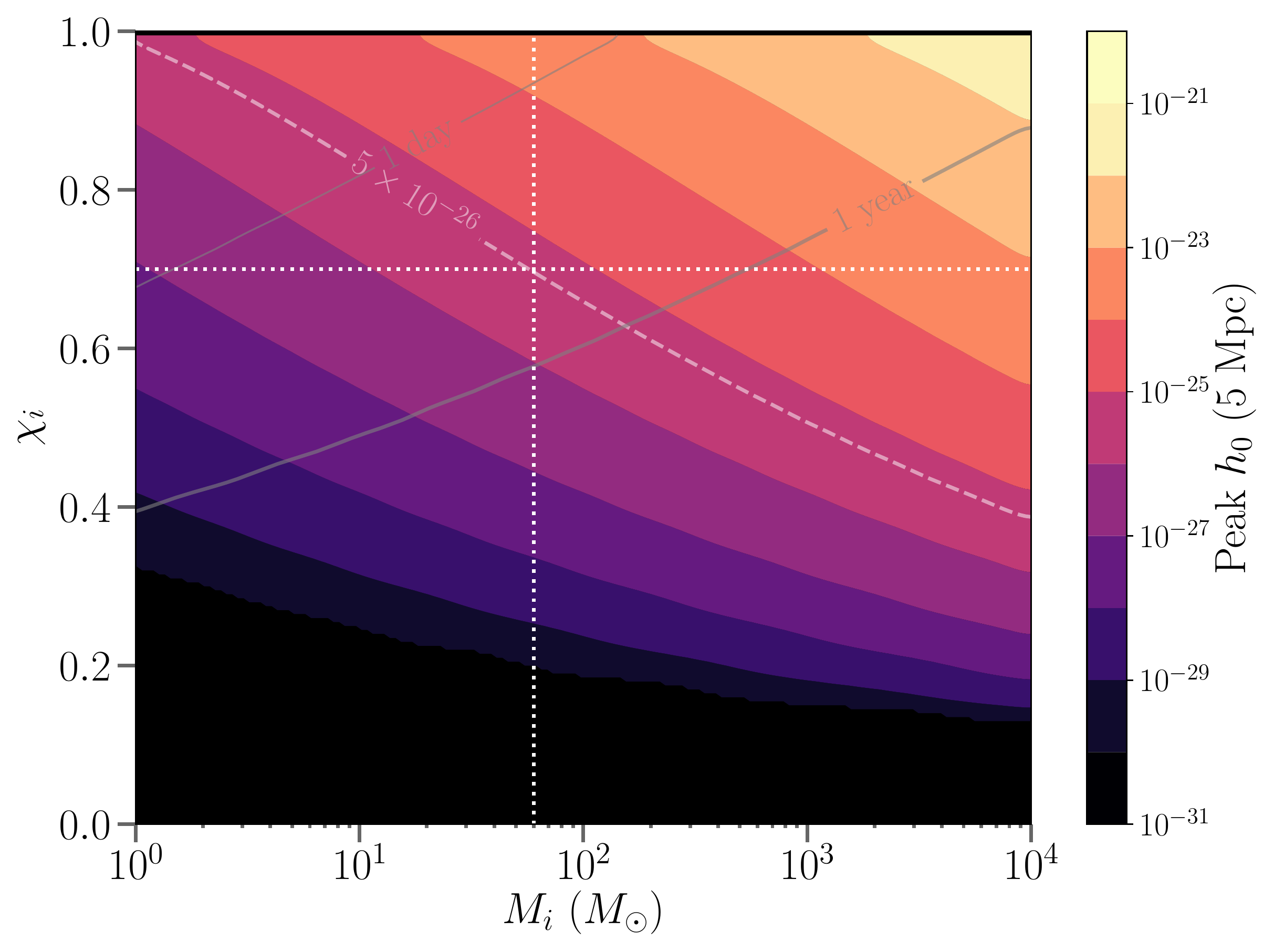}
\caption[Optimal strain amplitude vs initial BH parameters]{{\em Optimal strain amplitude vs initial BH parameters (scalar cloud)}.
Color gives the characteristic strain amplitude emitted by the best-possible cloud matched to a BH with the indicated initial mass ($x$-axis) and spin ($y$-axis).
Horizontal cuts yield the curves shown in the bottom panel of \fig{peak_mbh_chi}.
The intersection of the dotted white lines ($\mbh_i=\mbhex$, $\chi_i=\chiex$) corresponds to the peak of \fig{h0_fgw_alpha}.
Gray lines mark $\tinst=1\, {\rm day}$ and $\tinst=1\, {\rm yr}$ for reference.}
\label{fig:chi_mbh_h0}
\end{figure}

As suggested by \fig{fgw_mbh_h0} and \fig{fgw_chi_h0}, the characteristic GW amplitude emitted by a boson cloud as a function of frequency may show interesting structure as the initial BH mass and spin are varied.
However, in many situations, it suffices to know the expected amplitude of the {\em peak} emission from a given system.
This information is summarized in \fig{peak_mbh_chi}, which displays the characteristic amplitude and frequency for the optimal cloud as a function of BH mass, and for different values of the initial spin.
The curves in the bottom panel can be understood as constant-spin cuts of the full mass-spin plane shown in \fig{chi_mbh_h0}.
As for the other colormaps, the dotted white lines in that plot mark the values of our example BH ($\mbh_i=\mbhex$, $\chi_i=\chiex$), which can at best yield an amplitude of $h_0 = \hEx(\dex/r)$ (peak of \fig{h0_fgw_alpha}).
Gray lines mark representative values of the instability timescale of \eq{tinst} (see \sect{discuss_timescales}).
\fig{peak_mbh_chi} and \fig{chi_mbh_h0} once again reflects the fact that greater strains are obtained for heavier BHs with larger initial spins.
Some representative values are shown in Table~\ref{tab:example-bh_params}, where the bold row corresponds to the intersection of the dotted lines in \fig{fgw_mbh_h0}.

\begin{table}
\centering
  \caption{Parameters of optimal scalar cloud for representative BHs.
A ``k'' next to a value stands for ``$\times 10^3$''. The bold row corresponds to the intersection of the dotted lines in \fig{fgw_mbh_h0}.}
    \begin{ruledtabular}
    \begin{tabular}{c@{\quad} c@{\quad} c@{\quad} c@{\quad} c@{\quad} c@{\quad} c@{\quad} c@{\quad}}
    $\mbh_i$ & $\chi_i$ & $\mu$ & $\fs_i$ & $\fgw$ & $h_0$ & $\tinst$ & $\tgw$ \\
    {\tiny $\msun$} & & {\tiny $10^{-13}\,$eV} & & {\tiny Hz} & {\tiny $5{\rm Mpc}/r$} & {\tiny day} & {\tiny yr} \\
    \midrule
    3   & 0.90 & 122 & 0.273 & $5.8$k & $4\times10^{-26}$ & 0.1 & 2 \\
    10  & 0.90 & 36  & 0.273 & $1.7$k & $1\times10^{-25}$ & 0.3 & 6 \\
    {\bf 60}  & {\bf 0.70} & {\bf 4.0} & {\bf 0.179}  & $\mathbf{191}$ & $\mathbf{5 \times 10^{-26}}$ & {\bf 39} & $\mathbf{8}${\bf k} \\
    60  & 0.90 & 6.0 & 0.273 & $290$ & $7\times10^{-25}$ & 2 & 38 \\
    200 & 0.85 & 1.6 & 0.243 & 77  & $1\times10^{-24}$ & 12 & 511 \\
    300 & 0.95 & 1.4 & 0.311 & 66  & $8\times10^{-24}$ & 4 & 40\\
  \end{tabular}
  \label{tab:example-bh_params}
    \end{ruledtabular}
\end{table}

\subsubsection{Timescales} \label{sec:discuss_timescales}

The figures discussed so far provide important information about the expected strain as a function of frequency when searching for signals from a given BH, but it is important to also consider the timescales introduced in \sect{theory}.
There are two timescales associated with the gravitational signal: the time it takes to reach its peak amplitude, and its duration thereafter.

The signal-growth timescale depends strongly on $\fs$, as is illustrated in \fig{tau_alpha} (\red{orange} curve) for our example BH ($\mbh=\mbhex$, $\chi=\chiex$).
This high sensitivity on $\fs$ means that, when analyzing real data, it will be important to only consider values of the boson mass that could have yielded a detectable signal given the age of the BH being targeted. 
In particular, {\em strain upper-limits can only be meaningfully translated into boson-mass constraint if the BH is sufficiently old to support a cloud that would emit gravitational waves of such amplitude}.
If a search is carried out before such time, one should instead look for a weaker and still growing signal.
This will be especially important for young BHs.

For instance, for the BH in \fig{tau_alpha} this means that, to constrain the presence of the best-matching boson (\red{vertical dashed line}: $\fs\approx\alphaex$, i.e.~$\mu \approx \muex$), one must wait {\em at least} \red{1 month} from the moment the BH is born before looking for a GW signal in the data.
During that \red{first month}, the cloud is still growing and the signal has not reached its peak ($h_0\approx \hEx$, according to \fig{h0_fgw_alpha}), meaning it might be too weak and unstable for detection.
Thus, absence of a detectable signal during that initial period would {\em not} be evidence against the existence of the boson. 
The same is true for any other value of $\fs$, but the peak strains will be weaker (cf.~\fig{h0_fgw_alpha}) and the times required to reach them possibly much longer (if $\fs<\alphaex$, for our example with $\chi_i=\chiex$).

The second relevant timescale, the signal duration, is also strongly dependent on $\fs$.
This is also illustrated in \fig{tau_alpha} (\red{purple} curve) for our example BH ($\mbh_i=\mbhex$, $\chi_i=\chiex$).
For the boson that best matches this BH ($\fs\approx\alphaex$), the characteristic duration of the signal is \red{${\sim} 7.5 \times 10^3$ yr}.
Similar to the situation described above, absence of a detectable signal long after this would {\em not} constitute evidence against the existence of such a boson.
This is because, if one waits too long, the fastest energy level will have been depleted, and one should instead look for signals corresponding to the next level, cf.~\eq{best_level}.
This feature is especially important when targeting old BHs.

\begin{figure}
\centering
\includegraphics[width=\columnwidth]{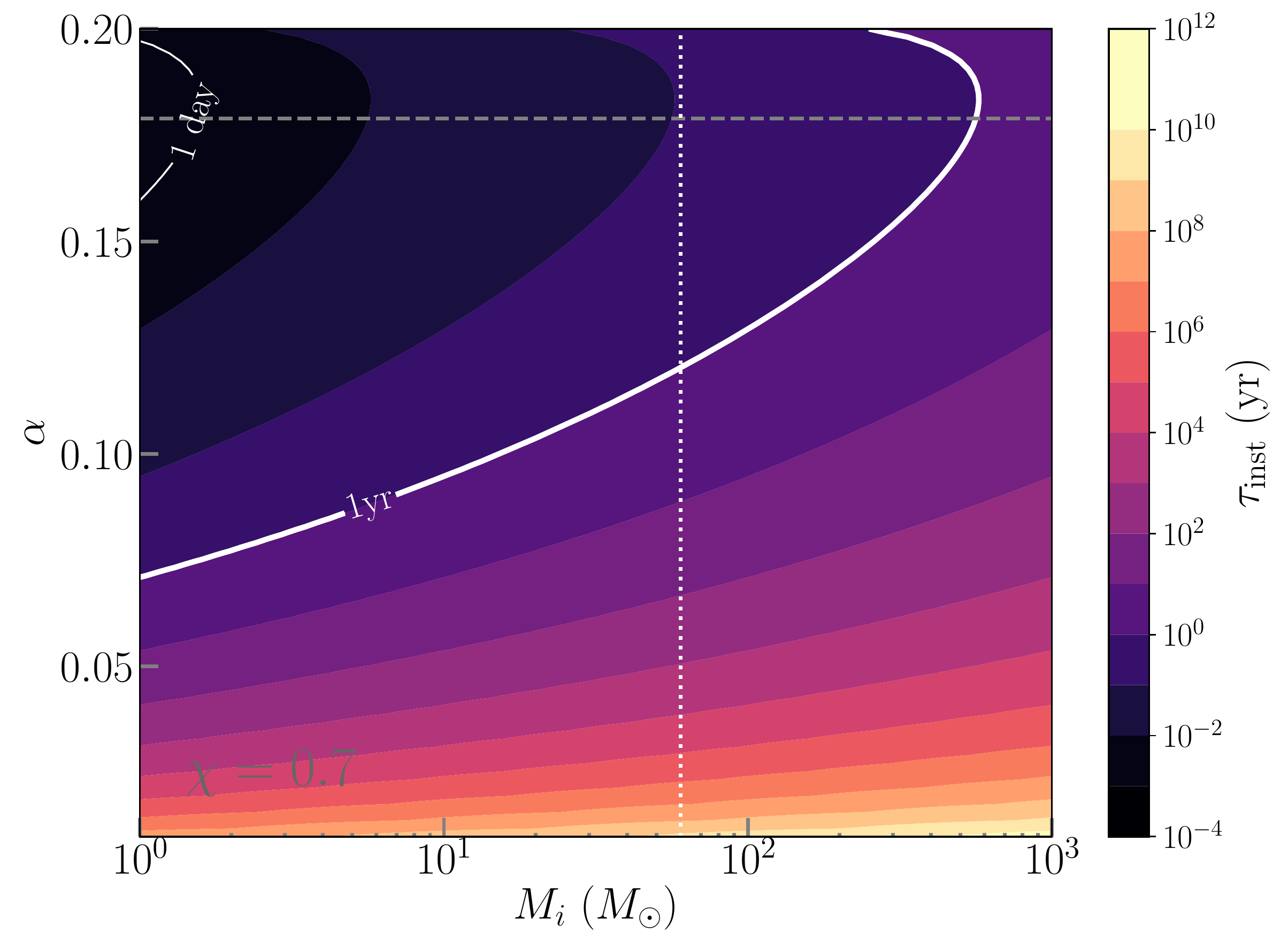}
\caption[Superradiant-instability timescale (scalar cloud]{{\em Superradiant-instability timescale (scalar cloud)}.
Color shows the characteristic growth time, \eq{tinst}, for a scalar cloud ($\lb=\mb=1,\, \nr=0$) as a function of BH mass ($x$-axis) and fine-structure constant ($y$-axis).
The BH is assumed to have an \red{initial} spin of $\chi=\chiex$.
The highlighted contours correspond to $\tau_{\rm inst}=1~{\rm day}$ (top, thin) and $\tau_{\rm inst}=1~{\rm yr}$ (bottom, thick).
The vertical dotted line gives the instability timescales for $\mbh=\mbhex$.
The horizontal dashed line corresponds to the value of $\fs$ that yields optimal GW emission for a BH with \red{initial} spin $\chi=\chiex$.
Note that values of $\fs>0.2$ preclude superradiance given this spin, cf.~\eq{sr2}.}
\label{fig:alpha_mbh_tinst}
\end{figure}

\begin{figure}
\centering
\includegraphics[width=\columnwidth]{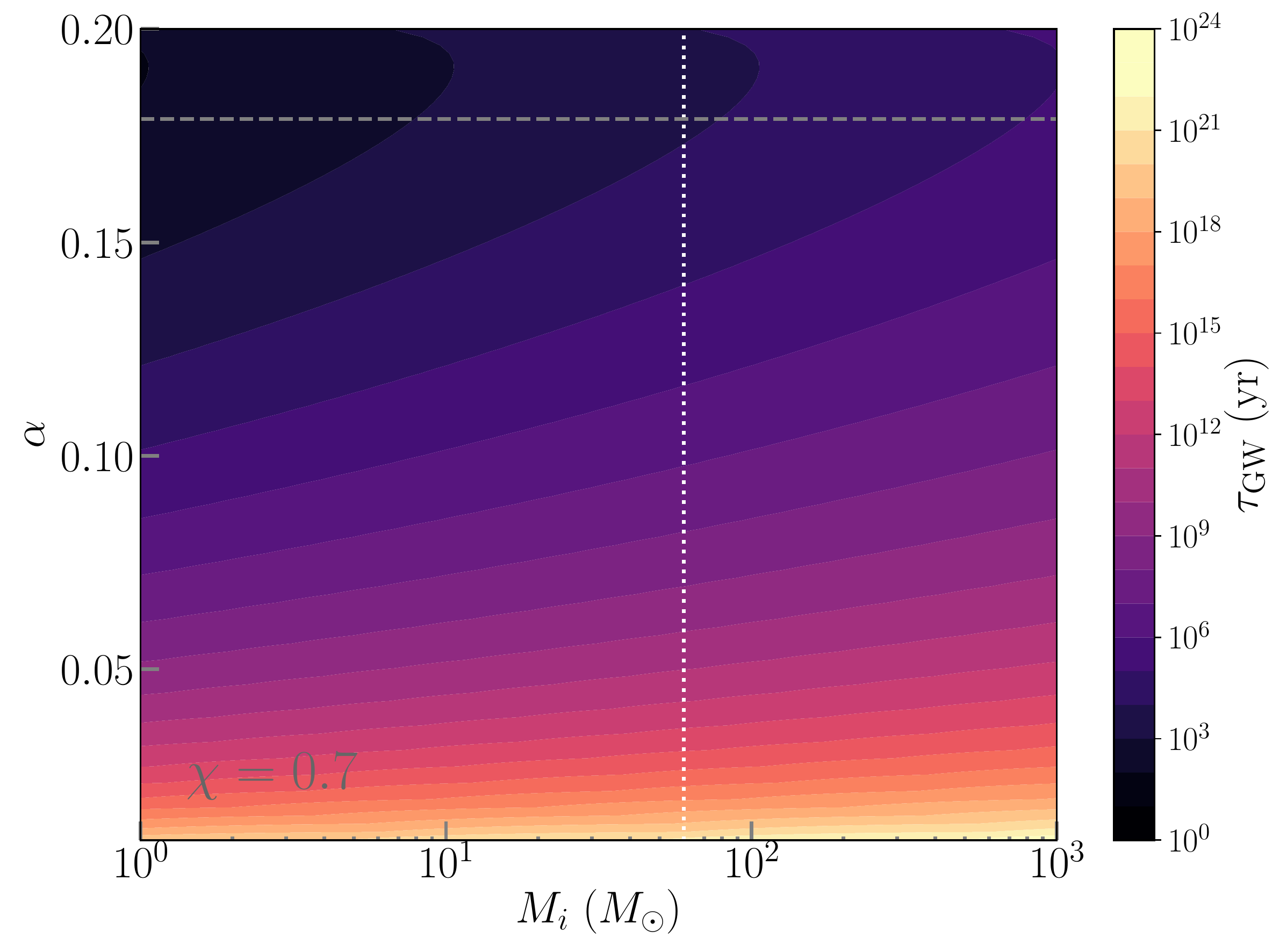}
\caption[Signal duration timescale (scalar cloud)]{{\em Signal duration timescale (scalar cloud)}.
Color shows the characteristic signal duration, \eq{tgw}, for a scalar cloud ($\lb=\mb=1,\, \nr=0$) as a function of BH mass ($x$-axis) and fine-structure constant ($y$-axis).
The BH is assumed to have an \red{initial} spin of $\chi=\chiex$.
The vertical dotted line gives the instability timescales for $\mbh=\mbhex$.
The horizontal dashed line corresponds to the value of $\fs$ that yields optimal GW emission for a BH with \red{initial} spin $\chi=\chiex$.
Note that values of $\fs>0.2$ preclude superradiance given this spin, cf.~\eq{sr2}.}
\label{fig:alpha_mbh_tgw}
\end{figure}

Both timescales are a function of BH mass, as reflected in \fig{alpha_mbh_tinst} and \fig{alpha_mbh_tgw}, and scale inversely with BH spin.
The color in \fig{alpha_mbh_tinst} and \fig{alpha_mbh_tgw} corresponds to the growth and duration timescales respectively, both assuming a spin of $\chi=\chiex$.
The horizontal dashed line marks $\fs=\alphaex$, the value of the fine-structure constant that yields peak emission for a BH with that spin.
Meanwhile, a vertical cut along the dotted lines ($\mbh=\mbhex$) would produce the orange and purple curves in \fig{tau_alpha}, respectively for  \fig{alpha_mbh_tinst} and \fig{alpha_mbh_tgw}.
Although both timescales vary widely for different $\fs$'s, for any given system ($\mbh$, $\chi$, $\fs$), $\tinst$ is always orders of magnitude shorter than $\tgw$ which allows the treatment of the cloud growth and signal emission as two different regimes, as explained in \sect{theory}.

\section{Directed searches} \label{sec:searches}

There are multiple observational signatures of BH superradiance that could be used to probe the boson-mass space \cite{Arvanitaki2011,Yoshino2014,Yoshino2015,Arvanitaki2015,Arvanitaki2017,Brito2017-letter,Brito2017,Baryakhtar2017,Cardoso2018,Baumann:2018vus,Hannuksela:2018izj,Zhang:2018kib,DAntonio:2018sff}.
Among all these, we will focus on the prospect for direct detection of the continuous gravitational waves expected from these sources (\sect{signal}).
In particular, we will restrict ourselves to searches {\em directed} at specific well-localized targets, rather than searches covering the whole sky (see, e.g., \cite{Riles2017} for a review of continuous-wave searches).
This means that we are interested in studying known (potential) BHs that could have the right mass, spin and age to possibly harbor a radiating boson cloud.
In order to apply existing search strategies, we would also like the cloud to be stable enough to make sure the signal lasts sufficiently long and evolves slowly enough to be considered ``persistent'' (we sharpen these criteria below).

Because the properties of the central BH can be measured \emph{a priori}, directed searches can potentially make unambiguous statements about the existence of ultralight bosons without relying on BH population models, which carry much uncertainty.
If a signal were found from a given target, detailed measurements of its morphology (see \sect{signal_waveform}) would provide invaluable information about the mass and dynamics of the new particle.
On the other hand, if a signal were \emph{not} found, knowledge of the BH parameters could allow us to place stringent constraints on the existence of bosons in the corresponding mass range.
Furthermore, having a specific sky location allows us to probe deeper in the noise, and explore a greater range of parameters to farther distances.
This comes at the price, of course, of the restriction to BHs that are already known, which may limit the use of the method in practice if no suitable BHs are discovered to target.

In the following, we introduce hidden Markov model (HMM) tracking as a well-suited method to carry out directed searches for these signals (\sect{methods}).
We evaluate its sensitivity with Monte-Carlo simulations and use the results to estimate the scalar-cloud detection horizons for future detectors (\sect{analysis}).
This discussion is agnostic as to the origin of the target BH, assuming only a known location and reasonably constrained intrinsic parameters.
The conclusions are, therefore, generally applicable to any known stellar-mass BH, but we devote special attention to remnants from compact-binary mergers and holes in x-ray binaries (\sect{targets}).
As we discuss below, vector signals present unique data-analysis and theoretical challenges, so we focus mainly (though not uniquely) on scalars.

\subsection{Search method} \label{sec:methods}

Hidden Markov model tracking is an efficient search strategy for detecting quasimonochromatic gravitational waves \cite{Suvorova2016,Sun2018}.
It was developed with rapidly-spinning neutron stars in mind, and has been applied in searches directed at several targets \cite{ScoX1ViterbiO1, gw170817,LVC:2018pmr,Sun:2018hmm}.
This strategy is well suited to searches for gravitational waves from boson clouds because its computational efficiency enables the coverage of a wide range of signal phase parameters and a grid of sky locations.
Furthermore, it allows small deviations from restrictive waveform models, unlike other coherent or semi-coherent search methods that rely on Taylor-series-based matched filters and are, thus, more model-restricted (e.g., \cite{Jaranowski1998,Dhurandhar2008}).
This makes it ideal to search for signals over a broad frequency band (cf.~\fig{h0_fgw_alpha}), even when the location of the source is only loosely known and when there is potential uncertainty in the signal morphology.

\subsubsection{Algorithm overview}

\newcommand{\tdrift}{T_{\rm drift}}
\newcommand{\tobs}{T_{\rm obs}}

The goal of HMM tracking is to find the most likely path that a putative signal takes in the time-frequency plane, contingent on the observed noisy data \cite{Suvorova2016,Sun2018}.
To do so, it divides the $f$-$t$ plane into pixels, assuming the signal is monochromatic over a period $\tdrift$ and splitting the frequency axis into bins of width $\Delta f = 1/(2 T_{\rm drift})$.
The signal power in each bin is then estimated coherently using the $\mathcal{F}$-statistic \cite{Jaranowski1998,F-stat2011}, a frequency-domain estimator that accounts for the motion of the Earth and is widely used in continuous-wave searches \cite{Riles2017}. 
At each time step $i$, this statistic is computed for each discrete frequency bin $j$ by coherently integrating over the time interval $(t_i, t_i + \tdrift$). Henceforth ``${f_0}_j$'' denotes the central value of the signal frequency, $f_0$, in the $j$th bin.%
\footnote{Here we follow the HMM-tracking literature by using ``$f_0$'' to denote the estimator for the (unknown) frequency of the signal, rather than ``$f$'' for frequency in general \cite{Suvorova2016,Sun2018}.}
If the total observation time is $T_{\rm obs}$, then the values of $\mathcal{F}(f_0)$ for the $N_T = \tobs/\tdrift$ blocks of duration $\tdrift$ are combined incoherently as described in \cite{Suvorova2016,Sun2018}.

Based on this information, the HMM algorithm computes the likelihood of different signal paths, assuming the signal can only transition between adjacent frequency bins from one time step to the next.
For application to boson signals, we assume the transition probability $A_{{f_0}_j {f_0}_k}$ between frequency bins ${f_0}_j$ and ${f_0}_k$ to be
\begin{equation}
\label{eq:trans_matrix}
A_{{f_0}_{j+1} {f_0}_j} = A_{{f_0}_j {f_0}_j} = \frac{1}{2}\, ,
\end{equation}
and to vanish otherwise (see \cite{Suvorova2016,Sun2018} for details).
The choice of \eq{trans_matrix} amounts to favoring signals with a positive frequency derivative, in agreement with the signal model of \sect{signal}.
We choose a uniform prior $\Pi_{{f_0}_j} = N_Q^{-1}$ on $f_0$ over the frequency band being searched, where $N_Q$ is the total number of frequency bins.
The result of the HMM tracking algorithm is summarized by a figure of merit representing the significance of the optimal path relative to all others (see \cite{ScoX1ViterbiO1,Sun2018}).
This quantity can then be treated as a regular (frequentist) detection statistic, and its background can be computed over several noise-only instantiations data to assign detection significances.

\subsubsection{Frequency-derivative tolerance}

Although, in principle, this method would be able to handle signals with arbitrary frequency evolutions, allowing for large frequency drifts ($\dot f$) comes with a significant reduction in sensitivity.
In order to allow for a maximum frequency derivative $\max(\dot{f})$ we must choose $\tdrift \leq \Delta f/|\max(\dot{f})|$, so as to guarantee that
\begin{equation}
\label{eq:int_T_drift}
\left|\int_t^{t+T_{\rm drift}} \mathrm{d} t' \dot{f}(t')\right| \leq \Delta f\, ,
\end{equation}
for $0<t<\tobs$ and where the frequency resolution is set to $\Delta f = (2\tdrift)^{-1}$, as mentioned above.
Therefore, tracking a signal with higher $\dot f$ requires reducing the coherent-integration time over which the $\mathcal{F}$ statistic is computed, which in turn diminishes the sensitivity of the search \cite{Sun2018}.

The implementation of the $\mathcal{F}$-statistic-based HMM used by the LIGO and Virgo Collaborations can currently track quasimonochromatic signals with derivatives of at most $\dot{f}\sim 10^{-8}$\,Hz\,s$^{-1}$ \cite{Sun2018}. 
This is more than enough to accommodate the majority of scalar signals [\eq{fdot_scalar}], but puts most (although not all) of the vector parameter space out of reach [\eq{fdot_vector}].
Detecting shorter-lived signals may be possible by extending the current method to track not only the signal frequency, but also its first time derivative with a two-dimensional HMM as in \cite{Sun2018}.
Such a strategy would also naturally handle noticeable decays in signal amplitude over the observation run.
The adaptation of the methods in \cite{Sun2018} to boson signals will be subject of future work.\footnote{The HMM tracking based on 1-s short Fourier transforms described in Ref.\cite{Sun:2018hmm} can be used to search for long-duration transient signals with timescales $\sim 10^2$\,s--$10^4$\,s. The timescale of a vector signal is much longer than that, and hence longer short Fourier transforms need to be used in the tracking. However, the Doppler modulation due to the motion of the Earth with respect to the solar system barycenter is not negligible when the length of short Fourier transforms is longer than $\sim 10$--$100$\,seconds \cite{Dhurandhar2008}.}

\begin{figure*}
	\centering
	\subfloat[Less variability\label{fig:sample-path_less}]{\includegraphics[trim={0.5cm 0.5cm 1cm 2.5cm},clip,width=\columnwidth]{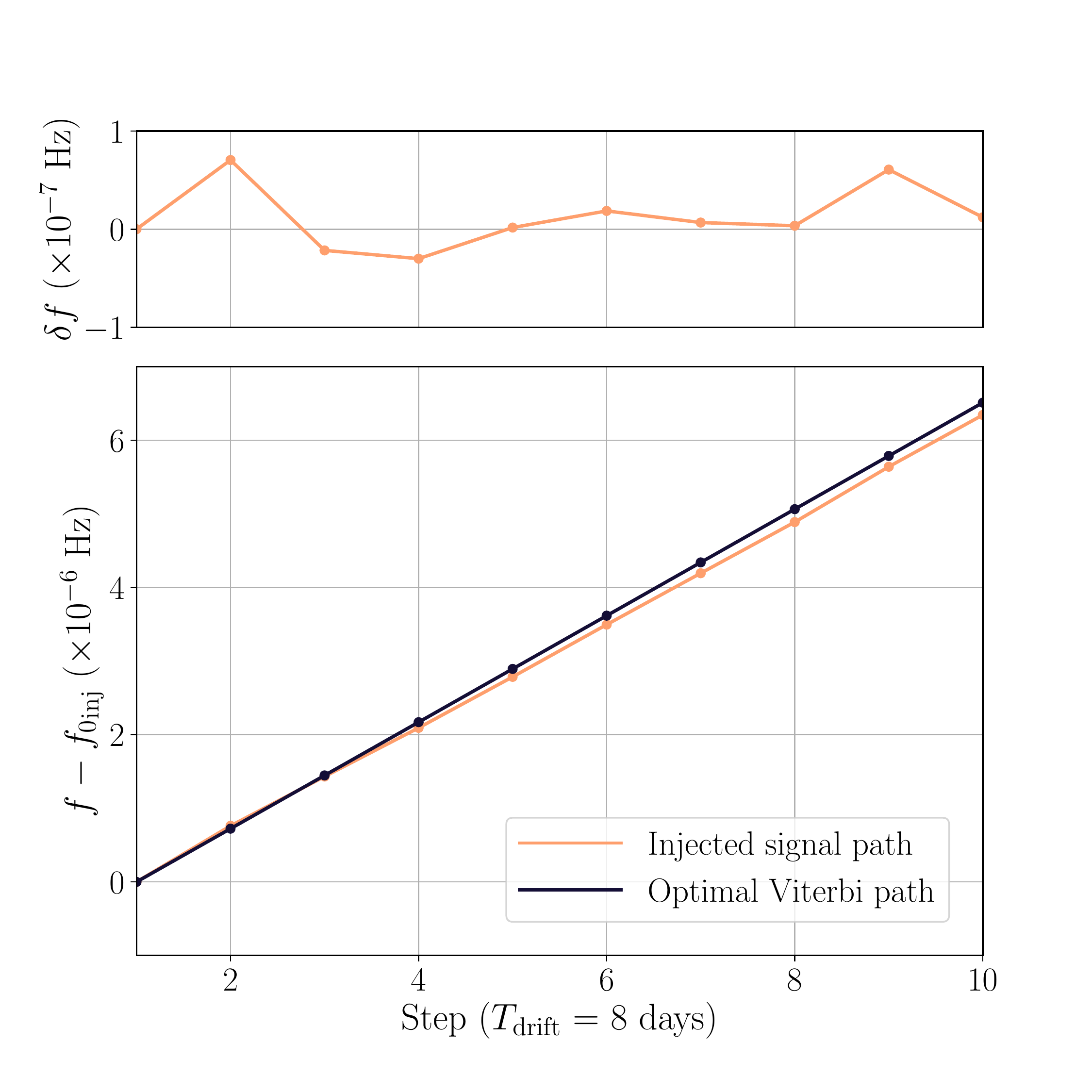}}
	\subfloat[More variability\label{fig:sample-path_more}]{\includegraphics[trim={0.5cm 0.5cm 1cm 2.5cm},clip,width=\columnwidth]{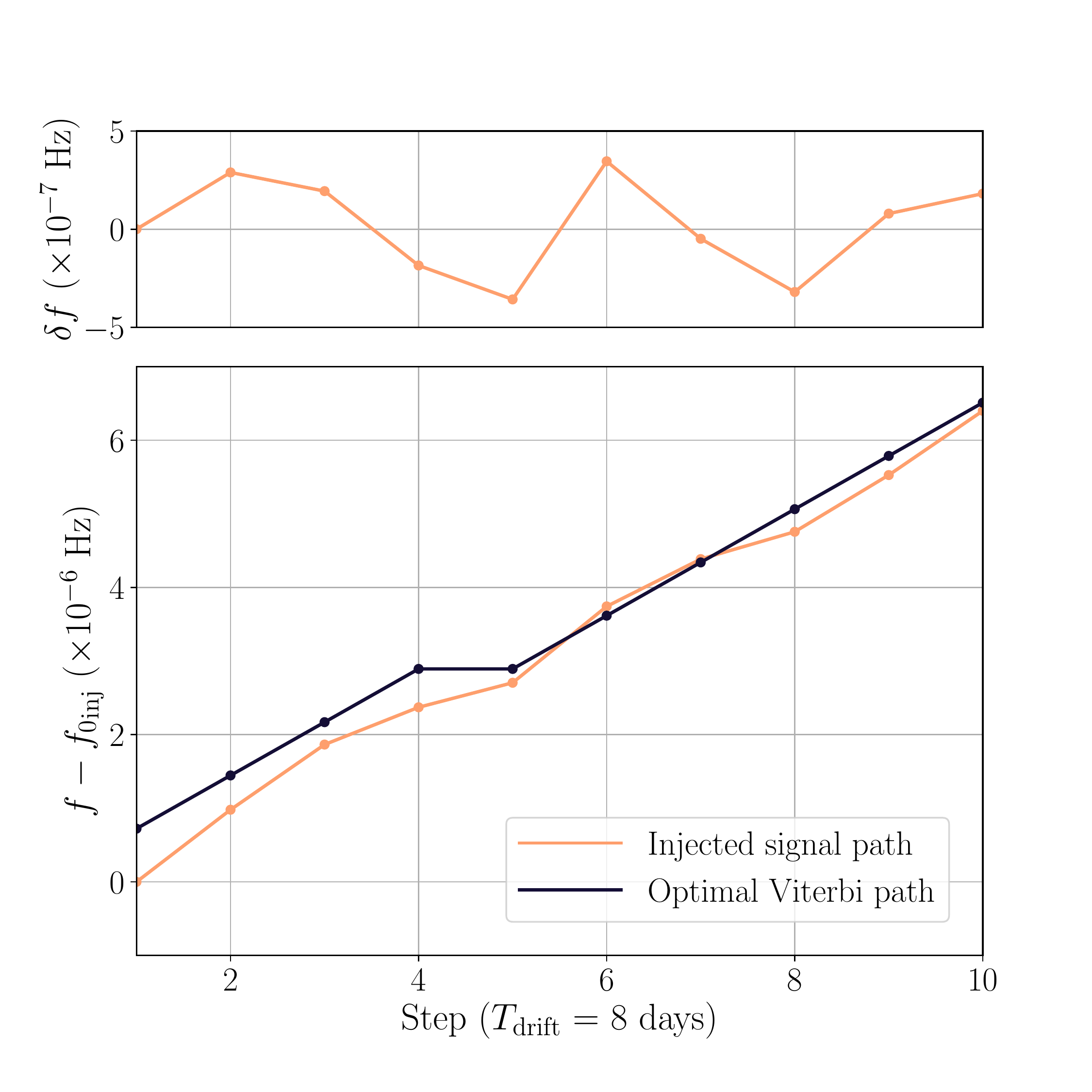}}
	\caption[HMM sample tracking paths]{{\em HMM sample tracking paths}. Injected $f_0(t)$ (light curves) and optimal Viterbi paths (dark curves) for the injected signals with (a) weaker random walk $|\delta f|\leq 0.1 \Delta f$ and (b) stronger random walk $|\delta f|\leq 0.5 \Delta f$. The top panels show the random walk $\delta f$ added to the injected signals at each step, which is too small to be seen by eye in the bottom panel of (a). The horizontal axis is in units of HMM steps with each step spanning for $T_{\rm drift}=8$\,d. Good matches are obtained in both (a) and (b) with $\varepsilon_{f_0} =0.16 \Delta f$ and $0.50 \Delta f$, respectively. Injection parameters are in Table~\ref{tab:inj-para} and the injected signal strain is $h_0 = 5\times 10^{-26}$.}
	\label{fig:sample-path}
\end{figure*}

\subsubsection{Computing cost}

The computing time for one central processing unit (CPU) over a total observing time $T_{\rm obs}$ in a frequency band from $f_{\rm min}$ to $f_{\rm max}$ is given by \cite{Sun2018}
\begin{equation}
\label{eq:computing-time}
\mathcal{T} =  2\kappa \beta N_{\rm ifo} T_{\rm drift} T_{\rm obs}T_{\rm SFT}^{-1}  N_{\rm sky}  (f_{\rm max} - f_{\rm min})\, ,
\end{equation}
where $T_{\rm SFT}$ is the length of the short Fourier transforms (SFTs) used to compute the $\mathcal{F}$-statistic \cite{F-stat2011}, $N_{\rm ifo}$ is the number of interferometers, $N_{\rm sky}$ is the number of sky locations, $\beta$ is the percentage of time that the interferometers collect data (``duty cycle''), and $\kappa$ is the time to compute the $\mathcal{F}$-statistic per template per SFT.
The value of $\kappa$ depends on $T_{\rm SFT}$ and the CPU architecture; we adopt the recent estimate that $\kappa=4\times 10^{-8}$\,s for $T_{\rm SFT}=1800$\,s.

We normally divide the full frequency band into multiple 1-Hz subbands to allow parallelized computing.
For example, if we have $10^2$ cores running in parallel, a search for $T_{\rm obs}=80$\,days over the frequency band spanning 100--200\,Hz in two detectors and with a fixed sky location ($N_{\rm sky} = 1$) takes about 7\,min to complete.
This estimation is consistent with the real cost of our simulations below.

\subsection{Sensitivity estimates} \label{sec:analysis}

We would like to study the sensitivity of ground-based detectors to continuous GW signals from boson clouds around known BHs.
For this purpose, we simulate signals consistent with the morphology described in \sect{signal_waveform} and study how well they can be recovered using the HMM tracking (\sect{sensitivity}).
We then translate expected strain sensitivities into detection horizons for boson signals with current and future ground-based detectors (\sect{horizons}).
Finally, we explore the impact of uncertainties in the source's sky location (\sect{skyloc}).

\subsubsection{Strain sensitivity}
\label{sec:sensitivity}

To study our sensitivity to waves from boson clouds, we inject synthetic signals with parameters consistent with the morphology described in \sect{signal_waveform} into simulated Gaussian noise corresponding to two aLIGO detectors at design sensitivity \cite{aLIGO_design_sensi}.
The signal frequency and frequency derivative were chosen to be roughly in agreement with an optimal scalar cloud around the example BH discussed in \sect{discuss}---that is, $\mbh_i=\mbhex$ and $\chi_i=\chiex$, consistent with the GW150914 remnant \cite{gw150914_pe}.
We choose source inclinations randomly such that $\cos \iota$ is uniformly distributed over the range $[-1,1]$, while we pick polarization angles $\psi$ and initial phases $\Phi_0$ uniformly over $[0,2\pi]$.
Unless otherwise stated, we fix the sky location to the values in Table~\ref{tab:inj-para}.
(See \sect{signal_waveform} for definitions of all the signal parameters.)
The HMM tracking is conducted with the settings shown in Table~\ref{tab:search-para}, directed at the true sky location of the injection. In this particular sample scenario, we choose $\tdrift =8$\,d assuming $\dot{f}^{(s)} \lesssim 10^{-12}$\,Hz/s, to satisfy \eq{int_T_drift}.
In a real search, we estimate $\dot{f}^{\rm (s)}$ using \eq{fdot_scalar} and choose the longest $\tdrift$ that satisfies \eq{int_T_drift}. 
On top of an overall positive frequency drift, we add a random frequency fluctuation $\delta f$ at each time step, in order to demonstrate that the HMM tracking is robust against small uncertainties in the signal model.
The choice of $\tdrift$ is independent of the unknown frequency fluctuation, which is assumed to be weaker compared to the secular phase evolution. We simulate the random frequency fluctuation for the purpose of demonstrating that the HMM tracking is robust against small uncertainties in the signal model. We vary the magnitude of $\delta f$, as well as the intrinsic amplitude $h_0$, for different injections.

\begin{table}
\centering 
\begin{ruledtabular}
	\caption[Boson injection parameters]{Injection parameters.}
	\label{tab:inj-para}
\begin{tabular}{l@{\quad} c@{\quad}c@{\quad}}
		Parameter  & Symbol& Value\\
		\midrule
		Initial Frequency & ${f_0}_{\rm inj}$ & 201.2\,Hz \\
		First derivative of ${f_0}_{\rm inj}$ & ${\dot{f}}_{\rm inj}$ & $1 \times 10^{-12}$\,Hz\,s$^{-1}$ \\
		Right ascension & $\ra$ & $23^{\rm h}23^ {\rm m}26.0^{\rm s}$\\
		Declination & $\dec$ & $58^{\circ}48' 0.0''$\\
		Inclination & $\cos\iota$ & $[-1,\,1]$\\
		Polarization & $\psi$ & $[0,\,2\pi]$\\
		Initial phase & $\Phi_0$ & $[0,\,2\pi]$\\
		Gaussian noise ASD &$S_h^{1/2}(f)$ & $4 \times 10^{-24}$\,Hz$^{-1/2}$ \\
	\end{tabular}
	\end{ruledtabular}
\end{table}

\begin{table}
	\begin{ruledtabular}
  \caption[Boson search parameters]{Search parameters.}
	\label{tab:search-para}
		\begin{tabular}{l@{\quad} c@{\quad}c@{\quad}}
		Parameter & Symbol & Value \\
		\midrule
		Search frequency band & $f$ &201--202\,Hz \\
		Coherent time & $T_{\rm drift}$ & 8\,d \\
		Bin size & $\Delta f$ & $7.23 \times 10^{-7}$\,Hz \\
		Total observing time &$T_{\rm obs}$ & 80\,d\\
		Number of steps & $N_T$ & 10\\
	\end{tabular}
	\end{ruledtabular}
\end{table}

To demonstrate that HMM can accurately reconstruct boson signals, \fig{sample-path} presents two tracking examples for injected signals with $h_0 = 5\times10^{-26}$ and parameters in Table~\ref{tab:inj-para}.
The frequency random walks are such that $|\delta f|\leq 0.1 \Delta f$ and $|\delta f|\leq 0.5 \Delta f$ for panels (a) and (b), respectively.
The optimal HMM paths (dark curves) match the injected path $f_0(t)$ (light curves) closely: the root-mean-square error (RMSE) between the optimal HMM paths and the actual signals are $\varepsilon = 1.16\times 10^{-7}\, {\rm Hz} = 0.16 \Delta f$ for panel (a) and $3.65\times 10^{-7}\, {\rm Hz} = 0.50 \Delta f$ for panel (b).
These small discrepancies are mostly due to the frequency discretization carried out by the HMM algorithm.

\begin{figure}
	\centering
	\includegraphics[trim={0cm 0cm 0cm 2cm},clip,width=\columnwidth]{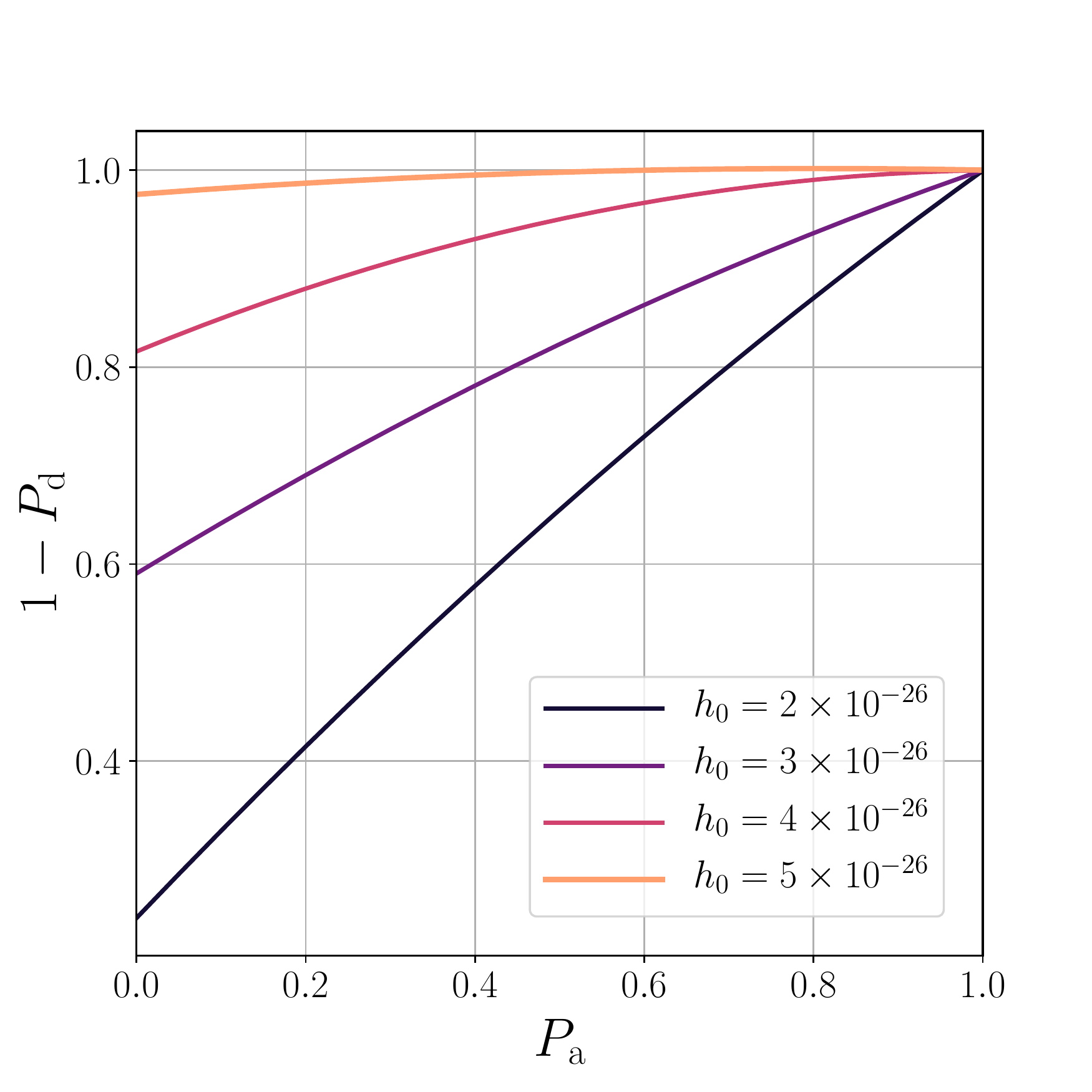}
	\caption{Receiver operator characteristic (ROC) curves for the injections with parameters in Table~\ref{tab:inj-para}. The four curves (from top to bottom) correspond to the four representative wave strains $h_0/10^{-26} = 5$, 4, 3, and 2. The horizontal and vertical axes indicate the false alarm probability $P_{\rm a}$ and detection probability $1-P_{\rm d}$, respectively. Each curve is based on 200 realizations with randomly chosen polarization and inclination angles and initial phase.}
	\label{fig:roc}
\end{figure}

We next quantify the efficiency of HMM tracking at detecting signals of different amplitudes.
For concreteness, we assume a small uncertainty in the signal model by setting $|\delta f|\leq 0.1 \Delta f$, as in \fig{sample-path_less}.
Figure \ref{fig:roc} shows the receiver operator characteristic (ROC) curves for injected signals with four values of $h_0$, ranging from $2 \times 10^{-26}$ to $5 \times 10^{-26}$.
For each signal amplitude, these curves show the detection probability ($1-P_{\rm d}$, where $P_{\rm d}$ is the false-dismissal probability) as a function of required false-alarm probability threshold ($P_{\rm a}$).
For instance, if we demand a false alarm probability $P_{\rm a}=1\%$, we can expect to detect a signal with $h_0=4 \times 10^{-26}$ ($5 \times 10^{-26}$) with $84\%$ ($98\%$) of the time.

The detection threshold in continuous-wave searches is traditionally defined to be $95\%$ false-dismissal rate at $1\%$ false-alarm probability \cite{Riles2017,ScoX1ViterbiO1}.
In our case, for an observation time of $T_{\rm obs} = 80$\,days with two aLIGO design detectors, this corresponds to a strain amplitude of $h_0^{95\%} = 4.7 \times 10^{-26}$ for unknown inclination.
Based on this, we will consider boson signals ``detectable'' if they reach an amplitude of $h_0^{95\%}$ or higher for the observation conditions.

From the empirical result that $h_0^{95\%} = 4.7 \times 10^{-26}$ obtained for the simulations above, it is straightforward to estimate how the sensitivity of the search would scale for different detector networks and observation times.
The sensitivity scaling will be given by \cite{Sun2018}
\begin{equation} \label{eq:h95scaling}
h_0^{95\%}(f) \propto N_{\rm ifo}^{-1/2} S_h(f)^{1/2}\left(\tdrift \tobs\right)^{-1/4}\, ,
\end{equation}
assuming a network of $N_{\rm ifo}$ detectors with power-spectral density (PSD) $S_h(f)$ at the signal frequency.%
\footnote{This requires that all detectors have comparable sensitivities given by $S_h(f)$; were this not the case, the $N_{\rm ifo}^{-1/2} S_h(f)^{1/2}$ factor would have to be replaced by the square-root of the effective PSD $S_{\rm eff}(f)$, which is itself given by the harmonic mean of the PSDs for each detector.}

\begin{figure}
	\centering
	\includegraphics[width=\columnwidth]{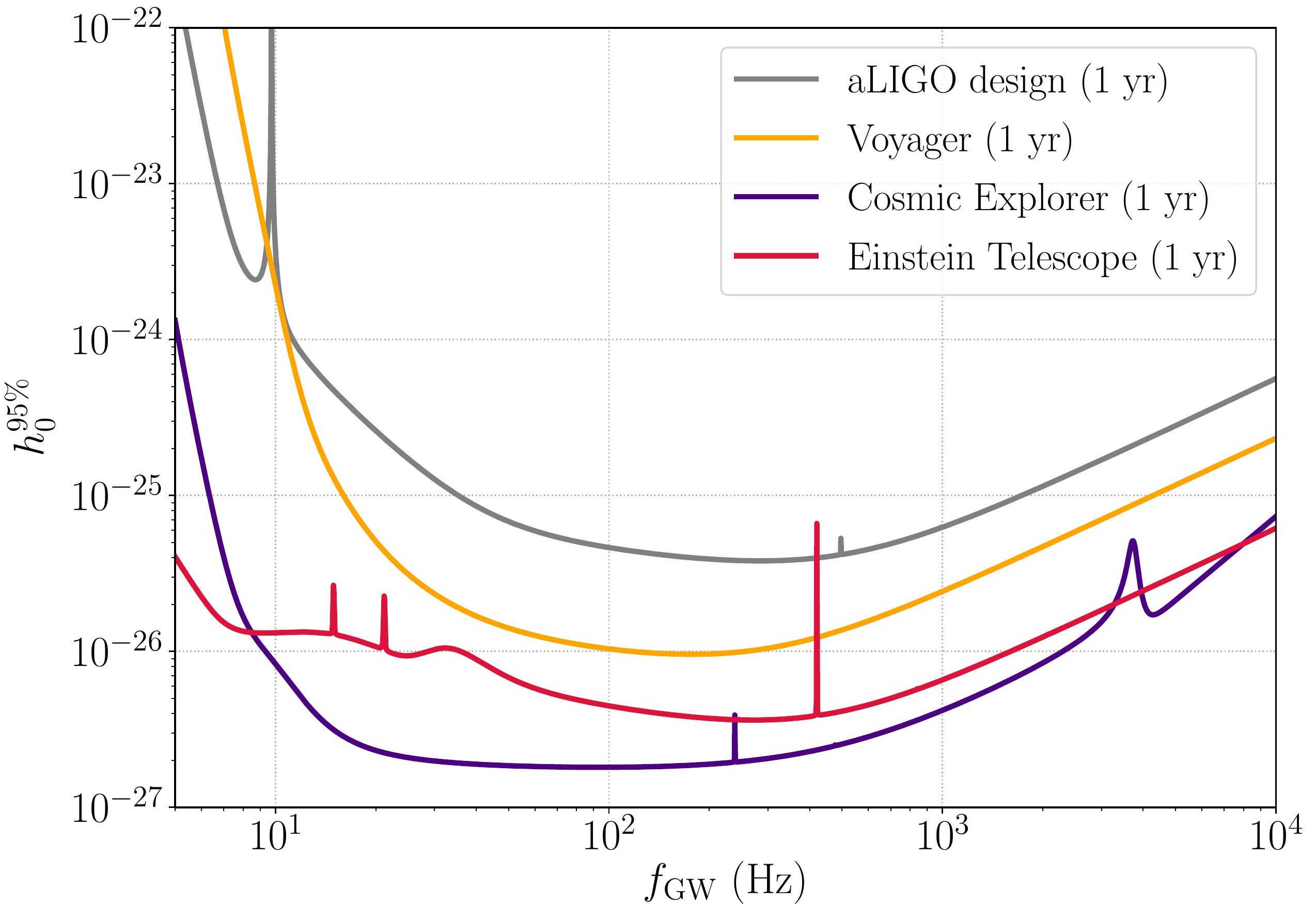}
	\caption[HMM sensitivity vs GW frequency for different detectors]{{\em Sensitivity vs GW frequency for different detectors.}
  Value of $h^{95\%}_0$ marginalized over source orientation for design aLIGO (gray), LIGO Voyager (yellow), Cosmic Explorer (purple) and the Einstein Telescope (red). All curves assume one year of continuous observation by a single detector of the indicated type.}
	\label{fig:h095_fgw}
\end{figure}

Using this, \fig{h095_fgw} presents projected 95\%-confidence strain upper limits, $h_0^{95\%}$, for different detectors as a function of GW frequency (assuming there is no detection).
We show results for aLIGO design sensitivity%
\footnote{For aLIGO, we use the latest-available sensitivity projections \cite{aLIGO_design_sensi}, which correspond to those in \cite{aLIGO} but with a reflectivity of 32.5\% in the signal recycling mirror (SRM).}
(gray), as well as proposed third-generation detectors: LIGO Voyager (yellow), Cosmic Explorer (purple) and Einstein Telescope (red) \cite{Evans:2016mbw,iswp2016,et-design,Sathyaprakash:2012jk}.
All curves in \fig{h095_fgw} were produced assuming $N_{\rm ifo}=1$ and $\tobs = 1\, $ year, but it is straightforward to rescale them for different configurations using \eq{h95scaling}.
In particular for the Einstein Telecope, we show the sensitivity of a 10 km, $90\degree$ interferometer in the ``D'' configuration---the sensitivity of the full triangular layout is up to 50\% better for a circularly polarized wave \cite{Freise:2008dk,Hild:2010id}.
Finally, note that these curves were obtained by effectively marginalizing over source orientation: they represent the value of $h_0^{95\%}$ marginalized over the distributions of $\cos\iota$ and $\psi$ in the ranges shown in Table~\ref{tab:inj-para}.
To obtain the values of $h_0^{95\%}$ corresponding to \emph{optimal} source orientation, one should divide the curves of \fig{h095_fgw} by a factor of ${\sim}2.8$ \cite{ScoX1ViterbiO1}.

The sky position of the source with respect to the detector does not impact the search sensitivity significantly because the variation due to the antenna pattern is averaged out when the integration time is much longer than a day.
We verify this by injecting signals with $h_0 = 5 \times 10^{-26}$ at different sky positions, and with all other parameters as in Table~\ref{tab:inj-para}.
As before, the HMM tracking is conducted with the settings shown in Table~\ref{tab:search-para}, \red{directed at the true sky location of the injection}.
The detection efficiencies for each sky location are listed in Table~\ref{tab:inj-sky}.
As before, each row is based on 200 realizations with randomly chosen $\cos \iota$, $\psi$, and $\Phi_0$.
The standard deviation of detection efficiencies at these eight sky positions is only 0.02.

\begin{table}
	\begin{ruledtabular}
		\caption[parameters]{Detection efficiency vs sky location ($h_0 = 5 \times 10^{-26}$)}
		\label{tab:inj-sky}
		\begin{tabular}{l@{\quad} r@{\quad}c@{\quad}}
			Right ascension & Declination & Detection efficiency \\
			\midrule
			23h 23m 26.0s & $58^{\circ}48' 0.0''$ &0.98\\
			23h 23m 26.0s & $-59^{\circ}35' 0.0''$ &0.97\\
			23h 23m 26.0s & $00^{\circ} 02' 0.0''$ &0.92\\
			23h 23m 26.0s & $88^{\circ}48' 0.0''$ &0.98\\
			23h 23m 26.0s & $-89^{\circ}18' 0.0''$ &0.98\\
			05h 23m 26.0s & $58^{\circ}48' 0.0''$ &0.98\\
			11h 23m 26.0s & $58^{\circ}48' 0.0''$ &0.98\\
			17h 23m 26.0s & $58^{\circ}48' 0.0''$ &0.99\\
		\end{tabular}
	\end{ruledtabular}
\end{table}

\subsubsection{Detection horizons for scalar clouds}
\label{sec:horizons}

\begin{figure*}%
	\centering
	\subfloat[aLIGO design\label{fig:horizon_aligo}]{\includegraphics[width=0.5\textwidth]{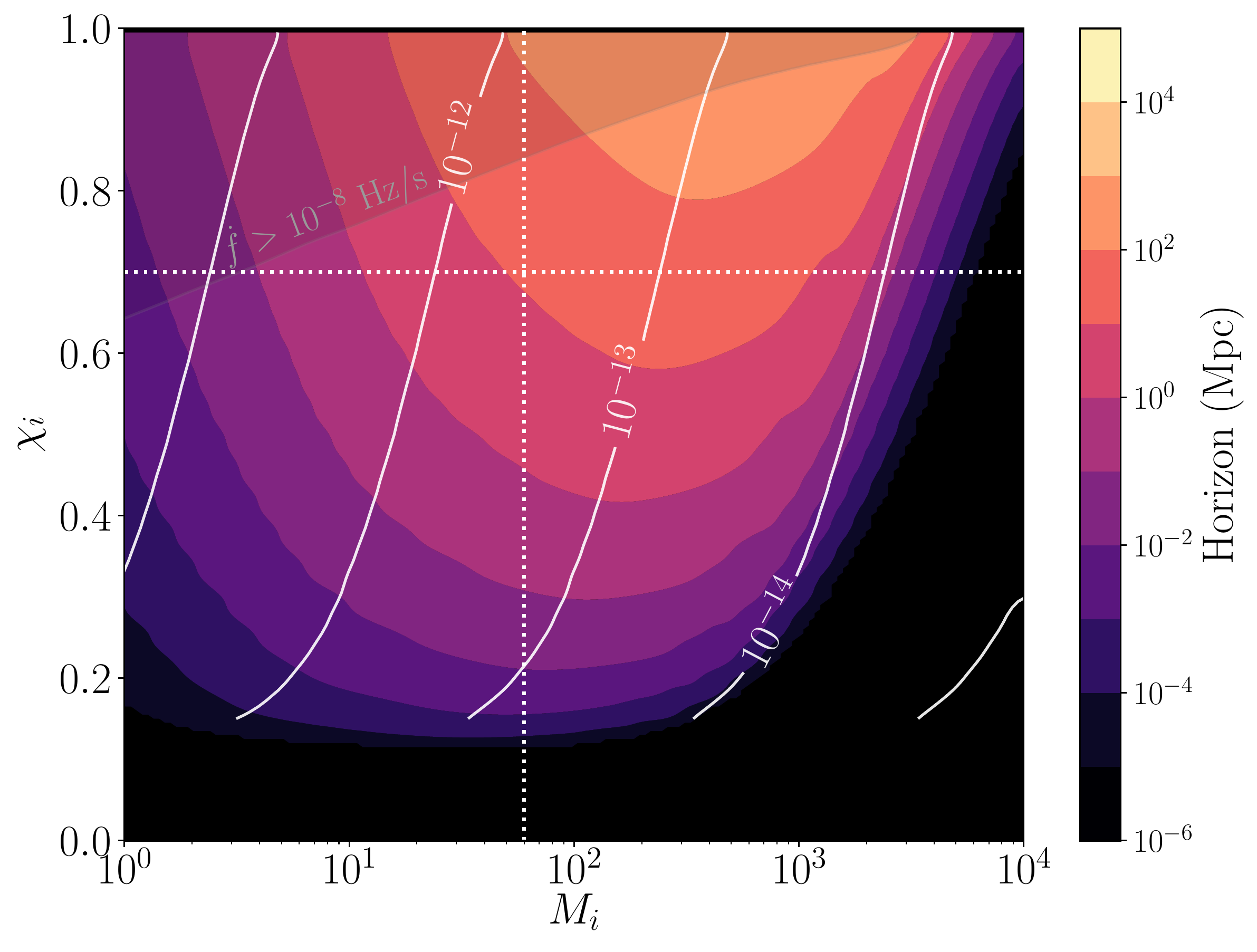}}\hfill
	\subfloat[LIGO Voyager\label{fig:horizon_voy}]{\includegraphics[width=0.5\textwidth]{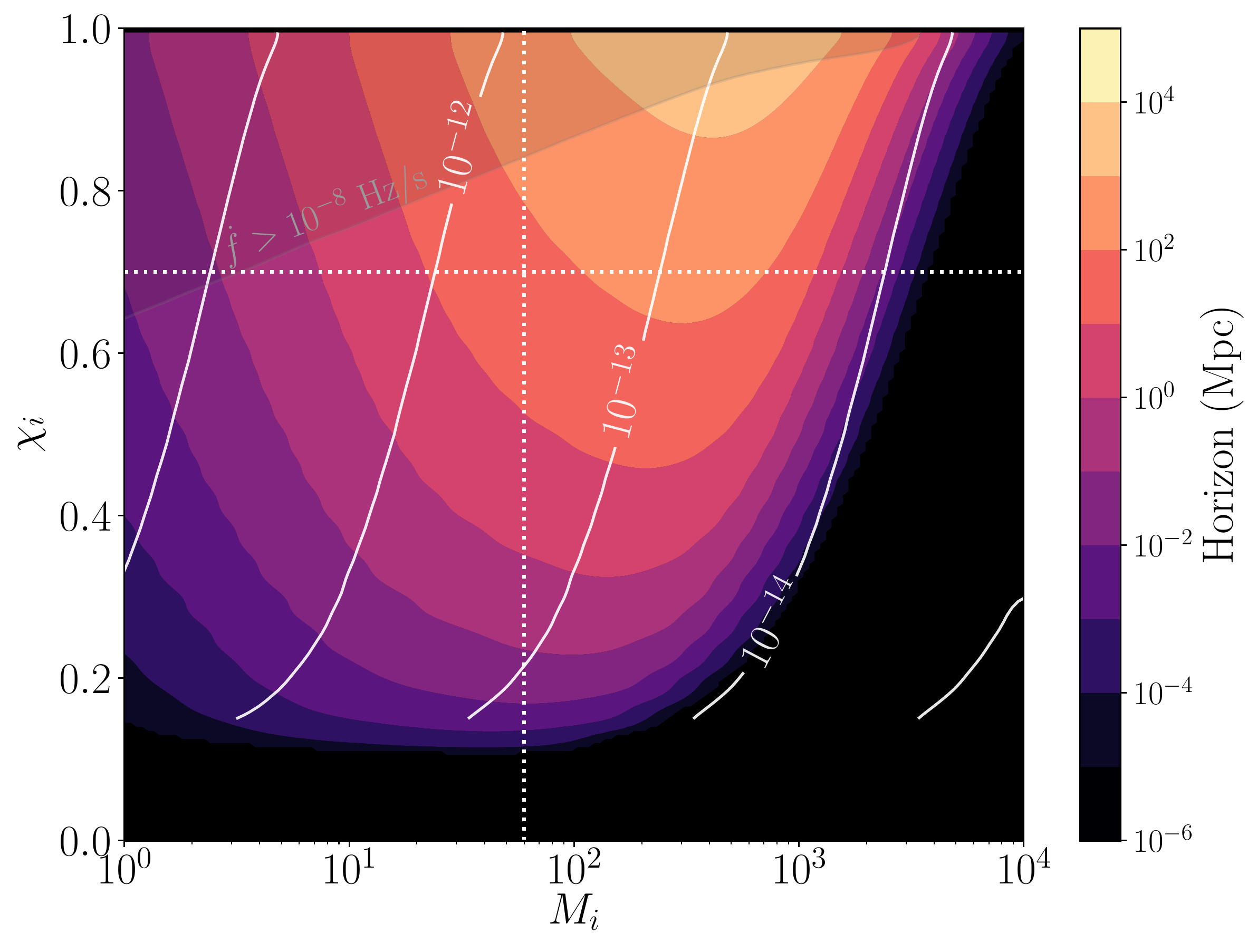}}\\
	\subfloat[Cosmic Explorer\label{fig:horizon_ce}]{\includegraphics[width=0.5\textwidth]{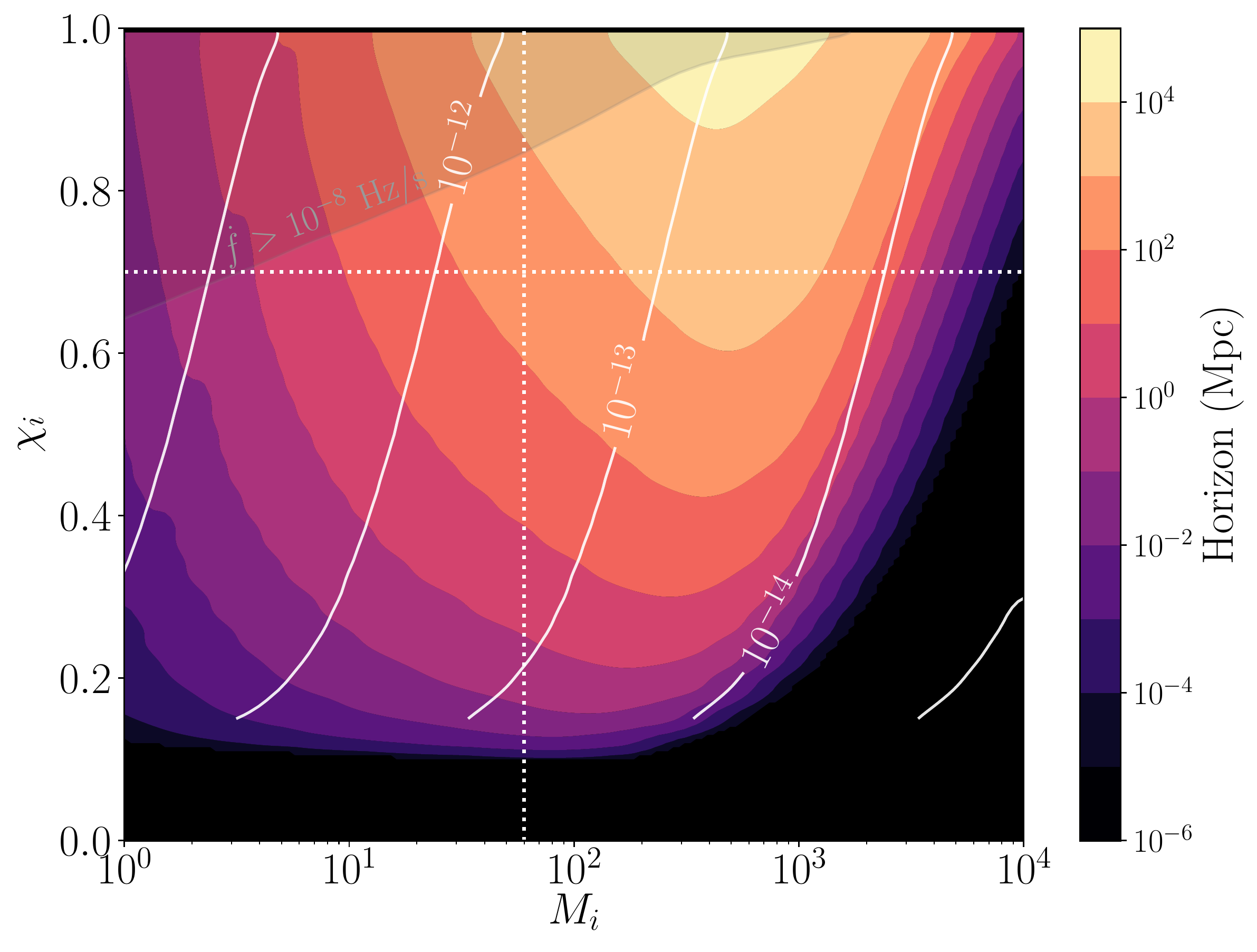}}\hfill
	\subfloat[Einstein Telescope\label{fig:horizon_et}]{\includegraphics[width=0.5\textwidth]{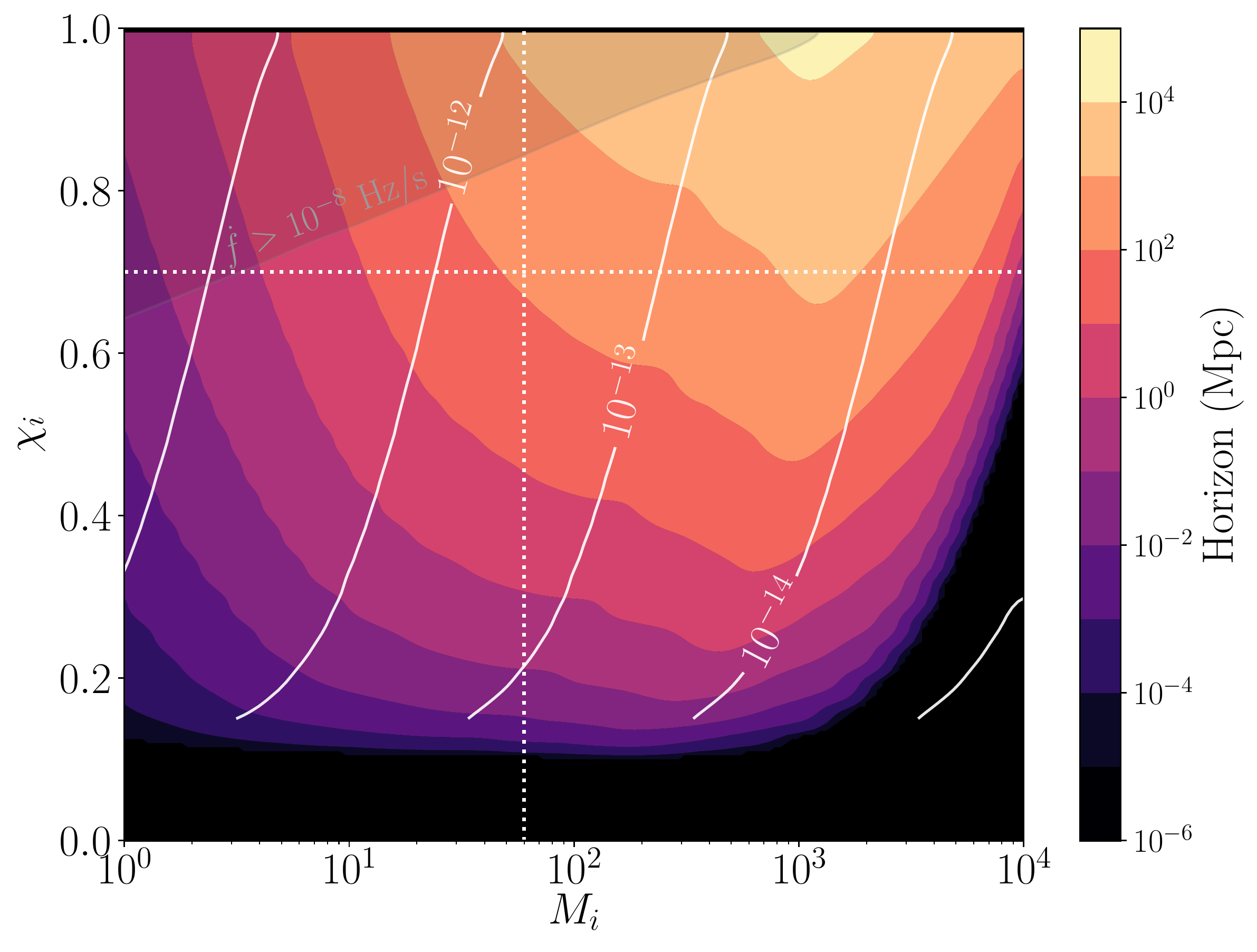}}
  \caption[Detection horizons for scalar clouds]{{\em Detection horizons for scalar clouds.}
 Maximum-detectable luminosity distances (color) for optimal scalar clouds ($\lb=\mb=1$, $\nr=0$, $\lgw=\mgw=2$) around BHs with the indicated initial mass ($x$-axis) and spin ($y$-axis) for different detectors.
  White contour lines indicate the values of the corresponding boson rest-energy $\mu/{\rm eV}$. 
  The shaded region (top-left) marks parameters that would yield signals evolving prohibitively fast ($\dot f_{\rm det} > 10^{-8}$) for existing search methods, based on \eq{fdot_scalar}.
  The dotted white lines highlight the mass and spin ($\mbhex$, $\chiex$) of the GW150914-like example discussed repeatedly in the main text.
  Values correspond to HMM tracking for one year of continuous observation by a single detector, accounting for signal redshifts and variability in maximum $\tdrift$ allowed by the expected signal, cf.~\eq{int_T_drift}.}
	\label{fig:horizons}
\end{figure*}

It is useful to translate the projected strain sensitivities of \fig{h095_fgw} into detection horizons for boson signals from BHs with different parameters.
The detection horizon is the farthest distance up to which we should expect to be able to detect an \emph{optimal} boson signal---namely, a signal from a boson cloud that perfectly matches its host BH (to maximize \emph{intrinsic} strain) and is optimally oriented with respect to the detector (to maximize \emph{measured} strain).
Consequently, horizons are a measure of how well we can do in the best-case scenario and are, thus, not generally representative of most detections (see e.g.~\cite{Chen:2017wpg} for an overview of distance measures in GW astronomy)---yet, they are a straightforward proxy for the reach of our instruments to this type of source.
We compute this quantity for \emph{scalar} clouds based on the results from \sect{discuss}; we defer computation of horizons for vector clouds until better numerical estimates of their intrinsic amplitudes become available and analysis methods suitable for higher frequency derivatives are developed.

Figure \ref{fig:horizons} shows the horizon luminosity distance (color) for scalar signals as a function of initial BH mass $M_i$ and spin $\chi_i$, for (a) design Advanced LIGO, (b) LIGO Voyager, (c) Cosmic Explorer, and (d) the Einstein Telescope.
White contours indicate the values of the boson rest-energy $\mu$ (eV) that we would be able to probe with a BH of that mass and spin.
The shaded region marks values for which we expect the signal to evolve too rapidly for current data analysis techniques to handle, based on \eq{fdot_scalar} and \eq{int_T_drift}.
Note that this varies slightly among plots due to minor differences in redshift.
In all cases we assume one year of uninterrupted observation by a single detector, as in \fig{h095_fgw}.

The horizon plots were obtained by finding the luminosity distance at which an optimal cloud for the given BH parameters would become barely detectable, i.e.~$h_0(d_L) = h_0^{95\%}(f_{\rm det})$ for detector-frame signal frequency $f_{\rm det}$.
In order to obtain the relevant value of $h_0^{95\%}$, we rescale the curves of \fig{h095_fgw} using \eq{h95scaling} to account for variations in $\tdrift$.
This is needed because the expected $\dot f_{\rm det}$ varies widely over the parameter space, affecting the maximum-allowed coherence time, cf.~\eq{int_T_drift}.%
\footnote{In an actual analysis, we might want to set a $\tdrift$ shorter than that implied by \eq{fdot_scalar} in order to allow for theoretical uncertainty in the predicted value of the frequency derivative.}
We also take into account the fact that both frequencies and frequency derivatives get redshifted as the signal makes its way to Earth,%
\footnote{Assuming standard $\Lambda$CDM cosmology with present parameters: $\Omega_m = 0.308$, $\Omega_\Lambda = 0.692$, $\Omega_k=0.0$, $h=0.678$ \cite{Ade:2015xua}.}
i.e.~$f_{\rm det} = f_{\rm src} (1+z)^{-1}$ and $\dot f_{\rm det} = \dot f_{\rm src}(1+z)^{-2}$ for a BH at redshift $z$ and source-frame frequency $f_{\rm src}$.
Finally, we rescale the curves in \fig{h095_fgw} to obtain values corresponding to optimal source orientation, as explained above.
All these different factors modulate the intrinsic strain inferred from \fig{chi_mbh_h0} to yield Fig.~\ref{fig:horizons}.

For all detectors, the horizon generally increases with initial BH mass and spin, as expected from \fig{chi_mbh_h0}.
Furthermore, higher masses and spins are expected to yield smaller $\dot f$'s, which enables longer coherent times (longer $\tdrift$'s) and, thus, slightly higher sensitivity, cf.~\eq{h95scaling}.
Yet, this tendency is offset by the fact that heavier systems yield lower frequencies (\fig{peak_mbh_chi}), causing the horizon to quickly drop as signals reach the lower end of the detector's sensitive band (cf.~\fig{h095_fgw}).
Moreover, signals from clouds around heavier BHs can more easily get redshifted out of the band.
At the other end of the spectrum, the instruments we consider tend to be more sensitive at higher frequencies, but these correspond to lower masses and, thus, lower radiated power (for a given $\chi_i$).
On the other hand, increasing the BH spin yields both higher GW amplitudes and, to an extent, frequencies.
Unfortunately, however, lower masses and higher spins also result in high $\dot f$'s that make much of that part of parameter space inaccessible to current methods (shaded regions).

All this means, roughly, that the farthest horizons will be obtained for BHs with masses in the range $10^2 \lesssim \mbh_i/\msun \lesssim 10^3$ and spin as high as possible, corresponding to boson masses within $10^{-14} \lesssim \mu/{\rm eV} \lesssim 10^{-12}$ (depending on $\chi_i$).
Even outside this range, these horizons are significantly more distant than the sources at which these searches are generally directed, which tend to lie within the Milky Way (see e.g.~\cite{Riles2017}).

As a concrete example, consider again a GW150914-like remnant with $\mbh_i=\mbhex$ and $\chi_i=\chiex$.
As we saw back in \sect{discuss_amplitude}, this BH would be best matched by a scalar boson with $\mu = 4\times 10^{-13}$ eV.
A scalar of that mass would yield an optimal cloud ($\lb=\mb=1$, $\nr=0$) that radiates gravitational waves ($\lgw=\mgw=2$) at a source-frame frequency of $\fex$ with characteristic amplitude $h_0 =\hEx (5\,{\rm Mpc}/d_L)$, corresponding to the peak in \fig{h0_fgw_alpha}.
From the intersection of the dotted lines in \fig{horizon_aligo}, we see that such a signal would be detectable, at most, up to 12 Mpc away with one aLIGO detector at design sensitivity observing continuously for 1 yr---or, equivalently, ${\sim}20$ Mpc for three such detectors.
This agrees with previous estimates in \cite{Arvanitaki2017}.

Prospects are even better for third-generation detectors, with farther horizons over most of the parameter space.
Third-generation detectors would offer significant improvements for mostly any target with $\mbh_i \lesssim 10^3 M_\odot$.
In particular, we find that Cosmic Explorer could reach ranges of over $10^4$ Mpc ($z\gtrsim 1.4$) for fast-spinning BHs over a range of masses (\fig{horizon_ce}).
The Einstein Telescope could also reach such distances, but for a more limited choice of parameters, and would have shorter reach for most of the sources we consider; on the other hand, this instrument would outperform all others at higher masses (lower frequencies).
Some representative values are presented in Table~\ref{tab:example-bh_horizons} to ease comparison between instruments.

We underscore that the horizons computed in this section (both in \fig{horizons} and Table~\ref{tab:example-bh_horizons}) correspond to single detectors and not to a network.
This is to facilitate rescaling for any specific configuration.
For instance, without taking into account detector orientation and signal polarization, the horizon for a GW150914-like remnant seen by a network of two Cosmic Explorers would be roughly ${\sim}\sqrt{2}\times300\,\text{Mpc} \approx 424\, \text{Mpc}$. 
For a network composed of one Cosmic Explorer and one Einstein Telescope in the full triangular ``D'' configuration, this would be instead ${\sim}\sqrt{(1.06\times100)^2 + 300^2}\,\text{Mpc}\approx 320\, \text{Mpc}$, where the factor of 1.06 accounts for the 6\% improvement in sensitivity to a linearly polarized wave when using the full triangular Einstein Telescope instead of the $90\degree$ configuration assumed in \fig{horizons} and Table~\ref{tab:example-bh_horizons} \cite{Freise:2008dk,Hild:2010id}.

\begin{figure}
	\centering
	\includegraphics[width=\columnwidth]{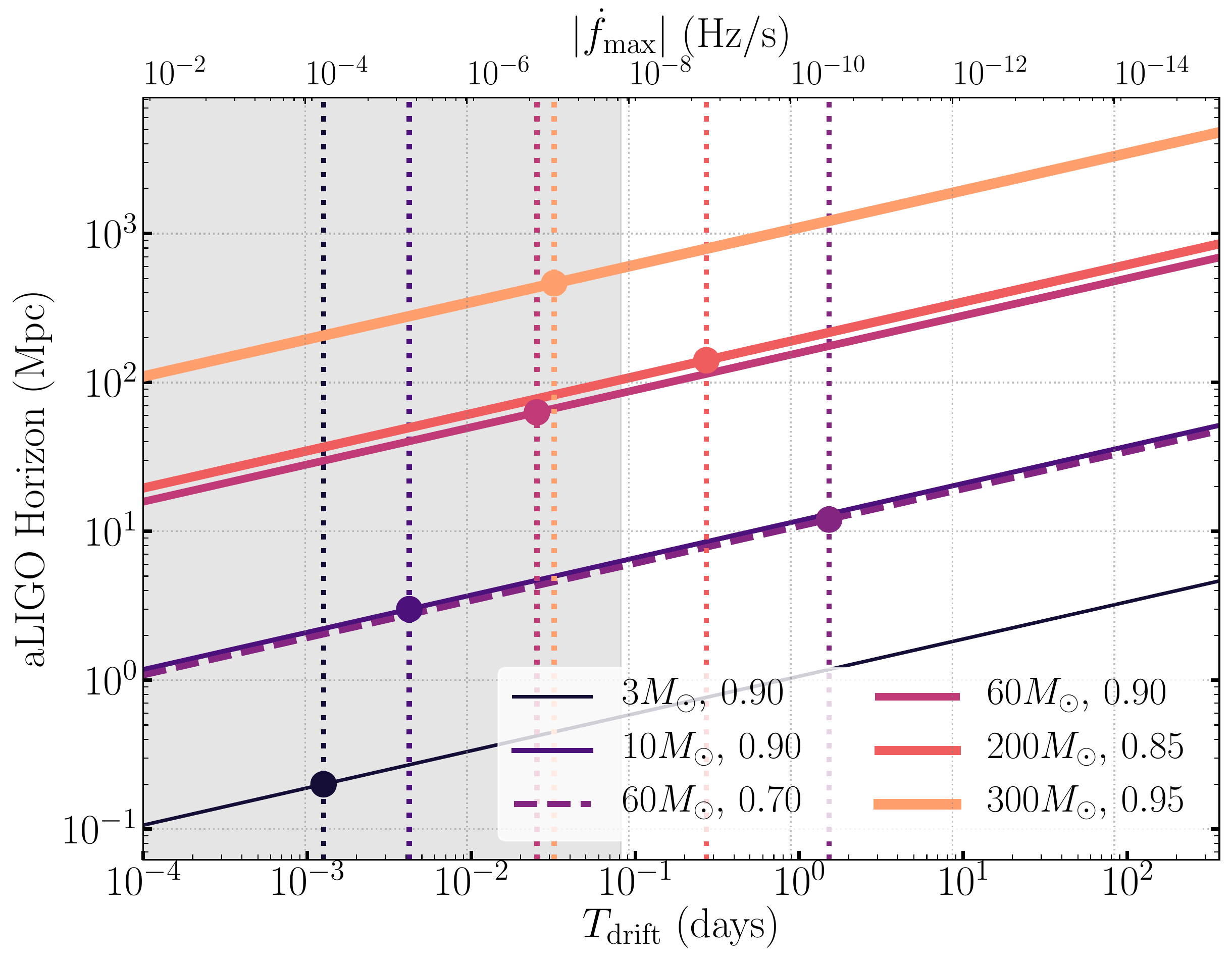}
	\caption[Horizon scaling with $\tdrift$]{{\em Horizon scaling with $\tdrift$}. Advanced LIGO horizon distance ($y$-axis) vs coherence window $\tdrift$ (bottom $x$-axis) and maximum allowed frequency derivative $|\dot{f}_\mathrm{max}|$ (top $x$-axis). Each curve corresponds to one of the systems in Table~\ref{tab:example-bh_params}, labeled by $(\mbh_i, \chi_i)$, and shows the $\tdrift^{1/4}$ scaling of \eq{h95scaling}. The dashed curve highlights the GW150914-like example. Vertical dotted lines mark the maximum $\tdrift$ allowed by the $\dot{\fgw}$ expected from each system, which is also the value assumed in Table~\ref{tab:example-bh_horizons} and \fig{horizons} (circles). We assume $\tobs=1\,\mathrm{yr}$. The shaded region cannot be explored with existing methods.}
	\label{fig:horizon_scaling}
\end{figure}

As mentioned above, for each BH mass and spin, our horizon computation assumes the coherent time window $\tdrift$ is set to the largest value that can accommodate the expected frequency derivative [cf.~\eq{int_T_drift}].
This choice is tailored to optimize sensitivity over the full parameter space [cf.~\eq{h95scaling}].
Instead, we could lengthen $\tdrift$ to slightly increase horizon distances, at the expense of losing all sensitivity to signals with $|\dot{\fgw}| > 1/(2\, \tdrift^2)$.
This would not be advisable except for targets somehow known not to follow \eq{fdot_scalar}.
\fig{horizon_scaling} shows how the horizon scales with $\tdrift$ for fixed $\tobs=1\,\mathrm{yr}$, as implied by \eq{h95scaling}, together with the maximum $|\dot{\fgw}|$ allowable for any given $\tdrift$ (top axis).
The different curves correspond to the representative systems of Table~\ref{tab:example-bh_params}, with vertical dotted lines marking the $\tdrift$ assumed in Table~\ref{tab:example-bh_horizons} and \fig{horizons}, i.e.~the highest $\tdrift$ compatible with \eq{fdot_scalar}.

Note that the horizons for $\tdrift=\tobs=1\,\mathrm{yr}$ shown in \fig{horizon_scaling} are the same as would be obtained in a fully coherent search of that duration, if we had enough computing resources to explicitly search over $\dot{\fgw}$ as well as $\fgw$.
This is the optimal sensitivity we could ever hope to achieve with 1 yr of data (scaling as $\sqrt{\tobs}$).
Even in this idealized case, the aLIGO range to a GW150914-like remnant would fall short of 100 Mpc.

\begin{table}
\centering
  \caption{Scalar-cloud horizons (Mpc) for representative BHs (boson and signal parameters shown in Table~\ref{tab:example-bh_params}).The bold row corresponds to the intersection of the dotted lines in \fig{horizons}.}
  \label{tab:example-bh_horizons}
    \begin{tabular}{c@{\quad} c@{\quad} c@{\quad} c@{\quad} c@{\quad} c@{\quad}}
    \hline
    \hline
    $\mbh_i$ ($\msun$) & $\chi_i$ & aLIGO & Voy & CE & ET \\
    \hline
    3   & 0.90  & 0.2           & 0.4           & 2             & 2             \\
    10  & 0.90  & 3             & 6             & 35            & 24            \\
    {\bf 60}  & {\bf 0.70}  & {\bf 12}          & {\bf 49}        & $\mathbf{3\times10^2}$ &$\mathbf{1\times10^2}$ \\
    60  & 0.90  & 60            & $2\times10^2$ & $1\times10^3$ & $7\times10^2$ \\
    200 & 0.85  & $2\times10^2$ & $6\times10^2$ & $5\times10^3$ & $1\times10^3$ \\
    300 & 0.95  & $5\times10^2$ & $2\times10^3$ & $2\times10^4$ & $4\times10^3$ \\
    \hline
    \hline
  \end{tabular}
\end{table}

\subsubsection{Effect of sky-location uncertainty}
\label{sec:skyloc}

\begin{figure*}
	\centering
	\includegraphics[width=\textwidth]{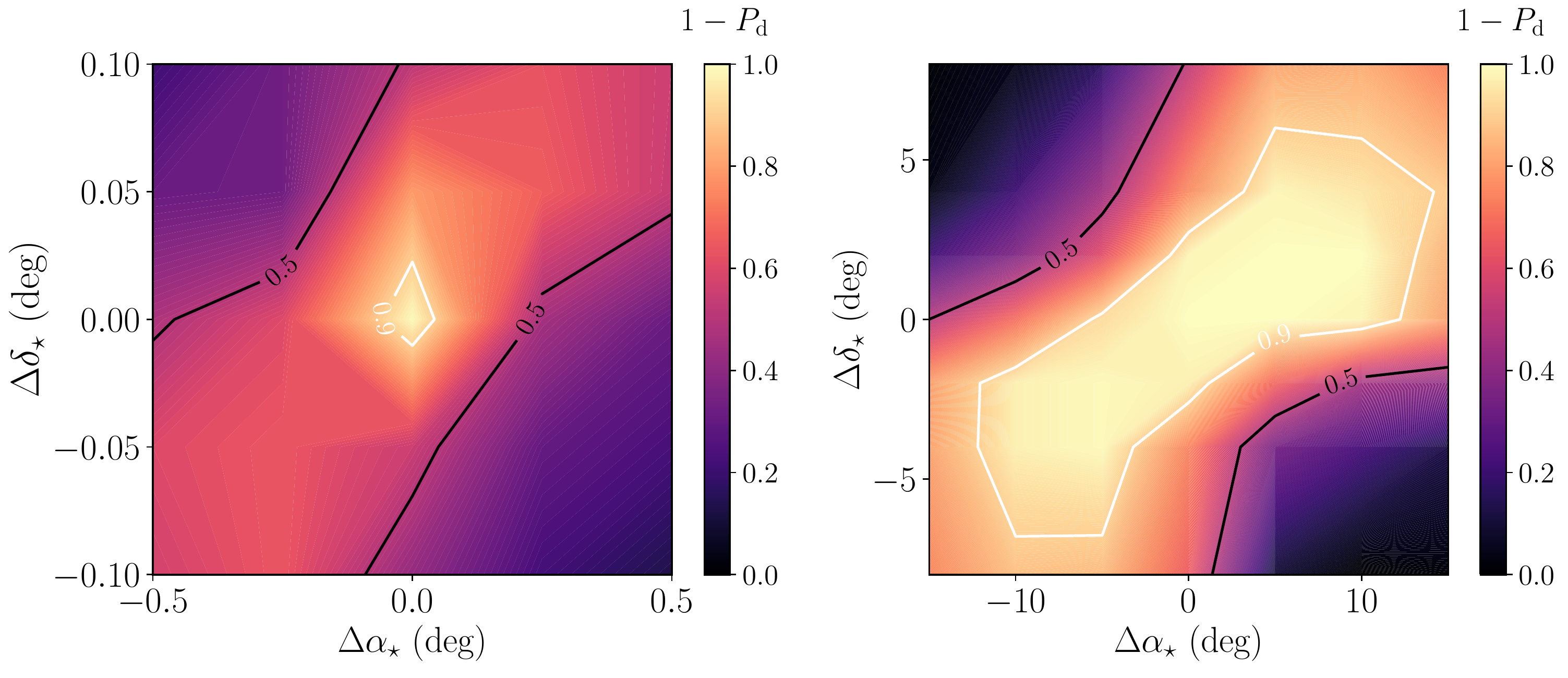}
	\caption[Sky resolution of boson searches]{{\em Sky resolution}.
  Color shows detection efficiency ($1-P_d$, for $P_d$ the false-dismissal probability) as a function of offsets in right ascension ($x$-axis) and declination ($y$-axis) with respect to the true location for injections with $h_0 = 5 \times 10^{-26}$ (left) and  $h_0 = 2 \times 10^{-25}$ (right).
  All other injection parameters are as in Table~\ref{tab:inj-para} and search settings are shown in Table~\ref{tab:search-para}.
  The left (right) plot was interpolated from a square grid with 5 (7) sky locations on each side.}
	\label{fig:sky-loc-sensitivity}
\end{figure*}

We would like to understand the effect of uncertainty in the source sky location on the HMM tracking, mainly motivated by the prospect of following up compact-binary mergers.
In order to find a continuous signal coming from some area in the sky, we would have to analyze the gravitational-wave data with the HMM multiple times assuming slightly different sky locations to tile the patch where the source is thought to lie.
The number of iterations (the number of ``templates'') needed, $N_{\rm sky}$, is determined both by the size of the target area and by the sky resolution of the analysis, which is in turn tied to the frequency resolution of the search and the amplitude of the signal.

To estimate $N_{\rm sky}$, we run two sets of simulations by injecting signals with $h_0=5 \times 10^{-26}$ and $h_0 = 2 \times 10^{-25}$ into simulated Gaussian noise for two aLIGO detectors at design sensitivity.
All other parameters, including the sky location, are as listed in Table~\ref{tab:inj-para}.
HMM searches are then conducted using the settings shown in Table~\ref{tab:search-para}, but for a grid of sky locations in the neighborhood of the injected signal.
In other words, for each injection amplitude ($h_0=5 \times 10^{-26}$ or $h_0 = 2 \times 10^{-25}$) and for each sky location assumed by the search, we inject a signal with random orientation and phase, but location fixed to the value in Table~\ref{tab:inj-para}; we repeat this 200 times to obtain detection probabilities for each of the search locations.

The results of this study are summarized in \fig{sky-loc-sensitivity} for both the soft (left panel) and loud (right panel) injections.
Color in these figures encodes the detection efficiency ($1-P_d$) for searches assuming a sky location indicated by their offset in right ascension ($\Delta \ra$) and declination ($\Delta \dec$) with respect to the true location ($\Delta \ra=\Delta \dec=0$).
The white (black) contours mark points at which signals were detected 90\% (50\%) of the time at \red{1\% false-alarm} probability.
Notice that, for the weaker signal, the 90\%-contour encloses an area of ${\sim}0.001\,{\rm deg}^2$ around the true location, while for the stronger signal this is roughly four orders of magnitude larger.
We may take the size of the 90\% contours as indicative of the spacing of the sky grid needed to capture a signal.

In an actual search, we need $N_{\rm sky} \sim 10^3$ sky templates per deg$^2$ to detect a weak signal near the detection limit. The sky resolution generally agrees with other coherent or semi-coherent CW search methods \cite{Brady1998,Brady2000}. Here we discuss the search feasibility given the required number of sky templates.
As a representative example, consider that the existing three-detector network (Advanced LIGO and Advanced Virgo) was able to localize the binary-neutron start merger GW170817 to a sky region spanning ${\sim}30$\,deg$^2$ with 90\% credibility \cite{gw170817}.
Based on \fig{sky-loc-sensitivity}, we would then need $N_{\rm sky} \sim 10^4$ to obtain 90\% detection efficiency of a signal with $h_0=5 \times 10^{-26}$ lying somewhere inside the GW170817 90\%-credible region; by contrast, $N_{\rm sky} \sim 3$ would suffice for a signal with $h_0=2 \times 10^{-25}$. 
Based on estimates for standard computing architectures and algorithm settings, finding a signal at 200 Hz with $h_0=5 \times 10^{-26}$ in a ${\sim}30$\,deg$^2$-region ($\tobs = 80$\,d, $N_{\rm ifo} = 2$) would take 5 days of computing on 1k CPUs, which is feasible but not cheap.
Note that $N_{\rm sky}$ scales as $f^2$ \cite{Brady1998,Brady2000,Wette2013,Wette2015}, so more templates will be required at higher frequencies.
Because computing cost scales directly with $N_{\rm sky}$, the burden will be vastly reduced once more gravitational detectors join the network and the sky locations reach the projected $\mathcal{O}(1\, {\rm deg}^2)$ \cite{osLRR2018}.

\subsection{Potential sources} \label{sec:targets}

The discussion thus far has been largely unconcerned with the kind of BH being targeted.
In this section, we flesh out the implications of the above conclusions for two types of promising sources: remnants from compact-binary mergers (\sect{remnants}) and BHs in x-ray binaries (\sect{xray}).

\subsubsection{Merger remnants}
\label{sec:remnants}

As pointed out before \cite{Arvanitaki2017,Baryakhtar2017}, nearby CBC remnants would be ideal targets for searches for gravitational signals from ultralight bosons.
Because we witness their birth firsthand, the age of remnant BHs is perfectly known and their mass and spin well constrained.
This would enable accurate estimation of the continuous-signal amplitude that should be expected for any given $\fs$ (\sect{discuss_amplitude}), allowing us to potentially place interesting constraints on the existence of matching bosons.
Furthermore, the location and orientation inferred from the initial chirp would allow us to take advantage of existing infrastructure for directed searches for continuous waves in LIGO and Virgo data (Secs. \ref{sec:methods} and \ref{sec:analysis}).

Ideally, we would follow up any and all mergers, as soon as a reasonable time has passed for the cloud to form (\sect{discuss_timescales}).
In practice, however, we may be limited by the uncertainty in the sky location.
Signals detected with only two instruments will be too loosely localized to allow for followup (e.g., LIGO's first detection was localized to sky region of ${\sim}260\,{\rm deg}^2$ \cite{gw150914}).
Fortunately, as we saw in \sect{skyloc}, the localization provided by a three-detector network would already be manageable with existing computational resources (e.g.~${\sim}28\,{\rm deg}^2$ for the binary neutron star \cite{gw170817}).
This will be further improved when more, and more sensitive, detectors join the effort: a network including LIGO India \cite{Iyer2011,Unnikrishnan:2013qwa} and KAGRA \cite{Somiya:2011np,Aso:2013eba}, on top of the three existing detectors operating at design sensitivity, is expected to routinely locate events to an area of order ${\sim}1\,{\rm deg}^2$ \cite{osLRR2018}.
Regardless of the number of instruments, events with an electromagnetic counterpart (e.g., mergers involving a neutron star) will always be sufficiently well localized.
The presence of an electromagnetic counterpart does not directly lower the detection threshold, but does significantly reduce the number of required sky templates and makes it practical to achieve the desired sensitivity with reasonable computing cost (Sec.~\ref{sec:analysis}).

Extracting information about bosons from one of these observations would also require good knowledge of the remnant distance and orientation.
This is required to translate strain ($h_0$) constraints into limits on radiated power ($\dot E_{\rm GW}$), which can then be turned into statements about the existence of a boson with a given mass [cf.~Eqs.~\eqref{eq:power} and \eqref{eq:h0}].
In particular, if the source is too far away, the signal from the hypothetical cloud would be undetectable at Earth, rendering constraints on its amplitude moot.
Thus, if the distance is not determined by other means (e.g.~association with a host galaxy), the implications for bosons will be contingent on the uncertainty in the luminosity distance inferred from the CBC observation.

In the case of a scalar cloud, for most remnant masses and spins, the source would have to be relatively close for the signal to be detectable by ground-based detectors (\fig{horizons}).
For a second-generation network at design sensitivity, the horizon would lie below $100\,{\rm Mpc}\,\times \sqrt{N_{\rm ifo}}$ for most signal parameters (\fig{horizon_aligo}).
Given that we have yet to observe a BH merger that close \cite{gw150914,gw151226,o1bbh,gw170104,gw170608,gw170814,gw170817}, this projection is not too auspicious.
Yet, note that the horizon can reach close to $10^3\,{\rm Mpc}\,\times \sqrt{N_{\rm ifo}}$ in some regions of parameter space---although taking advantage of this with existing algorithms would require a population CBCs yielding remnants with $\mbh \gtrsim 100\,\msun$ ($\dot f < 10^{-8}$ Hz/s).
Estimates of rates from BH population models were provided in \cite{Arvanitaki2017} based on the nonrelativistic approximation to the amplitude \eq{h0_scalar_approx}.

As we saw in \sect{horizons}, prospects are better for next-generation detectors, especially Cosmic Explorer (\fig{horizon_ce}).
Even then, a good reach to remnants with $\mbh\sim \mathcal{O}(10\msun)$ would require spins roughly $\gtrsim 0.85$, possibly less depending on the mass.
Although we have not yet observed any such events \cite{gw150914,gw151226,o1bbh,gw170104,gw170608,gw170814,gw170817}, numerical-relativity simulations routinely produce remnants with such spins \cite{Mroue:2013xna,Lovelace:2014twa,Scheel:2014ina,Chu:2015kft}.
Note that the horizons for boson signals are always significantly closer than those for compact-binary coalescences \cite{Evans:2016mbw}.

The vector case is slightly different.
Detection horizons are in principle considerably farther for vector clouds due to the intrinsically higher radiated power (see \sect{theory_gws}), making most remnant masses and spins accessible.
However, more radiated power also means shorter cloud lifetimes and, consequently, faster rates of change for the signal amplitude and frequency [cf.~\eq{fdot_vector}].
For much of the parameter space, the expected signal would then evolve too rapidly for existing continuous-wave algorithms to handle (see discussion in \sect{analysis}).
Therefore, the more powerful and quickly-evolving vector signals would currently not be detectable, effectively reducing our horizon to such sources. 
Detection rates for vectors taking this into account were estimated in \cite{Baryakhtar2017} by using the nonrelativistic approximation of \eq{h0_vector_approx}.

To get a sense of our potential reach to vector signals, we may use \eq{h95scaling} to obtain vector horizons starting from the scalar ones in \fig{horizons}.
To do this, note that $h_0^{\rm (v)}/h_0^{\rm (s)} \approx 5\times 10^3 (0.1/\alpha)^2$ by \eq{h0_scalar_approx} and \eq{h0_vector_approx}, while $\dot{\fgw}^{\rm (v)}/\dot{\fgw}^{\rm (s)} \approx 3\times 10^7 (0.1/\alpha)^4$ by \eq{fdot_scalar} and \eq{fdot_vector}.
Then, if we had the analysis infrastructure to handle the quickly-varying vector signals and could coherently search over $\fgw$ and $\dot{\fgw}$ (cf.~\fig{horizon_scaling}), our horizon for a vector signal from a GW150914-like remnant would be ${\sim}1575$ Mpc, instead of the ${\sim}12$ Mpc for scalar signals (Table \ref{tab:example-bh_horizons}).
This assumes $\alpha=0.1$ and $\tobs=1\,\mathrm{d}$ and $\tdrift= 23\,\mathrm{s}$, which is the highest value consistent with $\dot{\fgw}^{\rm (v)}$ for that system.

Even for scalar signals, the restriction to small frequency derivatives is quite detrimental, preventing us from accessing higher boson masses (lower BH masses).
Because the estimates of \eq{fdot_scalar} and \eq{fdot_vector} are only approximate, there is still sense in searching for signals with $\dot f > 10^{-8}$ Hz\,s$^{-1}$ in the shaded regions of \fig{horizons}---although a negative result would be harder to interpret as evidence against the existence of a boson in that mass range.
As suggested above, this is strong motivation to adapt analysis techniques that can handle quickly evolving continuous signals to make them suitable for boson searches---this is work in progress.

Finally, note that we expect to infer the remnant mass and spin from the CBC signal with enough precision to obtain a reasonably accurate prediction of the cloud GW amplitude for any given boson mass.
For instance, the mass and dimensionless spin of the GW150914 remnant were each measured with one-sided relative errors of under 10\% at 90\% credibility, which is sufficiently narrow to make a followup search possible (\sect{analysis}).
The characteristic magnitude of such errors is expected to be significantly reduced for detections at higher signal-to-noise ratio, which should be commonplace once the current network achieves design sensitivity and for next-generation detectors (see, e.g., \cite{Gaebel:2017zys,Vitale:2016icu,osLRR2018}).

In any case, a simplistic way to deal with parameter uncertainty would be to compute the optimal strain for a cloud around a BH corresponding to the upper bounds of the mass and spin credible intervals.
A value computed that way would itself be an upper limit on the boson strain, because this quantity scales directly with mass and spin (\fig{chi_mbh_h0}).
Alternatively, a rigorous statistical analysis would take in the full-dimensional posterior probability density on the BH parameters (intrinsic and extrinsic) and marginalize over all parameters to obtain a posterior on the expected boson strain as a function of $\fs$.
The development of this more sophisticated strategy is work in progress.

\subsubsection{X-ray binaries}
\label{sec:xray}

Another type of potentially interesting targets are known BHs in x-ray binaries (see e.g.~\cite{Remillard:2006fc,Middleton2016} for reviews treating such systems).
The relevance of x-ray binaries to this research program has been pointed out since the outset (e.g.~\cite{Arvanitaki2011,Arvanitaki2015,Yoshino2015,Baryakhtar2017}).
They have the advantage of being much closer and better-located in the sky compared to the CBC remnants, with good measurements of their mass and, in some cases, spin \cite{Miller:2014aaa}.
In fact, some limits on the boson-mass space have already been placed contingent on these measurements, \red{roughly excluding the mass interval $10^{-12} \lesssim \mu/{\rm eV} \lesssim 10^{-11}$ for scalars \cite{Arvanitaki2017,Cardoso2018} and $10^{-13} \lesssim \mu/{\rm eV} \lesssim 10^{-11}$ for vectors \cite{Baryakhtar2017,Cardoso2018}.}
Unfortunately, there is large uncertainty about the age and history of these systems, as well as important caveats about the systematics affecting their spin measurements \cite{Reynolds:2013qqa,McClintock:2013vwa}.
Furthermore, the effect of the active astrophysical environments surrounding these BHs is only understood at the order-of-magnitude level~\cite{Arvanitaki2015,Baryakhtar2017,Baumann:2018vus}.
For all these reasons, boson constraints derived from existing observations of x-ray binaries should be interpreted with caution.

There are also data-analysis challenges intrinsic to signals coming from sources in a binary system: the Doppler modulation due to the motion of the source within the binary causes the signal power to spread over multiple frequencies.
The signal must then be collected from ``orbital sidebands'' that span a frequency band 
\beq
B \approx 4\pi \frac{f_0 a_0}{cP}\, ,
\eeq
where $a_0$ is the BH's projected semimajor axis, $P$ is the orbital period, and $c$ is the speed of light \cite{Suvorova2016}.
Frequency-domain matched filters, like those presented in \cite{Suvorova2016, Suvorova2017}, can be applied to sum up the distributed signal power using (imperfect) knowledge of the orbital parameters.
Those methods would generally demand $B \lesssim 0.5$\,Hz in order to achieve the required sensitivity (see \sect{sensitivity}).
It becomes prohibitively expensive to detect a weak signal from a binary if the orbital parameters (e.g., $P$, $a_0$, and time of passage through the orbit's ascending node $T_P$) are poorly measured.
More details can be found in Sec. III B of Ref.~\cite{Suvorova2017}.

As an example, consider the nearby Cygnus X-1 binary, which has been proposed as an interesting target for boson searches \cite{Yoshino2015}.
If we take the source parameters in Table~\ref{tab:cygX1-paras} and assume $f_0 \sim 500$\,Hz based on \eq{fgw_approx}, then the power of a signal from Cygnus X-1 would span a frequency band $B\approx 0.3$\,Hz, which is acceptable using existing methods.
The strain amplitude estimated given the source distance is $\sim 10^{-22}$--$10^{-21}$, possibly detectable with aLIGO even with imperfect knowledge of the binary orbit \cite{Suvorova2016}.
However, the non-negligible uncertainty in $a_0$ and the limited knowledge of $T_P$ would require searching a large number of templates. 

Unfortunately, for most of the interesting x-ray systems, the orbital parameters are not well measured electromagnetically, and the sidebands in most of the high-mass x-ray binaries would be broader than ${\sim}1$\,Hz. 
BHs in low-mass x-ray binaries (i.e., the companion star is less massive and hence $B$ is generally narrower) with well-measured parameters are likely better candidates.
Due to these considerations, only a handful of potential sources will be of interest and targets in x-ray binaries will need to be chosen carefully.
Future improved electromagnetic measurements of x-ray systems would benefit the analysis by improving overall sensitivity and result interpretability.

\begin{table}
\begin{ruledtabular}
	\caption{Cygnus X-1 parameters.}
	\label{tab:cygX1-paras}
\begin{tabular}{l@{\quad} c@{\quad}c@{\quad}c@{\quad}}
		Parameter  & Symbol& Value & Ref.\\
		\midrule
		Mass ($\msun$) & $M$& $14.8\pm1.0$ & \cite{Orosz2011,Middleton2016}\\
		Spin & $\chi$& $\geq 0.95$& \cite{Middleton2016}\\
		Right ascension & $\ra$& $19^{\rm h}58^{\rm m}22^{\rm s}$ & \cite{Reid2011}\\
		Declination  & $\dec$ & $35^{\circ}12'0.6''$ & \cite{Reid2011}\\
		Inclination (deg) & $\iota$& $27.1\pm0.8$&\cite{Orosz2011}\\
		Distance (kpc) & $r$& $1.86^{+0.12}_{-0.11}$ &\cite{Middleton2016}\\
		Orbital period (days) & $P$& 5.6 &\cite{Iorio2008,Orosz2011}\\
		Proj.\ semimajor axis (l-s) & $a_0$& $25.56^{+3.15}_{-3.11}$&\cite{Orosz2011}\\
	\end{tabular}
\end{ruledtabular}
\end{table}

\section{Conclusion}\label{sec:conclusion}

Black-hole superradiance could be the key that allows gravitational-wave detectors to uncover evidence of ultralight bosons, thus bringing particle physics within the reach of gravitational-wave science.
In this paper, we explored the prospect for achieving this goal by looking for the continuous gravitational signals expected from scalar and vector clouds, using searches directed at known BHs.

We began by reviewing the physics of boson clouds (\sect{theory}) and examined in detail the properties of continuous signals from clouds around a known BH (\sect{signal}).
In doing so, we hoped to provide a bridge between the theory and data-analysis literatures.
We then used numerical techniques, combined with the latest analytic results, to compute the features of gravitational waves emitted by scalar clouds around BHs with different initial parameters (Figs.~\ref{fig:h0_fgw_alpha}--\ref{fig:alpha_mbh_tgw}).

We put forward the use of hidden Markov model (HMM) tracking \cite{Suvorova2016,Sun2018,ScoX1ViterbiO1, gw170817} to carry out directed searches for boson signals (\sect{methods}).
This strategy is well suited to searches for gravitational waves from boson clouds because its computational efficiency enables the coverage of a wide range of signal parameters, and because it does not rely on restrictive waveform models.
This makes it ideal to search for signals over a broad frequency band (cf.~\fig{h0_fgw_alpha}), even when the location of the source is only loosely known and when there is potential uncertainty in the signal morphology.
We demonstrated this through a series of Monte-Carlo simulations (\sect{analysis}).

From our simulations, we obtained an empirical estimate of the sensitivity of directed searches to boson signals in data from future ground-based detectors: aLIGO design, LIGO Voyager, Cosmic Explorer and the Einstein telescope (\fig{h095_fgw}).
For scalar clouds, we translated the expected strain sensitivities into detection horizons for those four detectors (\fig{horizons}), assuming one year of observation by a single detector.
We found that, for a second-generation network at design sensitivity, the horizon would lie below $100\,{\rm Mpc}\,\times \sqrt{N_{\rm ifo}}$ for most signal parameters; prospects are better for next-generation detectors, especially Cosmic Explorer, for which horizons could reach up to ${\sim}10^5$\, Mpc.
Generally speaking, these horizons lie much farther than the sources at which continuous-wave searches are generally directed \cite{Riles2017}, but significantly closer than horizons for compact-binary coalescences \cite{Evans:2016mbw}.
Some representative values are shown in Table~\ref{tab:example-bh_params}, and their scaling with the drift time, one of the primary algorithm settings, is shown \fig{horizon_scaling}.

In computing signal amplitudes from scalar clouds, we numerically solved the evolution equations governing cloud growth and made use of numerical estimates from BH perturbation theory to obtain the radiated power \cite{Brito2017}.
Furthermore, to estimate horizons, we incorporated the effect of redshifts on the signal frequency and frequency derivative.
We also took into account that the settings of the search algorithm should be varied across parameter space for optimal performance.
This allowed us to obtain sensitivity estimates that should be more reliable than previously published projections.

Finally, we discussed implications for the followup of remnants from compact-binary coalescences (\sect{remnants}), as well as BHs in x-ray binaries (\sect{xray}).
We explored the impact of uncertainties in the source's sky location, and showed that HMM tracking will be able to efficiently cover the localization credible-regions obtained from CBC signals with a network of at least three detectors.
We also discussed the challenges intrinsic to vector signals, that make their analysis difficult in spite of their higher radiated power.
We emphasized the strong motivation to extend existing search techniques to handle signals with higher frequency derivatives, so as to bring a significant portion of the scalar and vector signal space into reach.
The implementation of such techniques, as well as development of statistical strategies to rigorously handle uncertainty in BH parameters, is work in progress.

\begin{acknowledgments}
The authors thank Evan Hall for useful input regarding the sensitivities of present and future detectors.
M.I.\ and R.B.\ thank Asimina Arvanitaki, Masha Baryakhtar, William East and Robert Lasenby for organizing the meeting ``Searching for New Particles with Black Hole Superradiance'' held at the Perimeter Institute for Theoretical Physics in May 2018.
Research at Perimeter Institute is supported by the Government of Canada through Industry Canada and by the Province of Ontario through the Ministry of Economic Development \& Innovation.
M.I.\ and L.S.\ are members of the LIGO Laboratory.
LIGO was constructed by the California Institute of Technology and Massachusetts Institute of Technology with funding from the National Science Foundation and operates under cooperative agreement PHY--0757058.
Support for this work was provided by NASA through the NASA Hubble Fellowship grant \#HST--HF2--51410.001--A awarded by the Space Telescope Science Institute, which is operated by the Association of Universities for Research in Astronomy, Inc., for NASA, under contract NAS5--26555.
L.S.\ was supported by an Australian Research Training Program Stipend Scholarship and the Albert Shimmins Fund at earlier stages of this project.
The research was also supported by Australian Research Council (ARC) Discovery Project DP170103625 and the ARC Centre of Excellence for Gravitational Wave Discovery CE170100004. 
R.B. acknowledges financial support from the European Union's Horizon 2020 research and innovation programme under the Marie Sk\l odowska-Curie grant agreement No. 792862.
This paper carries LIGO Document Number \dcc.
\end{acknowledgments}

\appendix
\section{Frequency drift} \label{app:fdot}

The gravitational self-energy of the cloud affects the boson's eigenfrequencies and, consequently, the gravitational wave frequency.
As the cloud dissipates due to gravitational emission, this causes an increase in the emitted signal frequency, similar to what happens in a compact-binary coalescence.
Therefore, we may treat the system adiabatically to obtain the frequency drift from the radiated power.
This computation was presented in \cite{Baryakhtar2017} for vectors, and we reproduce it here for scalars.

The gravitational self-energy of a bound state per particle is given by
\beq
U_{c} = -G\frac{m_b}{M_c}\int{\frac{\rho(r,\theta,\phi) \,m(r)}{r} \text{d}^{3}\mathbf{x}} \,,
\eeq
where $m_b$ is the mass of the boson field, $M_c$ the overall mass in the cloud, $\rho(r,\theta,\phi)$ is its density and $m(r)$ is the the mass of cloud enclosed in the radius $r$, namely
\beq
m(r) = \int_{0}^{r} \rho(r,\theta,\phi) \text{ d}^{3}\mathbf{x}\,.
\eeq
 The  rate of change of the GW frequency can then estimated by~\cite{Baryakhtar2017}
\beq
\dot \fgw \simeq \frac{1}{2\pi\hbar} \times 2 \dot U_{c}\, .
\eeq
As the cloud dissipates, the total mass of the system decreases, causing the binding energy to increase ($U_c<0$) and the GW frequency to increase.

The dominant scalar field mode can be approximated by the $\ell=m=1$ hydrogen wave function, with a density $\rho$ given by Eq.(11) in Ref.~\cite{Brito2014}.
After some algebra one finds that, at leading order in $\fs$,
\beq
\dot\fgw \simeq \frac{93}{1024}\frac{c\,\alpha^3}{\pi G}\frac{\dot{E}_{\rm GW}}{M^2}
=\frac{93}{1024} \frac{c^3\alpha^3}{\pi G}\frac{M_c}{M^2} \frac{1}{\tgw} \, ,
\eeq
where we have used $\dot M_c = - \dot{E}_{\rm GW} c^{-2}$ and $ \dot{E}_{\rm GW}=M_c c^2/\tau_{\rm GW}$.
Using \eq{mc} to approximate $M_c \sim \alpha M_i \chi_i$ (valid in the limit $\alpha\ll 1$ for $\mb=1$) and \eq{tgw_scalar} to write $\tgw \sim GM^2/(0.025M_c\alpha^{14}c^3)$, we get
\beq
\dot \fgw \simeq 3\times 10^{-14}\, {\rm Hz/ s} \left(\frac{10M_{\odot}}{M}\right)^2 \left(\frac{\alpha}{0.1}\right)^{19} \chi_i^2\, .
\eeq

\bibliography{gw}

\begin{thebibliography}{116}%
\makeatletter
\providecommand \@ifxundefined [1]{%
 \@ifx{#1\undefined}
}%
\providecommand \@ifnum [1]{%
 \ifnum #1\expandafter \@firstoftwo
 \else \expandafter \@secondoftwo
 \fi
}%
\providecommand \@ifx [1]{%
 \ifx #1\expandafter \@firstoftwo
 \else \expandafter \@secondoftwo
 \fi
}%
\providecommand \natexlab [1]{#1}%
\providecommand \enquote  [1]{``#1''}%
\providecommand \bibnamefont  [1]{#1}%
\providecommand \bibfnamefont [1]{#1}%
\providecommand \citenamefont [1]{#1}%
\providecommand \href@noop [0]{\@secondoftwo}%
\providecommand \href [0]{\begingroup \@sanitize@url \@href}%
\providecommand \@href[1]{\@@startlink{#1}\@@href}%
\providecommand \@@href[1]{\endgroup#1\@@endlink}%
\providecommand \@sanitize@url [0]{\catcode `\\12\catcode `\$12\catcode
  `\&12\catcode `\#12\catcode `\^12\catcode `\_12\catcode `\%12\relax}%
\providecommand \@@startlink[1]{}%
\providecommand \@@endlink[0]{}%
\providecommand \url  [0]{\begingroup\@sanitize@url \@url }%
\providecommand \@url [1]{\endgroup\@href {#1}{\urlprefix }}%
\providecommand \urlprefix  [0]{URL }%
\providecommand \Eprint [0]{\href }%
\providecommand \doibase [0]{http://dx.doi.org/}%
\providecommand \selectlanguage [0]{\@gobble}%
\providecommand \bibinfo  [0]{\@secondoftwo}%
\providecommand \bibfield  [0]{\@secondoftwo}%
\providecommand \translation [1]{[#1]}%
\providecommand \BibitemOpen [0]{}%
\providecommand \bibitemStop [0]{}%
\providecommand \bibitemNoStop [0]{.\EOS\space}%
\providecommand \EOS [0]{\spacefactor3000\relax}%
\providecommand \BibitemShut  [1]{\csname bibitem#1\endcsname}%
\let\auto@bib@innerbib\@empty
\bibitem [{\citenamefont {Aasi}\ \emph {et~al.}(2015)\citenamefont {Aasi} \emph
  {et~al.}}]{aLIGO}%
  \BibitemOpen
  \bibfield  {author} {\bibinfo {author} {\bibfnamefont {J.}~\bibnamefont
  {Aasi}} \emph {et~al.} (\bibinfo {collaboration} {LIGO Scientific}),\
  }\bibfield  {title} {\enquote {\bibinfo {title} {{Advanced LIGO}},}\ }\href
  {\doibase 10.1088/0264-9381/32/7/074001} {\bibfield  {journal} {\bibinfo
  {journal} {Class. Quant. Grav.}\ }\textbf {\bibinfo {volume} {32}},\ \bibinfo
  {pages} {074001} (\bibinfo {year} {2015})},\ \Eprint
  {http://arxiv.org/abs/1411.4547} {arXiv:1411.4547 [gr-qc]} \BibitemShut
  {NoStop}%
\bibitem [{\citenamefont {Acernese}\ \emph {et~al.}(2015)\citenamefont
  {Acernese} \emph {et~al.}}]{Virgo}%
  \BibitemOpen
  \bibfield  {author} {\bibinfo {author} {\bibfnamefont {F.}~\bibnamefont
  {Acernese}} \emph {et~al.},\ }\bibfield  {title} {\enquote {\bibinfo {title}
  {{Advanced Virgo: a second-generation interferometric gravitational wave
  detector}},}\ }\href {\doibase 10.1088/0264-9381/32/2/024001} {\bibfield
  {journal} {\bibinfo  {journal} {Class. Quant. Grav.}\ }\textbf {\bibinfo
  {volume} {32}},\ \bibinfo {pages} {024001} (\bibinfo {year}
  {2015})}\BibitemShut {NoStop}%
\bibitem [{\citenamefont {Abbott}\ \emph
  {et~al.}(2016{\natexlab{a}})\citenamefont {Abbott} \emph
  {et~al.}}]{gw150914}%
  \BibitemOpen
  \bibfield  {author} {\bibinfo {author} {\bibfnamefont {B.~P.}\ \bibnamefont
  {Abbott}} \emph {et~al.} (\bibinfo {collaboration} {LIGO Scientific
  Collaboration, Virgo Collaboration}),\ }\bibfield  {title} {\enquote
  {\bibinfo {title} {{Observation of Gravitational Waves from a Binary Black
  Hole Merger}},}\ }\href {\doibase 10.1103/PhysRevLett.116.061102} {\bibfield
  {journal} {\bibinfo  {journal} {Phys. Rev. Lett.}\ }\textbf {\bibinfo
  {volume} {116}},\ \bibinfo {pages} {061102} (\bibinfo {year}
  {2016}{\natexlab{a}})}\BibitemShut {NoStop}%
\bibitem [{\citenamefont {Abbott}\ \emph
  {et~al.}(2016{\natexlab{b}})\citenamefont {Abbott} \emph
  {et~al.}}]{gw151226}%
  \BibitemOpen
  \bibfield  {author} {\bibinfo {author} {\bibfnamefont {B.~P.}\ \bibnamefont
  {Abbott}} \emph {et~al.} (\bibinfo {collaboration} {LIGO Scientific
  Collaboration, Virgo Collaboration}),\ }\bibfield  {title} {\enquote
  {\bibinfo {title} {{GW151226: Observation of Gravitational Waves from a
  22-Solar-Mass Binary Black Hole Coalescence}},}\ }\href {\doibase
  10.1103/PhysRevLett.116.241103} {\bibfield  {journal} {\bibinfo  {journal}
  {Phys. Rev. Lett.}\ }\textbf {\bibinfo {volume} {116}},\ \bibinfo {pages}
  {241103} (\bibinfo {year} {2016}{\natexlab{b}})}\BibitemShut {NoStop}%
\bibitem [{\citenamefont {Abbott}\ \emph
  {et~al.}(2016{\natexlab{c}})\citenamefont {Abbott} \emph {et~al.}}]{o1bbh}%
  \BibitemOpen
  \bibfield  {author} {\bibinfo {author} {\bibfnamefont {B.~P.}\ \bibnamefont
  {Abbott}} \emph {et~al.} (\bibinfo {collaboration} {LIGO Scientific
  Collaboration, Virgo Collaboration}),\ }\bibfield  {title} {\enquote
  {\bibinfo {title} {{Binary Black Hole Mergers in the First Advanced LIGO
  Observing Run}},}\ }\href {\doibase 10.1103/PhysRevX.6.041015} {\bibfield
  {journal} {\bibinfo  {journal} {Phys. Rev. X}\ }\textbf {\bibinfo {volume}
  {6}},\ \bibinfo {pages} {041015} (\bibinfo {year}
  {2016}{\natexlab{c}})}\BibitemShut {NoStop}%
\bibitem [{\citenamefont {Abbott}\ \emph
  {et~al.}(2017{\natexlab{a}})\citenamefont {Abbott} \emph
  {et~al.}}]{gw170104}%
  \BibitemOpen
  \bibfield  {author} {\bibinfo {author} {\bibfnamefont {B.~P.}\ \bibnamefont
  {Abbott}} \emph {et~al.} (\bibinfo {collaboration} {LIGO Scientific
  Collaboration, Virgo Collaboration}),\ }\bibfield  {title} {\enquote
  {\bibinfo {title} {{GW170104: Observation of a 50-Solar-Mass Binary Black
  Hole Coalescence at Redshift 0.2}},}\ }\href {\doibase
  10.1103/PhysRevLett.118.221101} {\bibfield  {journal} {\bibinfo  {journal}
  {Phys. Rev. Lett.}\ }\textbf {\bibinfo {volume} {118}},\ \bibinfo {pages}
  {221101} (\bibinfo {year} {2017}{\natexlab{a}})}\BibitemShut {NoStop}%
\bibitem [{\citenamefont {Abbott}\ \emph
  {et~al.}(2017{\natexlab{b}})\citenamefont {Abbott} \emph
  {et~al.}}]{gw170608}%
  \BibitemOpen
  \bibfield  {author} {\bibinfo {author} {\bibfnamefont {B.~P.}\ \bibnamefont
  {Abbott}} \emph {et~al.} (\bibinfo {collaboration} {LIGO Scientific
  Collaboration, Virgo Collaboration}),\ }\bibfield  {title} {\enquote
  {\bibinfo {title} {{GW170608: Observation of a 19 Solar-mass Binary Black
  Hole Coalescence}},}\ }\href {\doibase 10.3847/2041-8213/aa9f0c} {\bibfield
  {journal} {\bibinfo  {journal} {Astrophys. J.}\ }\textbf {\bibinfo {volume}
  {851}},\ \bibinfo {pages} {L35} (\bibinfo {year}
  {2017}{\natexlab{b}})}\BibitemShut {NoStop}%
\bibitem [{\citenamefont {Abbott}\ \emph
  {et~al.}(2017{\natexlab{c}})\citenamefont {Abbott} \emph
  {et~al.}}]{gw170814}%
  \BibitemOpen
  \bibfield  {author} {\bibinfo {author} {\bibfnamefont {B.~P.}\ \bibnamefont
  {Abbott}} \emph {et~al.} (\bibinfo {collaboration} {LIGO Scientific
  Collaboration, Virgo Collaboration}),\ }\bibfield  {title} {\enquote
  {\bibinfo {title} {{GW170814: A Three-Detector Observation of Gravitational
  Waves from a Binary Black Hole Coalescence}},}\ }\href {\doibase
  10.1103/PhysRevLett.119.141101} {\bibfield  {journal} {\bibinfo  {journal}
  {Phys. Rev. Lett.}\ }\textbf {\bibinfo {volume} {119}},\ \bibinfo {pages}
  {141101} (\bibinfo {year} {2017}{\natexlab{c}})}\BibitemShut {NoStop}%
\bibitem [{\citenamefont {Abbott}\ \emph
  {et~al.}(2017{\natexlab{d}})\citenamefont {Abbott} \emph
  {et~al.}}]{gw170817}%
  \BibitemOpen
  \bibfield  {author} {\bibinfo {author} {\bibfnamefont {B.~P.}\ \bibnamefont
  {Abbott}} \emph {et~al.},\ }\bibfield  {title} {\enquote {\bibinfo {title}
  {{GW170817: Observation of Gravitational Waves from a Binary Neutron Star
  Inspiral}},}\ }\href {\doibase 10.1103/PhysRevLett.119.161101} {\bibfield
  {journal} {\bibinfo  {journal} {Phys. Rev. Lett.}\ }\textbf {\bibinfo
  {volume} {119}},\ \bibinfo {pages} {161101} (\bibinfo {year}
  {2017}{\natexlab{d}})}\BibitemShut {NoStop}%
\bibitem [{\citenamefont {Abbott}\ \emph
  {et~al.}(2016{\natexlab{d}})\citenamefont {Abbott} \emph
  {et~al.}}]{gw150914_tgr}%
  \BibitemOpen
  \bibfield  {author} {\bibinfo {author} {\bibfnamefont {B.~P.}\ \bibnamefont
  {Abbott}} \emph {et~al.} (\bibinfo {collaboration} {LIGO Scientific
  Collaboration, Virgo Collaboration}),\ }\bibfield  {title} {\enquote
  {\bibinfo {title} {{Tests of General Relativity with GW150914}},}\ }\href
  {\doibase 10.1103/PhysRevLett.116.221101} {\bibfield  {journal} {\bibinfo
  {journal} {Phys. Rev. Lett.}\ }\textbf {\bibinfo {volume} {116}},\ \bibinfo
  {pages} {221101} (\bibinfo {year} {2016}{\natexlab{d}})}\BibitemShut
  {NoStop}%
\bibitem [{\citenamefont {Yunes}\ \emph {et~al.}(2016)\citenamefont {Yunes},
  \citenamefont {Yagi},\ and\ \citenamefont {Pretorius}}]{Yunes:2016jcc}%
  \BibitemOpen
  \bibfield  {author} {\bibinfo {author} {\bibfnamefont {Nicolas}\ \bibnamefont
  {Yunes}}, \bibinfo {author} {\bibfnamefont {Kent}\ \bibnamefont {Yagi}}, \
  and\ \bibinfo {author} {\bibfnamefont {Frans}\ \bibnamefont {Pretorius}},\
  }\bibfield  {title} {\enquote {\bibinfo {title} {{Theoretical Physics
  Implications of the Binary Black-Hole Mergers GW150914 and GW151226}},}\
  }\href {\doibase 10.1103/PhysRevD.94.084002} {\bibfield  {journal} {\bibinfo
  {journal} {Phys. Rev.}\ }\textbf {\bibinfo {volume} {D94}},\ \bibinfo {pages}
  {084002} (\bibinfo {year} {2016})}\BibitemShut {NoStop}%
\bibitem [{\citenamefont {Abbott}\ \emph
  {et~al.}(2018{\natexlab{a}})\citenamefont {Abbott} \emph
  {et~al.}}]{gw170817_pe}%
  \BibitemOpen
  \bibfield  {author} {\bibinfo {author} {\bibfnamefont {B.~P.}\ \bibnamefont
  {Abbott}} \emph {et~al.} (\bibinfo {collaboration} {LIGO Scientific
  Collaboration, Virgo Collaboration}),\ }\href@noop {} {\enquote {\bibinfo
  {title} {{Properties of the binary neutron star merger GW170817}},}\ }
  (\bibinfo {year} {2018}{\natexlab{a}}),\ \Eprint
  {http://arxiv.org/abs/1805.11579} {arXiv:1805.11579 [gr-qc]} \BibitemShut
  {NoStop}%
\bibitem [{\citenamefont {Abbott}\ \emph
  {et~al.}(2018{\natexlab{b}})\citenamefont {Abbott} \emph
  {et~al.}}]{gw170817_eos}%
  \BibitemOpen
  \bibfield  {author} {\bibinfo {author} {\bibfnamefont {B.~P.}\ \bibnamefont
  {Abbott}} \emph {et~al.} (\bibinfo {collaboration} {LIGO Scientific
  Collaboration, Virgo Collaboration}),\ }\href@noop {} {\enquote {\bibinfo
  {title} {{GW170817: Measurements of neutron star radii and equation of
  state}},}\ } (\bibinfo {year} {2018}{\natexlab{b}}),\ \Eprint
  {http://arxiv.org/abs/1805.11581} {arXiv:1805.11581 [gr-qc]} \BibitemShut
  {NoStop}%
\bibitem [{\citenamefont {Arvanitaki}\ and\ \citenamefont
  {Dubovsky}(2011)}]{Arvanitaki2011}%
  \BibitemOpen
  \bibfield  {author} {\bibinfo {author} {\bibfnamefont {Asimina}\ \bibnamefont
  {Arvanitaki}}\ and\ \bibinfo {author} {\bibfnamefont {Sergei}\ \bibnamefont
  {Dubovsky}},\ }\bibfield  {title} {\enquote {\bibinfo {title} {{Exploring the
  string axiverse with precision black hole physics}},}\ }\href {\doibase
  10.1103/PhysRevD.83.044026} {\bibfield  {journal} {\bibinfo  {journal}
  {Physical Review D}\ }\textbf {\bibinfo {volume} {83}},\ \bibinfo {pages}
  {044026} (\bibinfo {year} {2011})}\BibitemShut {NoStop}%
\bibitem [{\citenamefont {Yoshino}\ and\ \citenamefont
  {Kodama}(2014)}]{Yoshino2014}%
  \BibitemOpen
  \bibfield  {author} {\bibinfo {author} {\bibfnamefont {Hirotaka}\
  \bibnamefont {Yoshino}}\ and\ \bibinfo {author} {\bibfnamefont {Hideo}\
  \bibnamefont {Kodama}},\ }\bibfield  {title} {\enquote {\bibinfo {title}
  {{Gravitational radiation from an axion cloud around a black hole:
  Superradiant phase}},}\ }\href {\doibase 10.1093/ptep/ptu029} {\bibfield
  {journal} {\bibinfo  {journal} {Progress of Theoretical and Experimental
  Physics}\ }\textbf {\bibinfo {volume} {2014}},\ \bibinfo {pages} {43E02--0}
  (\bibinfo {year} {2014})}\BibitemShut {NoStop}%
\bibitem [{\citenamefont {Yoshino}\ and\ \citenamefont
  {Kodama}(2015{\natexlab{a}})}]{Yoshino2015}%
  \BibitemOpen
  \bibfield  {author} {\bibinfo {author} {\bibfnamefont {Hirotaka}\
  \bibnamefont {Yoshino}}\ and\ \bibinfo {author} {\bibfnamefont {Hideo}\
  \bibnamefont {Kodama}},\ }\bibfield  {title} {\enquote {\bibinfo {title}
  {{Probing the string axiverse by gravitational waves from Cygnus X-1}},}\
  }\href {\doibase 10.1093/ptep/ptv067} {\bibfield  {journal} {\bibinfo
  {journal} {Progress of Theoretical and Experimental Physics}\ }\textbf
  {\bibinfo {volume} {2015}} (\bibinfo {year} {2015}{\natexlab{a}}),\
  10.1093/ptep/ptv067}\BibitemShut {NoStop}%
\bibitem [{\citenamefont {Arvanitaki}\ \emph {et~al.}(2015)\citenamefont
  {Arvanitaki}, \citenamefont {Baryakhtar},\ and\ \citenamefont
  {Huang}}]{Arvanitaki2015}%
  \BibitemOpen
  \bibfield  {author} {\bibinfo {author} {\bibfnamefont {Asimina}\ \bibnamefont
  {Arvanitaki}}, \bibinfo {author} {\bibfnamefont {Masha}\ \bibnamefont
  {Baryakhtar}}, \ and\ \bibinfo {author} {\bibfnamefont {Xinlu}\ \bibnamefont
  {Huang}},\ }\bibfield  {title} {\enquote {\bibinfo {title} {{Discovering the
  QCD axion with black holes and gravitational waves}},}\ }\href {\doibase
  10.1103/PhysRevD.91.084011} {\bibfield  {journal} {\bibinfo  {journal}
  {Physical Review D}\ }\textbf {\bibinfo {volume} {91}},\ \bibinfo {pages}
  {084011} (\bibinfo {year} {2015})}\BibitemShut {NoStop}%
\bibitem [{\citenamefont {Arvanitaki}\ \emph {et~al.}(2017)\citenamefont
  {Arvanitaki}, \citenamefont {Baryakhtar}, \citenamefont {Dimopoulos},
  \citenamefont {Dubovsky},\ and\ \citenamefont {Lasenby}}]{Arvanitaki2017}%
  \BibitemOpen
  \bibfield  {author} {\bibinfo {author} {\bibfnamefont {Asimina}\ \bibnamefont
  {Arvanitaki}}, \bibinfo {author} {\bibfnamefont {Masha}\ \bibnamefont
  {Baryakhtar}}, \bibinfo {author} {\bibfnamefont {Savas}\ \bibnamefont
  {Dimopoulos}}, \bibinfo {author} {\bibfnamefont {Sergei}\ \bibnamefont
  {Dubovsky}}, \ and\ \bibinfo {author} {\bibfnamefont {Robert}\ \bibnamefont
  {Lasenby}},\ }\bibfield  {title} {\enquote {\bibinfo {title} {{Black hole
  mergers and the QCD axion at Advanced LIGO}},}\ }\href {\doibase
  10.1103/PhysRevD.95.043001} {\bibfield  {journal} {\bibinfo  {journal}
  {Physical Review D}\ }\textbf {\bibinfo {volume} {95}},\ \bibinfo {pages}
  {043001} (\bibinfo {year} {2017})}\BibitemShut {NoStop}%
\bibitem [{\citenamefont {Brito}\ \emph
  {et~al.}(2017{\natexlab{a}})\citenamefont {Brito}, \citenamefont {{Ghosh}},
  \citenamefont {{Barausse}}, \citenamefont {{Berti}}, \citenamefont
  {{Cardoso}}, \citenamefont {{Dvorkin}}, \citenamefont {{Klein}},\ and\
  \citenamefont {{Pani}}}]{Brito2017-letter}%
  \BibitemOpen
  \bibfield  {author} {\bibinfo {author} {\bibfnamefont {Richard}\ \bibnamefont
  {Brito}}, \bibinfo {author} {\bibfnamefont {S.}~\bibnamefont {{Ghosh}}},
  \bibinfo {author} {\bibfnamefont {E.}~\bibnamefont {{Barausse}}}, \bibinfo
  {author} {\bibfnamefont {E.}~\bibnamefont {{Berti}}}, \bibinfo {author}
  {\bibfnamefont {V.}~\bibnamefont {{Cardoso}}}, \bibinfo {author}
  {\bibfnamefont {I.}~\bibnamefont {{Dvorkin}}}, \bibinfo {author}
  {\bibfnamefont {A.}~\bibnamefont {{Klein}}}, \ and\ \bibinfo {author}
  {\bibfnamefont {P.}~\bibnamefont {{Pani}}},\ }\bibfield  {title} {\enquote
  {\bibinfo {title} {{Stochastic and Resolvable Gravitational Waves from
  Ultralight Bosons}},}\ }\href {\doibase 10.1103/PhysRevLett.119.131101}
  {\bibfield  {journal} {\bibinfo  {journal} {Physical Review Letters}\
  }\textbf {\bibinfo {volume} {119}},\ \bibinfo {pages} {131101} (\bibinfo
  {year} {2017}{\natexlab{a}})}\BibitemShut {NoStop}%
\bibitem [{\citenamefont {Brito}\ \emph
  {et~al.}(2017{\natexlab{b}})\citenamefont {Brito}, \citenamefont {{Ghosh}},
  \citenamefont {{Barausse}}, \citenamefont {{Berti}}, \citenamefont
  {{Cardoso}}, \citenamefont {{Dvorkin}}, \citenamefont {{Klein}},\ and\
  \citenamefont {{Pani}}}]{Brito2017}%
  \BibitemOpen
  \bibfield  {author} {\bibinfo {author} {\bibfnamefont {Richard}\ \bibnamefont
  {Brito}}, \bibinfo {author} {\bibfnamefont {S.}~\bibnamefont {{Ghosh}}},
  \bibinfo {author} {\bibfnamefont {E.}~\bibnamefont {{Barausse}}}, \bibinfo
  {author} {\bibfnamefont {E.}~\bibnamefont {{Berti}}}, \bibinfo {author}
  {\bibfnamefont {V.}~\bibnamefont {{Cardoso}}}, \bibinfo {author}
  {\bibfnamefont {I.}~\bibnamefont {{Dvorkin}}}, \bibinfo {author}
  {\bibfnamefont {A.}~\bibnamefont {{Klein}}}, \ and\ \bibinfo {author}
  {\bibfnamefont {P.}~\bibnamefont {{Pani}}},\ }\bibfield  {title} {\enquote
  {\bibinfo {title} {{Gravitational wave searches for ultralight bosons with
  LIGO and LISA}},}\ }\href {\doibase 10.1103/PhysRevD.96.064050} {\bibfield
  {journal} {\bibinfo  {journal} {Physical Review D}\ }\textbf {\bibinfo
  {volume} {96}},\ \bibinfo {pages} {064050} (\bibinfo {year}
  {2017}{\natexlab{b}})}\BibitemShut {NoStop}%
\bibitem [{\citenamefont {Baryakhtar}\ \emph {et~al.}(2017)\citenamefont
  {Baryakhtar}, \citenamefont {Lasenby},\ and\ \citenamefont
  {Teo}}]{Baryakhtar2017}%
  \BibitemOpen
  \bibfield  {author} {\bibinfo {author} {\bibfnamefont {Masha}\ \bibnamefont
  {Baryakhtar}}, \bibinfo {author} {\bibfnamefont {Robert}\ \bibnamefont
  {Lasenby}}, \ and\ \bibinfo {author} {\bibfnamefont {Mae}\ \bibnamefont
  {Teo}},\ }\bibfield  {title} {\enquote {\bibinfo {title} {{Black hole
  superradiance signatures of ultralight vectors}},}\ }\href {\doibase
  10.1103/PhysRevD.96.035019} {\bibfield  {journal} {\bibinfo  {journal} {Phys.
  Rev. D}\ }\textbf {\bibinfo {volume} {96}},\ \bibinfo {pages} {035019}
  (\bibinfo {year} {2017})}\BibitemShut {NoStop}%
\bibitem [{\citenamefont {{Peccei}}\ and\ \citenamefont
  {{Quinn}}(1977{\natexlab{a}})}]{Peccei1977}%
  \BibitemOpen
  \bibfield  {author} {\bibinfo {author} {\bibfnamefont {R.~D.}\ \bibnamefont
  {{Peccei}}}\ and\ \bibinfo {author} {\bibfnamefont {H.~R.}\ \bibnamefont
  {{Quinn}}},\ }\bibfield  {title} {\enquote {\bibinfo {title} {{CP
  conservation in the presence of pseudoparticles}},}\ }\href {\doibase
  10.1103/PhysRevLett.38.1440} {\bibfield  {journal} {\bibinfo  {journal}
  {Physical Review Letters}\ }\textbf {\bibinfo {volume} {38}},\ \bibinfo
  {pages} {1440--1443} (\bibinfo {year} {1977}{\natexlab{a}})}\BibitemShut
  {NoStop}%
\bibitem [{\citenamefont {{Peccei}}\ and\ \citenamefont
  {{Quinn}}(1977{\natexlab{b}})}]{Peccei1977PhRvD}%
  \BibitemOpen
  \bibfield  {author} {\bibinfo {author} {\bibfnamefont {R.~D.}\ \bibnamefont
  {{Peccei}}}\ and\ \bibinfo {author} {\bibfnamefont {H.~R.}\ \bibnamefont
  {{Quinn}}},\ }\bibfield  {title} {\enquote {\bibinfo {title} {{Constraints
  imposed by CP conservation in the presence of pseudoparticles}},}\ }\href
  {\doibase 10.1103/PhysRevD.16.1791} {\bibfield  {journal} {\bibinfo
  {journal} {Physical Review D}\ }\textbf {\bibinfo {volume} {16}},\ \bibinfo
  {pages} {1791--1797} (\bibinfo {year} {1977}{\natexlab{b}})}\BibitemShut
  {NoStop}%
\bibitem [{\citenamefont {{Weinberg}}(1978)}]{Weinberg1978}%
  \BibitemOpen
  \bibfield  {author} {\bibinfo {author} {\bibfnamefont {Steven}\ \bibnamefont
  {{Weinberg}}},\ }\bibfield  {title} {\enquote {\bibinfo {title} {{A new light
  boson?}}}\ }\href {\doibase 10.1103/PhysRevLett.40.223} {\bibfield  {journal}
  {\bibinfo  {journal} {Physical Review Letters}\ }\textbf {\bibinfo {volume}
  {40}},\ \bibinfo {pages} {223--226} (\bibinfo {year} {1978})}\BibitemShut
  {NoStop}%
\bibitem [{\citenamefont {Arvanitaki}\ \emph {et~al.}(2010)\citenamefont
  {Arvanitaki}, \citenamefont {Dimopoulos}, \citenamefont {Dubovsky},
  \citenamefont {Kaloper},\ and\ \citenamefont
  {March-Russell}}]{Arvanitaki2010}%
  \BibitemOpen
  \bibfield  {author} {\bibinfo {author} {\bibfnamefont {Asimina}\ \bibnamefont
  {Arvanitaki}}, \bibinfo {author} {\bibfnamefont {Savas}\ \bibnamefont
  {Dimopoulos}}, \bibinfo {author} {\bibfnamefont {Sergei}\ \bibnamefont
  {Dubovsky}}, \bibinfo {author} {\bibfnamefont {Nemanja}\ \bibnamefont
  {Kaloper}}, \ and\ \bibinfo {author} {\bibfnamefont {John}\ \bibnamefont
  {March-Russell}},\ }\bibfield  {title} {\enquote {\bibinfo {title} {{String
  axiverse}},}\ }\href {\doibase 10.1103/PhysRevD.81.123530} {\bibfield
  {journal} {\bibinfo  {journal} {Physical Review D}\ }\textbf {\bibinfo
  {volume} {81}},\ \bibinfo {pages} {123530} (\bibinfo {year}
  {2010})}\BibitemShut {NoStop}%
\bibitem [{\citenamefont {Goodsell}\ \emph {et~al.}(2009)\citenamefont
  {Goodsell}, \citenamefont {Jaeckel}, \citenamefont {Redondo},\ and\
  \citenamefont {Ringwald}}]{Goodsell2009}%
  \BibitemOpen
  \bibfield  {author} {\bibinfo {author} {\bibfnamefont {Mark}\ \bibnamefont
  {Goodsell}}, \bibinfo {author} {\bibfnamefont {Joerg}\ \bibnamefont
  {Jaeckel}}, \bibinfo {author} {\bibfnamefont {Javier}\ \bibnamefont
  {Redondo}}, \ and\ \bibinfo {author} {\bibfnamefont {Andreas}\ \bibnamefont
  {Ringwald}},\ }\bibfield  {title} {\enquote {\bibinfo {title} {{Naturally
  Light Hidden Photons in LARGE Volume String Compactifications}},}\ }\href
  {\doibase 10.1088/1126-6708/2009/11/027} {\bibfield  {journal} {\bibinfo
  {journal} {JHEP}\ }\textbf {\bibinfo {volume} {11}},\ \bibinfo {pages} {027}
  (\bibinfo {year} {2009})}\BibitemShut {NoStop}%
\bibitem [{\citenamefont {Jaeckel}\ and\ \citenamefont
  {Ringwald}(2010)}]{Jaeckel:2010ni}%
  \BibitemOpen
  \bibfield  {author} {\bibinfo {author} {\bibfnamefont {Joerg}\ \bibnamefont
  {Jaeckel}}\ and\ \bibinfo {author} {\bibfnamefont {Andreas}\ \bibnamefont
  {Ringwald}},\ }\bibfield  {title} {\enquote {\bibinfo {title} {{The
  Low-Energy Frontier of Particle Physics}},}\ }\href {\doibase
  10.1146/annurev.nucl.012809.104433} {\bibfield  {journal} {\bibinfo
  {journal} {Ann. Rev. Nucl. Part. Sci.}\ }\textbf {\bibinfo {volume} {60}},\
  \bibinfo {pages} {405--437} (\bibinfo {year} {2010})}\BibitemShut {NoStop}%
\bibitem [{\citenamefont {Essig}\ \emph {et~al.}(2013)\citenamefont {Essig}
  \emph {et~al.}}]{Essig:2013lka}%
  \BibitemOpen
  \bibfield  {author} {\bibinfo {author} {\bibfnamefont {Rouven}\ \bibnamefont
  {Essig}} \emph {et~al.},\ }\bibfield  {title} {\enquote {\bibinfo {title}
  {{Working Group Report: New Light Weakly Coupled Particles}},}\ }in\ \href
  {http://www.slac.stanford.edu/econf/C1307292/docs/
  IntensityFrontier/NewLight-17.pdf} {\emph {\bibinfo {booktitle}
  {{Proceedings, 2013 Community Summer Study on the Future of U.S. Particle
  Physics: Snowmass on the Mississippi (CSS2013): Minneapolis, MN, USA, July
  29-August 6, 2013}}}}\ (\bibinfo {year} {2013})\BibitemShut {NoStop}%
\bibitem [{\citenamefont {Hui}\ \emph {et~al.}(2017)\citenamefont {Hui},
  \citenamefont {Ostriker}, \citenamefont {Tremaine},\ and\ \citenamefont
  {Witten}}]{Hui:2016ltb}%
  \BibitemOpen
  \bibfield  {author} {\bibinfo {author} {\bibfnamefont {Lam}\ \bibnamefont
  {Hui}}, \bibinfo {author} {\bibfnamefont {Jeremiah~P.}\ \bibnamefont
  {Ostriker}}, \bibinfo {author} {\bibfnamefont {Scott}\ \bibnamefont
  {Tremaine}}, \ and\ \bibinfo {author} {\bibfnamefont {Edward}\ \bibnamefont
  {Witten}},\ }\bibfield  {title} {\enquote {\bibinfo {title} {{Ultralight
  scalars as cosmological dark matter}},}\ }\href {\doibase
  10.1103/PhysRevD.95.043541} {\bibfield  {journal} {\bibinfo  {journal} {Phys.
  Rev.}\ }\textbf {\bibinfo {volume} {D95}},\ \bibinfo {pages} {043541}
  (\bibinfo {year} {2017})}\BibitemShut {NoStop}%
\bibitem [{\citenamefont {Kim}\ and\ \citenamefont
  {Carosi}(2010)}]{Kim:2008hd}%
  \BibitemOpen
  \bibfield  {author} {\bibinfo {author} {\bibfnamefont {Jihn~E.}\ \bibnamefont
  {Kim}}\ and\ \bibinfo {author} {\bibfnamefont {Gianpaolo}\ \bibnamefont
  {Carosi}},\ }\bibfield  {title} {\enquote {\bibinfo {title} {{Axions and the
  Strong CP Problem}},}\ }\href {\doibase 10.1103/RevModPhys.82.557} {\bibfield
   {journal} {\bibinfo  {journal} {Rev. Mod. Phys.}\ }\textbf {\bibinfo
  {volume} {82}},\ \bibinfo {pages} {557--602} (\bibinfo {year} {2010})},\
  \Eprint {http://arxiv.org/abs/0807.3125} {arXiv:0807.3125 [hep-ph]}
  \BibitemShut {NoStop}%
\bibitem [{\citenamefont {Raffelt}(2006)}]{Raffelt2006}%
  \BibitemOpen
  \bibfield  {author} {\bibinfo {author} {\bibfnamefont {Georg~G.}\
  \bibnamefont {Raffelt}},\ }\bibfield  {title} {\enquote {\bibinfo {title}
  {{Astrophysical Axion Bounds}},}\ }\href {\doibase
  10.1007/978-3-540-73518-2_3} {\bibfield  {journal} {\bibinfo  {journal}
  {Lect. Notes Phys.}\ }\textbf {\bibinfo {volume} {741}},\ \bibinfo {pages}
  {51--71} (\bibinfo {year} {2006})}\BibitemShut {NoStop}%
\bibitem [{\citenamefont {{Penrose}}(1969)}]{Penrose1969}%
  \BibitemOpen
  \bibfield  {author} {\bibinfo {author} {\bibfnamefont {Roger}\ \bibnamefont
  {{Penrose}}},\ }\bibfield  {title} {\enquote {\bibinfo {title}
  {{Gravitational Collapse: the Role of General Relativity}},}\ }\href@noop {}
  {\bibfield  {journal} {\bibinfo  {journal} {Nuovo Cimento Rivista Serie}\
  }\textbf {\bibinfo {volume} {1}} (\bibinfo {year} {1969})}\BibitemShut
  {NoStop}%
\bibitem [{\citenamefont {Press}\ and\ \citenamefont
  {Teukolsky}(1972)}]{Press1972}%
  \BibitemOpen
  \bibfield  {author} {\bibinfo {author} {\bibfnamefont {William~H.}\
  \bibnamefont {Press}}\ and\ \bibinfo {author} {\bibfnamefont {Saul~A.}\
  \bibnamefont {Teukolsky}},\ }\bibfield  {title} {\enquote {\bibinfo {title}
  {{Floating Orbits, Superradiant Scattering and the Black-hole Bomb}},}\
  }\href {\doibase 10.1038/238211a0} {\bibfield  {journal} {\bibinfo  {journal}
  {Nature}\ }\textbf {\bibinfo {volume} {238}},\ \bibinfo {pages} {211--212}
  (\bibinfo {year} {1972})}\BibitemShut {NoStop}%
\bibitem [{\citenamefont {Bekenstein}(1973)}]{Bekenstein1973}%
  \BibitemOpen
  \bibfield  {author} {\bibinfo {author} {\bibfnamefont {Jacob~D.}\
  \bibnamefont {Bekenstein}},\ }\bibfield  {title} {\enquote {\bibinfo {title}
  {{Extraction of Energy and Charge from a Black Hole}},}\ }\href {\doibase
  10.1103/PhysRevD.7.949} {\bibfield  {journal} {\bibinfo  {journal} {Phys.
  Rev. D}\ }\textbf {\bibinfo {volume} {7}},\ \bibinfo {pages} {949--953}
  (\bibinfo {year} {1973})}\BibitemShut {NoStop}%
\bibitem [{\citenamefont {Brito}\ \emph
  {et~al.}(2015{\natexlab{a}})\citenamefont {Brito}, \citenamefont {Cardoso},\
  and\ \citenamefont {Pani}}]{Brito2014}%
  \BibitemOpen
  \bibfield  {author} {\bibinfo {author} {\bibfnamefont {Richard}\ \bibnamefont
  {Brito}}, \bibinfo {author} {\bibfnamefont {Vitor}\ \bibnamefont {Cardoso}},
  \ and\ \bibinfo {author} {\bibfnamefont {Paolo}\ \bibnamefont {Pani}},\
  }\bibfield  {title} {\enquote {\bibinfo {title} {{Black holes as particle
  detectors: evolution of superradiant instabilities}},}\ }\href {\doibase
  10.1088/0264-9381/32/13/134001} {\bibfield  {journal} {\bibinfo  {journal}
  {Class. Quant. Grav.}\ }\textbf {\bibinfo {volume} {32}},\ \bibinfo {pages}
  {134001} (\bibinfo {year} {2015}{\natexlab{a}})}\BibitemShut {NoStop}%
\bibitem [{\citenamefont {Brito}\ \emph
  {et~al.}(2015{\natexlab{b}})\citenamefont {Brito}, \citenamefont {Cardoso},\
  and\ \citenamefont {Pani}}]{Brito2015}%
  \BibitemOpen
  \bibfield  {author} {\bibinfo {author} {\bibfnamefont {Richard}\ \bibnamefont
  {Brito}}, \bibinfo {author} {\bibfnamefont {Vitor}\ \bibnamefont {Cardoso}},
  \ and\ \bibinfo {author} {\bibfnamefont {Paolo}\ \bibnamefont {Pani}},\
  }\href {\doibase 10.1007/978-3-319-19000-6} {\emph {\bibinfo {title}
  {{Superradiance}}}},\ \bibinfo {series} {Lecture Notes in Physics}, Vol.\
  \bibinfo {volume} {906}\ (\bibinfo  {publisher} {Springer International
  Publishing},\ \bibinfo {address} {Cham},\ \bibinfo {year} {2015})\BibitemShut
  {NoStop}%
\bibitem [{\citenamefont {Plascencia}\ and\ \citenamefont
  {Urbano}(2018)}]{Plascencia:2017kca}%
  \BibitemOpen
  \bibfield  {author} {\bibinfo {author} {\bibfnamefont {Alexis~D.}\
  \bibnamefont {Plascencia}}\ and\ \bibinfo {author} {\bibfnamefont {Alfredo}\
  \bibnamefont {Urbano}},\ }\bibfield  {title} {\enquote {\bibinfo {title}
  {{Black hole superradiance and polarization-dependent bending of light}},}\
  }\href {\doibase 10.1088/1475-7516/2018/04/059} {\bibfield  {journal}
  {\bibinfo  {journal} {JCAP}\ }\textbf {\bibinfo {volume} {1804}},\ \bibinfo
  {pages} {059} (\bibinfo {year} {2018})},\ \Eprint
  {http://arxiv.org/abs/1711.08298} {arXiv:1711.08298 [gr-qc]} \BibitemShut
  {NoStop}%
\bibitem [{\citenamefont {Cardoso}\ \emph {et~al.}(2018)\citenamefont
  {Cardoso}, \citenamefont {Dias}, \citenamefont {Hartnett}, \citenamefont
  {Middleton}, \citenamefont {Pani},\ and\ \citenamefont
  {Santos}}]{Cardoso2018}%
  \BibitemOpen
  \bibfield  {author} {\bibinfo {author} {\bibfnamefont {Vitor}\ \bibnamefont
  {Cardoso}}, \bibinfo {author} {\bibfnamefont {Oscar~J.C.}\ \bibnamefont
  {Dias}}, \bibinfo {author} {\bibfnamefont {Gavin~S.}\ \bibnamefont
  {Hartnett}}, \bibinfo {author} {\bibfnamefont {Matthew}\ \bibnamefont
  {Middleton}}, \bibinfo {author} {\bibfnamefont {Paolo}\ \bibnamefont {Pani}},
  \ and\ \bibinfo {author} {\bibfnamefont {Jorge~E.}\ \bibnamefont {Santos}},\
  }\bibfield  {title} {\enquote {\bibinfo {title} {{Constraining the mass of
  dark photons and axion-like particles through black-hole superradiance}},}\
  }\href {\doibase 10.1088/1475-7516/2018/03/043} {\bibfield  {journal}
  {\bibinfo  {journal} {J. Cosmol. Astropart. Phys.}\ }\textbf {\bibinfo
  {volume} {2018}},\ \bibinfo {pages} {043--043} (\bibinfo {year}
  {2018})}\BibitemShut {NoStop}%
\bibitem [{\citenamefont {Baumann}\ \emph {et~al.}(2018)\citenamefont
  {Baumann}, \citenamefont {Chia},\ and\ \citenamefont
  {Porto}}]{Baumann:2018vus}%
  \BibitemOpen
  \bibfield  {author} {\bibinfo {author} {\bibfnamefont {Daniel}\ \bibnamefont
  {Baumann}}, \bibinfo {author} {\bibfnamefont {Horng~Sheng}\ \bibnamefont
  {Chia}}, \ and\ \bibinfo {author} {\bibfnamefont {Rafael~A.}\ \bibnamefont
  {Porto}},\ }\href@noop {} {\enquote {\bibinfo {title} {{Probing Ultralight
  Bosons with Binary Black Holes}},}\ } (\bibinfo {year} {2018}),\ \Eprint
  {http://arxiv.org/abs/1804.03208} {arXiv:1804.03208 [gr-qc]} \BibitemShut
  {NoStop}%
\bibitem [{\citenamefont {Hannuksela}\ \emph {et~al.}(2018)\citenamefont
  {Hannuksela}, \citenamefont {Brito}, \citenamefont {Berti},\ and\
  \citenamefont {Li}}]{Hannuksela:2018izj}%
  \BibitemOpen
  \bibfield  {author} {\bibinfo {author} {\bibfnamefont {Otto~A.}\ \bibnamefont
  {Hannuksela}}, \bibinfo {author} {\bibfnamefont {Richard}\ \bibnamefont
  {Brito}}, \bibinfo {author} {\bibfnamefont {Emanuele}\ \bibnamefont {Berti}},
  \ and\ \bibinfo {author} {\bibfnamefont {Tjonnie G.~F.}\ \bibnamefont {Li}},\
  }\href@noop {} {\enquote {\bibinfo {title} {{Probing the existence of
  ultralight bosons with a single gravitational-wave measurement}},}\ }
  (\bibinfo {year} {2018}),\ \Eprint {http://arxiv.org/abs/1804.09659}
  {arXiv:1804.09659 [astro-ph.HE]} \BibitemShut {NoStop}%
\bibitem [{\citenamefont {Zhang}\ and\ \citenamefont
  {Yang}(2018)}]{Zhang:2018kib}%
  \BibitemOpen
  \bibfield  {author} {\bibinfo {author} {\bibfnamefont {Jun}\ \bibnamefont
  {Zhang}}\ and\ \bibinfo {author} {\bibfnamefont {Huan}\ \bibnamefont
  {Yang}},\ }\href@noop {} {\enquote {\bibinfo {title} {{Gravitational floating
  orbits around hairy black holes}},}\ } (\bibinfo {year} {2018}),\ \Eprint
  {http://arxiv.org/abs/1808.02905} {arXiv:1808.02905 [gr-qc]} \BibitemShut
  {NoStop}%
\bibitem [{\citenamefont {Reynolds}(2014)}]{Reynolds:2013qqa}%
  \BibitemOpen
  \bibfield  {author} {\bibinfo {author} {\bibfnamefont {Christopher~S.}\
  \bibnamefont {Reynolds}},\ }\bibfield  {title} {\enquote {\bibinfo {title}
  {{Measuring Black Hole Spin using X-ray Reflection Spectroscopy}},}\ }\href
  {\doibase 10.1007/s11214-013-0006-6} {\bibfield  {journal} {\bibinfo
  {journal} {Space Sci. Rev.}\ }\textbf {\bibinfo {volume} {183}},\ \bibinfo
  {pages} {277--294} (\bibinfo {year} {2014})}\BibitemShut {NoStop}%
\bibitem [{\citenamefont {McClintock}\ \emph {et~al.}(2014)\citenamefont
  {McClintock}, \citenamefont {Narayan},\ and\ \citenamefont
  {Steiner}}]{McClintock:2013vwa}%
  \BibitemOpen
  \bibfield  {author} {\bibinfo {author} {\bibfnamefont {Jeffrey~E.}\
  \bibnamefont {McClintock}}, \bibinfo {author} {\bibfnamefont {Ramesh}\
  \bibnamefont {Narayan}}, \ and\ \bibinfo {author} {\bibfnamefont {James~F.}\
  \bibnamefont {Steiner}},\ }\bibfield  {title} {\enquote {\bibinfo {title}
  {{Black Hole Spin via Continuum Fitting and the Role of Spin in Powering
  Transient Jets}},}\ }\href {\doibase 10.1007/s11214-013-0003-9} {\bibfield
  {journal} {\bibinfo  {journal} {Space Sci. Rev.}\ }\textbf {\bibinfo {volume}
  {183}},\ \bibinfo {pages} {295--322} (\bibinfo {year} {2014})}\BibitemShut
  {NoStop}%
\bibitem [{\citenamefont {Ternov}\ \emph {et~al.}(1978)\citenamefont {Ternov},
  \citenamefont {Khalilov}, \citenamefont {Chizhov},\ and\ \citenamefont
  {Gaina}}]{Ternov:1978gq}%
  \BibitemOpen
  \bibfield  {author} {\bibinfo {author} {\bibfnamefont {I.~M.}\ \bibnamefont
  {Ternov}}, \bibinfo {author} {\bibfnamefont {V.~R.}\ \bibnamefont
  {Khalilov}}, \bibinfo {author} {\bibfnamefont {G.~A.}\ \bibnamefont
  {Chizhov}}, \ and\ \bibinfo {author} {\bibfnamefont {Alex~B.}\ \bibnamefont
  {Gaina}},\ }\bibfield  {title} {\enquote {\bibinfo {title} {{Finite motion of
  massive particles in the Kerr and Schwarzschild fields}},}\ }\href {\doibase
  10.1007/BF00894575} {\bibfield  {journal} {\bibinfo  {journal} {Sov. Phys.
  J.}\ }\textbf {\bibinfo {volume} {21}},\ \bibinfo {pages} {1200--1204}
  (\bibinfo {year} {1978})},\ \bibinfo {note} {[Izv. Vuz.
  Fiz.21N9,109(1978)]}\BibitemShut {NoStop}%
\bibitem [{\citenamefont {Zouros}\ and\ \citenamefont
  {Eardley}(1979)}]{Zouros1979}%
  \BibitemOpen
  \bibfield  {author} {\bibinfo {author} {\bibfnamefont {Theodoros~J.M.}\
  \bibnamefont {Zouros}}\ and\ \bibinfo {author} {\bibfnamefont {Douglas~M.}\
  \bibnamefont {Eardley}},\ }\bibfield  {title} {\enquote {\bibinfo {title}
  {{Instabilities of massive scalar perturbations of a rotating black hole}},}\
  }\href {\doibase 10.1016/0003-4916(79)90237-9} {\bibfield  {journal}
  {\bibinfo  {journal} {Ann. Phys. (N. Y).}\ }\textbf {\bibinfo {volume}
  {118}},\ \bibinfo {pages} {139--155} (\bibinfo {year} {1979})}\BibitemShut
  {NoStop}%
\bibitem [{\citenamefont {Detweiler}(1980)}]{Detweiler1980}%
  \BibitemOpen
  \bibfield  {author} {\bibinfo {author} {\bibfnamefont {Steven}\ \bibnamefont
  {Detweiler}},\ }\bibfield  {title} {\enquote {\bibinfo {title} {{Klein-Gordon
  equation and rotating black holes}},}\ }\href {\doibase
  10.1103/PhysRevD.22.2323} {\bibfield  {journal} {\bibinfo  {journal} {Phys.
  Rev. D}\ }\textbf {\bibinfo {volume} {22}},\ \bibinfo {pages} {2323--2326}
  (\bibinfo {year} {1980})}\BibitemShut {NoStop}%
\bibitem [{\citenamefont {Dolan}(2007)}]{Dolan2007}%
  \BibitemOpen
  \bibfield  {author} {\bibinfo {author} {\bibfnamefont {Sam~R.}\ \bibnamefont
  {Dolan}},\ }\bibfield  {title} {\enquote {\bibinfo {title} {{Instability of
  the massive Klein-Gordon field on the Kerr spacetime}},}\ }\href {\doibase
  10.1103/PhysRevD.76.084001} {\bibfield  {journal} {\bibinfo  {journal} {Phys.
  Rev. D}\ }\textbf {\bibinfo {volume} {76}},\ \bibinfo {pages} {084001}
  (\bibinfo {year} {2007})}\BibitemShut {NoStop}%
\bibitem [{\citenamefont {Witek}\ \emph {et~al.}(2013)\citenamefont {Witek},
  \citenamefont {Cardoso}, \citenamefont {Ishibashi},\ and\ \citenamefont
  {Sperhake}}]{Witek2013}%
  \BibitemOpen
  \bibfield  {author} {\bibinfo {author} {\bibfnamefont {Helvi}\ \bibnamefont
  {Witek}}, \bibinfo {author} {\bibfnamefont {Vitor}\ \bibnamefont {Cardoso}},
  \bibinfo {author} {\bibfnamefont {Akihiro}\ \bibnamefont {Ishibashi}}, \ and\
  \bibinfo {author} {\bibfnamefont {Ulrich}\ \bibnamefont {Sperhake}},\
  }\bibfield  {title} {\enquote {\bibinfo {title} {{Superradiant instabilities
  in astrophysical systems}},}\ }\href {\doibase 10.1103/PhysRevD.87.043513}
  {\bibfield  {journal} {\bibinfo  {journal} {Phys. Rev. D}\ }\textbf {\bibinfo
  {volume} {87}},\ \bibinfo {pages} {043513} (\bibinfo {year}
  {2013})}\BibitemShut {NoStop}%
\bibitem [{\citenamefont {East}(2017)}]{East2017}%
  \BibitemOpen
  \bibfield  {author} {\bibinfo {author} {\bibfnamefont {William~E.}\
  \bibnamefont {East}},\ }\bibfield  {title} {\enquote {\bibinfo {title}
  {{Superradiant instability of massive vector fields around spinning black
  holes in the relativistic regime}},}\ }\href {\doibase
  10.1103/PhysRevD.96.024004} {\bibfield  {journal} {\bibinfo  {journal} {Phys.
  Rev. D}\ }\textbf {\bibinfo {volume} {96}} (\bibinfo {year} {2017}),\
  10.1103/PhysRevD.96.024004}\BibitemShut {NoStop}%
\bibitem [{\citenamefont {Teukolsky}(2014)}]{Teukolsky2015}%
  \BibitemOpen
  \bibfield  {author} {\bibinfo {author} {\bibfnamefont {Saul~A.}\ \bibnamefont
  {Teukolsky}},\ }\bibfield  {title} {\enquote {\bibinfo {title} {{The Kerr
  Metric}},}\ }\href {\doibase 10.1088/0264-9381/32/12/124006} {\bibfield
  {journal} {\bibinfo  {journal} {Class. Quantum Gravity}\ }\textbf {\bibinfo
  {volume} {32}},\ \bibinfo {pages} {124006} (\bibinfo {year}
  {2014})}\BibitemShut {NoStop}%
\bibitem [{\citenamefont {Zel'dovich}(1971)}]{zeldovich1}%
  \BibitemOpen
  \bibfield  {author} {\bibinfo {author} {\bibfnamefont {Ya.~B.}\ \bibnamefont
  {Zel'dovich}},\ }\href@noop {} {\bibfield  {journal} {\bibinfo  {journal}
  {Pis'ma Zh. Eksp. Teor. Fiz.}\ }\textbf {\bibinfo {volume} {14}},\ \bibinfo
  {pages} {270 [JETP Lett. {\bf14}, 180 (1971)]} (\bibinfo {year}
  {1971})}\BibitemShut {NoStop}%
\bibitem [{\citenamefont {Zel'dovich}(1972)}]{zeldovich2}%
  \BibitemOpen
  \bibfield  {author} {\bibinfo {author} {\bibfnamefont {Ya.~B.}\ \bibnamefont
  {Zel'dovich}},\ }\href@noop {} {\bibfield  {journal} {\bibinfo  {journal}
  {Zh. Eksp. Teor. Fiz}\ }\textbf {\bibinfo {volume} {62}},\ \bibinfo {pages}
  {2076 [Sov.Phys. JETP {\bf 35}, 1085 (1972)]} (\bibinfo {year}
  {1972})}\BibitemShut {NoStop}%
\bibitem [{\citenamefont {Bekenstein}\ and\ \citenamefont
  {Schiffer}(1998)}]{Bekenstein:1998nt}%
  \BibitemOpen
  \bibfield  {author} {\bibinfo {author} {\bibfnamefont {Jacob~D.}\
  \bibnamefont {Bekenstein}}\ and\ \bibinfo {author} {\bibfnamefont {Marcelo}\
  \bibnamefont {Schiffer}},\ }\bibfield  {title} {\enquote {\bibinfo {title}
  {{The Many faces of superradiance}},}\ }\href {\doibase
  10.1103/PhysRevD.58.064014} {\bibfield  {journal} {\bibinfo  {journal} {Phys.
  Rev.}\ }\textbf {\bibinfo {volume} {D58}},\ \bibinfo {pages} {064014}
  (\bibinfo {year} {1998})}\BibitemShut {NoStop}%
\bibitem [{\citenamefont {Damour}\ \emph {et~al.}(1976)\citenamefont {Damour},
  \citenamefont {Deruelle},\ and\ \citenamefont {Ruffini}}]{Damour:1976kh}%
  \BibitemOpen
  \bibfield  {author} {\bibinfo {author} {\bibfnamefont {T.}~\bibnamefont
  {Damour}}, \bibinfo {author} {\bibfnamefont {N.}~\bibnamefont {Deruelle}}, \
  and\ \bibinfo {author} {\bibfnamefont {R.}~\bibnamefont {Ruffini}},\
  }\bibfield  {title} {\enquote {\bibinfo {title} {{On Quantum Resonances in
  Stationary Geometries}},}\ }\href {\doibase 10.1007/BF02725534} {\bibfield
  {journal} {\bibinfo  {journal} {Lett. Nuovo Cim.}\ }\textbf {\bibinfo
  {volume} {15}},\ \bibinfo {pages} {257--262} (\bibinfo {year}
  {1976})}\BibitemShut {NoStop}%
\bibitem [{\citenamefont {Furuhashi}\ and\ \citenamefont
  {Nambu}(2004)}]{Furuhashi:2004jk}%
  \BibitemOpen
  \bibfield  {author} {\bibinfo {author} {\bibfnamefont {Hironobu}\
  \bibnamefont {Furuhashi}}\ and\ \bibinfo {author} {\bibfnamefont {Yasusada}\
  \bibnamefont {Nambu}},\ }\bibfield  {title} {\enquote {\bibinfo {title}
  {{Instability of massive scalar fields in Kerr-Newman space-time}},}\ }\href
  {\doibase 10.1143/PTP.112.983} {\bibfield  {journal} {\bibinfo  {journal}
  {Prog. Theor. Phys.}\ }\textbf {\bibinfo {volume} {112}},\ \bibinfo {pages}
  {983--995} (\bibinfo {year} {2004})},\ \Eprint
  {http://arxiv.org/abs/gr-qc/0402037} {arXiv:gr-qc/0402037 [gr-qc]}
  \BibitemShut {NoStop}%
\bibitem [{\citenamefont {Strafuss}\ and\ \citenamefont
  {Khanna}(2005)}]{Strafuss:2004qc}%
  \BibitemOpen
  \bibfield  {author} {\bibinfo {author} {\bibfnamefont {Matthew~J.}\
  \bibnamefont {Strafuss}}\ and\ \bibinfo {author} {\bibfnamefont {Gaurav}\
  \bibnamefont {Khanna}},\ }\bibfield  {title} {\enquote {\bibinfo {title}
  {{Massive scalar field instability in Kerr spacetime}},}\ }\href {\doibase
  10.1103/PhysRevD.71.024034} {\bibfield  {journal} {\bibinfo  {journal} {Phys.
  Rev.}\ }\textbf {\bibinfo {volume} {D71}},\ \bibinfo {pages} {024034}
  (\bibinfo {year} {2005})},\ \Eprint {http://arxiv.org/abs/gr-qc/0412023}
  {arXiv:gr-qc/0412023 [gr-qc]} \BibitemShut {NoStop}%
\bibitem [{\citenamefont {East}\ and\ \citenamefont
  {Pretorius}(2017)}]{East:2017ovw}%
  \BibitemOpen
  \bibfield  {author} {\bibinfo {author} {\bibfnamefont {William~E.}\
  \bibnamefont {East}}\ and\ \bibinfo {author} {\bibfnamefont {Frans}\
  \bibnamefont {Pretorius}},\ }\bibfield  {title} {\enquote {\bibinfo {title}
  {{Superradiant Instability and Backreaction of Massive Vector Fields around
  Kerr Black Holes}},}\ }\href {\doibase 10.1103/PhysRevLett.119.041101}
  {\bibfield  {journal} {\bibinfo  {journal} {Phys. Rev. Lett.}\ }\textbf
  {\bibinfo {volume} {119}},\ \bibinfo {pages} {041101} (\bibinfo {year}
  {2017})}\BibitemShut {NoStop}%
\bibitem [{\citenamefont {Herdeiro}\ and\ \citenamefont
  {Radu}(2017)}]{Herdeiro:2017phl}%
  \BibitemOpen
  \bibfield  {author} {\bibinfo {author} {\bibfnamefont {Carlos A.~R.}\
  \bibnamefont {Herdeiro}}\ and\ \bibinfo {author} {\bibfnamefont {Eugen}\
  \bibnamefont {Radu}},\ }\bibfield  {title} {\enquote {\bibinfo {title}
  {{Dynamical Formation of Kerr Black Holes with Synchronized Hair: An Analytic
  Model}},}\ }\href {\doibase 10.1103/PhysRevLett.119.261101} {\bibfield
  {journal} {\bibinfo  {journal} {Phys. Rev. Lett.}\ }\textbf {\bibinfo
  {volume} {119}},\ \bibinfo {pages} {261101} (\bibinfo {year}
  {2017})}\BibitemShut {NoStop}%
\bibitem [{\citenamefont {Pani}\ \emph {et~al.}(2012)\citenamefont {Pani},
  \citenamefont {Cardoso}, \citenamefont {Gualtieri}, \citenamefont {Berti},\
  and\ \citenamefont {Ishibashi}}]{Pani:2012bp}%
  \BibitemOpen
  \bibfield  {author} {\bibinfo {author} {\bibfnamefont {Paolo}\ \bibnamefont
  {Pani}}, \bibinfo {author} {\bibfnamefont {Vitor}\ \bibnamefont {Cardoso}},
  \bibinfo {author} {\bibfnamefont {Leonardo}\ \bibnamefont {Gualtieri}},
  \bibinfo {author} {\bibfnamefont {Emanuele}\ \bibnamefont {Berti}}, \ and\
  \bibinfo {author} {\bibfnamefont {Akihiro}\ \bibnamefont {Ishibashi}},\
  }\bibfield  {title} {\enquote {\bibinfo {title} {{Perturbations of slowly
  rotating black holes: massive vector fields in the Kerr metric}},}\ }\href
  {\doibase 10.1103/PhysRevD.86.104017} {\bibfield  {journal} {\bibinfo
  {journal} {Phys. Rev.}\ }\textbf {\bibinfo {volume} {D86}},\ \bibinfo {pages}
  {104017} (\bibinfo {year} {2012})}\BibitemShut {NoStop}%
\bibitem [{\citenamefont {Frolov}\ \emph {et~al.}(2018)\citenamefont {Frolov},
  \citenamefont {Krtou}, \citenamefont {Kubizak},\ and\ \citenamefont
  {Santos}}]{Frolov:2018ezx}%
  \BibitemOpen
  \bibfield  {author} {\bibinfo {author} {\bibfnamefont {Valeri~P.}\
  \bibnamefont {Frolov}}, \bibinfo {author} {\bibfnamefont {Pavel}\
  \bibnamefont {Krtou}}, \bibinfo {author} {\bibfnamefont {David}\ \bibnamefont
  {Kubizak}}, \ and\ \bibinfo {author} {\bibfnamefont {Jorge~E.}\ \bibnamefont
  {Santos}},\ }\bibfield  {title} {\enquote {\bibinfo {title} {{Massive Vector
  Fields in Rotating Black-Hole Spacetimes: Separability and Quasinormal
  Modes}},}\ }\href {\doibase 10.1103/PhysRevLett.120.231103} {\bibfield
  {journal} {\bibinfo  {journal} {Phys. Rev. Lett.}\ }\textbf {\bibinfo
  {volume} {120}},\ \bibinfo {pages} {231103} (\bibinfo {year}
  {2018})}\BibitemShut {NoStop}%
\bibitem [{\citenamefont {Dolan}(2018)}]{Dolan:2018dqv}%
  \BibitemOpen
  \bibfield  {author} {\bibinfo {author} {\bibfnamefont {Sam~R.}\ \bibnamefont
  {Dolan}},\ }\href@noop {} {\enquote {\bibinfo {title} {{Instability of the
  Proca field on Kerr spacetime}},}\ } (\bibinfo {year} {2018}),\ \Eprint
  {http://arxiv.org/abs/1806.01604} {arXiv:1806.01604 [gr-qc]} \BibitemShut
  {NoStop}%
\bibitem [{\citenamefont {East}(2018)}]{East:2018glu}%
  \BibitemOpen
  \bibfield  {author} {\bibinfo {author} {\bibfnamefont {William~E.}\
  \bibnamefont {East}},\ }\href@noop {} {\enquote {\bibinfo {title} {{Massive
  Boson Superradiant Instability of Black Holes: Nonlinear Growth, Saturation,
  and Gravitational Radiation}},}\ } (\bibinfo {year} {2018}),\ \Eprint
  {http://arxiv.org/abs/1807.00043} {arXiv:1807.00043 [gr-qc]} \BibitemShut
  {NoStop}%
\bibitem [{\citenamefont {Yoshino}\ and\ \citenamefont
  {Kodama}(2012)}]{Yoshino:2012kn}%
  \BibitemOpen
  \bibfield  {author} {\bibinfo {author} {\bibfnamefont {Hirotaka}\
  \bibnamefont {Yoshino}}\ and\ \bibinfo {author} {\bibfnamefont {Hideo}\
  \bibnamefont {Kodama}},\ }\bibfield  {title} {\enquote {\bibinfo {title}
  {{Bosenova collapse of axion cloud around a rotating black hole}},}\ }\href
  {\doibase 10.1143/PTP.128.153} {\bibfield  {journal} {\bibinfo  {journal}
  {Prog. Theor. Phys.}\ }\textbf {\bibinfo {volume} {128}},\ \bibinfo {pages}
  {153--190} (\bibinfo {year} {2012})}\BibitemShut {NoStop}%
\bibitem [{\citenamefont {Yoshino}\ and\ \citenamefont
  {Kodama}(2015{\natexlab{b}})}]{Yoshino:2015nsa}%
  \BibitemOpen
  \bibfield  {author} {\bibinfo {author} {\bibfnamefont {Hirotaka}\
  \bibnamefont {Yoshino}}\ and\ \bibinfo {author} {\bibfnamefont {Hideo}\
  \bibnamefont {Kodama}},\ }\bibfield  {title} {\enquote {\bibinfo {title}
  {{The bosenova and axiverse}},}\ }\href {\doibase
  10.1088/0264-9381/32/21/214001} {\bibfield  {journal} {\bibinfo  {journal}
  {Class. Quant. Grav.}\ }\textbf {\bibinfo {volume} {32}},\ \bibinfo {pages}
  {214001} (\bibinfo {year} {2015}{\natexlab{b}})}\BibitemShut {NoStop}%
\bibitem [{\citenamefont {Klimenko}\ \emph {et~al.}(2008)\citenamefont
  {Klimenko}, \citenamefont {Yakushin}, \citenamefont {Mercer},\ and\
  \citenamefont {Mitselmakher}}]{Klimenko:2008fu}%
  \BibitemOpen
  \bibfield  {author} {\bibinfo {author} {\bibfnamefont {S.}~\bibnamefont
  {Klimenko}}, \bibinfo {author} {\bibfnamefont {I.}~\bibnamefont {Yakushin}},
  \bibinfo {author} {\bibfnamefont {A.}~\bibnamefont {Mercer}}, \ and\ \bibinfo
  {author} {\bibfnamefont {Guenakh}\ \bibnamefont {Mitselmakher}},\ }\bibfield
  {title} {\enquote {\bibinfo {title} {{Coherent method for detection of
  gravitational wave bursts}},}\ }\href {\doibase
  10.1088/0264-9381/25/11/114029} {\bibfield  {journal} {\bibinfo  {journal}
  {Class. Quant. Grav.}\ }\textbf {\bibinfo {volume} {25}},\ \bibinfo {pages}
  {114029} (\bibinfo {year} {2008})}\BibitemShut {NoStop}%
\bibitem [{\citenamefont {Klimenko}\ \emph {et~al.}(2016)\citenamefont
  {Klimenko} \emph {et~al.}}]{Klimenko:2015ypf}%
  \BibitemOpen
  \bibfield  {author} {\bibinfo {author} {\bibfnamefont {S.}~\bibnamefont
  {Klimenko}} \emph {et~al.},\ }\bibfield  {title} {\enquote {\bibinfo {title}
  {{Method for detection and reconstruction of gravitational wave transients
  with networks of advanced detectors}},}\ }\href {\doibase
  10.1103/PhysRevD.93.042004} {\bibfield  {journal} {\bibinfo  {journal} {Phys.
  Rev.}\ }\textbf {\bibinfo {volume} {D93}},\ \bibinfo {pages} {042004}
  (\bibinfo {year} {2016})}\BibitemShut {NoStop}%
\bibitem [{\citenamefont {Cornish}\ and\ \citenamefont
  {Littenberg}(2014)}]{Cornish2014}%
  \BibitemOpen
  \bibfield  {author} {\bibinfo {author} {\bibfnamefont {Neil~J.}\ \bibnamefont
  {Cornish}}\ and\ \bibinfo {author} {\bibfnamefont {Tyson~B.}\ \bibnamefont
  {Littenberg}},\ }\bibfield  {title} {\enquote {\bibinfo {title} {{BayesWave:
  Bayesian Inference for Gravitational Wave Bursts and Instrument Glitches}},}\
  }\href {\doibase 10.1088/0264-9381/32/13/135012} {\bibfield  {journal}
  {\bibinfo  {journal} {Class. Quantum Gravity}\ }\textbf {\bibinfo {volume}
  {32}},\ \bibinfo {pages} {135012} (\bibinfo {year} {2014})}\BibitemShut
  {NoStop}%
\bibitem [{\citenamefont {Thorne}(1987)}]{Thorne1987}%
  \BibitemOpen
  \bibfield  {author} {\bibinfo {author} {\bibfnamefont {Kip~S.}\ \bibnamefont
  {Thorne}},\ }\bibfield  {title} {\enquote {\bibinfo {title}
  {{Gravitation}},}\ }in\ \href
  {http://adsabs.harvard.edu/abs/1987thyg.book..330T} {\emph {\bibinfo
  {booktitle} {Three hundred years of gravitation}}},\ \bibinfo {editor}
  {edited by\ \bibinfo {editor} {\bibfnamefont {Sthephen~W.}\ \bibnamefont
  {Hawking}}\ and\ \bibinfo {editor} {\bibfnamefont {Werner}\ \bibnamefont
  {Israel}}}\ (\bibinfo  {publisher} {Cambridge University Press},\ \bibinfo
  {address} {Cambridge},\ \bibinfo {year} {1987})\ Chap.~\bibinfo {chapter}
  {9}, pp.\ \bibinfo {pages} {330 -- 458}\BibitemShut {NoStop}%
\bibitem [{\citenamefont {Anderson}\ \emph {et~al.}(2001)\citenamefont
  {Anderson}, \citenamefont {Brady}, \citenamefont {Creighton},\ and\
  \citenamefont {Flanagan}}]{Anderson2001}%
  \BibitemOpen
  \bibfield  {author} {\bibinfo {author} {\bibfnamefont {Warren}\ \bibnamefont
  {Anderson}}, \bibinfo {author} {\bibfnamefont {Patrick}\ \bibnamefont
  {Brady}}, \bibinfo {author} {\bibfnamefont {Jolien}\ \bibnamefont
  {Creighton}}, \ and\ \bibinfo {author} {\bibfnamefont {{\'{E}}anna}\
  \bibnamefont {Flanagan}},\ }\bibfield  {title} {\enquote {\bibinfo {title}
  {{Excess power statistic for detection of burst sources of gravitational
  radiation}},}\ }\href {\doibase 10.1103/PhysRevD.63.042003} {\bibfield
  {journal} {\bibinfo  {journal} {Phys. Rev. D}\ }\textbf {\bibinfo {volume}
  {63}},\ \bibinfo {pages} {042003} (\bibinfo {year} {2001})}\BibitemShut
  {NoStop}%
\bibitem [{\citenamefont {Teukolsky}(1973)}]{Teukolsky1973}%
  \BibitemOpen
  \bibfield  {author} {\bibinfo {author} {\bibfnamefont {Saul~A.}\ \bibnamefont
  {Teukolsky}},\ }\bibfield  {title} {\enquote {\bibinfo {title}
  {{Perturbations of a Rotating Black Hole. I. Fundamental Equations for
  Gravitational, Electromagnetic, and Neutrino-Field Perturbations}},}\ }\href
  {\doibase 10.1086/152444} {\bibfield  {journal} {\bibinfo  {journal}
  {Astrophys. J.}\ }\textbf {\bibinfo {volume} {185}},\ \bibinfo {pages} {635}
  (\bibinfo {year} {1973})}\BibitemShut {NoStop}%
\bibitem [{\citenamefont {Leaver}(1985)}]{Leaver1985}%
  \BibitemOpen
  \bibfield  {author} {\bibinfo {author} {\bibfnamefont {E.~W.}\ \bibnamefont
  {Leaver}},\ }\bibfield  {title} {\enquote {\bibinfo {title} {{An Analytic
  Representation for the Quasi-Normal Modes of Kerr Black Holes}},}\ }\href
  {\doibase 10.1098/rspa.1985.0119} {\bibfield  {journal} {\bibinfo  {journal}
  {Proc. R. Soc. A Math. Phys. Eng. Sci.}\ }\textbf {\bibinfo {volume} {402}},\
  \bibinfo {pages} {285--298} (\bibinfo {year} {1985})}\BibitemShut {NoStop}%
\bibitem [{\citenamefont {Berti}\ \emph {et~al.}(2006)\citenamefont {Berti},
  \citenamefont {Cardoso},\ and\ \citenamefont {Casals}}]{Berti2006}%
  \BibitemOpen
  \bibfield  {author} {\bibinfo {author} {\bibfnamefont {Emanuele}\
  \bibnamefont {Berti}}, \bibinfo {author} {\bibfnamefont {Vitor}\ \bibnamefont
  {Cardoso}}, \ and\ \bibinfo {author} {\bibfnamefont {Marc}\ \bibnamefont
  {Casals}},\ }\bibfield  {title} {\enquote {\bibinfo {title} {{Eigenvalues and
  eigenfunctions of spin-weighted spheroidal harmonics in four and higher
  dimensions}},}\ }\href {\doibase 10.1103/PhysRevD.73.024013} {\bibfield
  {journal} {\bibinfo  {journal} {Phys. Rev. D}\ }\textbf {\bibinfo {volume}
  {73}},\ \bibinfo {pages} {024013} (\bibinfo {year} {2006})}\BibitemShut
  {NoStop}%
\bibitem [{\citenamefont {{Blandford}}\ and\ \citenamefont
  {{Teukolsky}}(1976)}]{Blandford1976}%
  \BibitemOpen
  \bibfield  {author} {\bibinfo {author} {\bibfnamefont {R.}~\bibnamefont
  {{Blandford}}}\ and\ \bibinfo {author} {\bibfnamefont {S.~A.}\ \bibnamefont
  {{Teukolsky}}},\ }\bibfield  {title} {\enquote {\bibinfo {title}
  {{Arrival-time analysis for a pulsar in a binary system.}}}\ }\href {\doibase
  10.1086/154315} {\bibfield  {journal} {\bibinfo  {journal} {\apj}\ }\textbf
  {\bibinfo {volume} {205}},\ \bibinfo {pages} {580--591} (\bibinfo {year}
  {1976})}\BibitemShut {NoStop}%
\bibitem [{\citenamefont {Taylor}\ and\ \citenamefont
  {Weisberg}(1989)}]{Taylor1989}%
  \BibitemOpen
  \bibfield  {author} {\bibinfo {author} {\bibfnamefont {J.~H.}\ \bibnamefont
  {Taylor}}\ and\ \bibinfo {author} {\bibfnamefont {J.~M.}\ \bibnamefont
  {Weisberg}},\ }\bibfield  {title} {\enquote {\bibinfo {title} {{Further
  experimental tests of relativistic gravity using the binary pulsar PSR 1913 +
  16}},}\ }\href {\doibase 10.1086/167917} {\bibfield  {journal} {\bibinfo
  {journal} {Astrophys. J.}\ }\textbf {\bibinfo {volume} {345}},\ \bibinfo
  {pages} {434} (\bibinfo {year} {1989})}\BibitemShut {NoStop}%
\bibitem [{\citenamefont {Riles}(2017)}]{Riles2017}%
  \BibitemOpen
  \bibfield  {author} {\bibinfo {author} {\bibfnamefont {Keith}\ \bibnamefont
  {Riles}},\ }\bibfield  {title} {\enquote {\bibinfo {title} {{Recent searches
  for continuous gravitational waves}},}\ }\href {\doibase
  10.1142/S021773231730035X} {\bibfield  {journal} {\bibinfo  {journal} {Mod.
  Phys. Lett. A}\ }\textbf {\bibinfo {volume} {32}},\ \bibinfo {pages}
  {1730035} (\bibinfo {year} {2017})}\BibitemShut {NoStop}%
\bibitem [{\citenamefont {D'Antonio}\ \emph {et~al.}(2018)\citenamefont
  {D'Antonio} \emph {et~al.}}]{DAntonio:2018sff}%
  \BibitemOpen
  \bibfield  {author} {\bibinfo {author} {\bibfnamefont {S.}~\bibnamefont
  {D'Antonio}} \emph {et~al.},\ }\bibfield  {title} {\enquote {\bibinfo {title}
  {{A semi-coherent analysis method to search for continuous gravitational
  waves emitted by ultra-light boson clouds around spinning black holes}},}\
  }\href@noop {} {\  (\bibinfo {year} {2018})},\ \Eprint
  {http://arxiv.org/abs/1809.07202} {arXiv:1809.07202 [gr-qc]} \BibitemShut
  {NoStop}%
\bibitem [{\citenamefont {Suvorova}\ \emph {et~al.}(2016)\citenamefont
  {Suvorova}, \citenamefont {Sun}, \citenamefont {Melatos}, \citenamefont
  {Moran},\ and\ \citenamefont {Evans}}]{Suvorova2016}%
  \BibitemOpen
  \bibfield  {author} {\bibinfo {author} {\bibfnamefont {S.}~\bibnamefont
  {Suvorova}}, \bibinfo {author} {\bibfnamefont {L.}~\bibnamefont {Sun}},
  \bibinfo {author} {\bibfnamefont {A.}~\bibnamefont {Melatos}}, \bibinfo
  {author} {\bibfnamefont {W.}~\bibnamefont {Moran}}, \ and\ \bibinfo {author}
  {\bibfnamefont {Robin~J.}\ \bibnamefont {Evans}},\ }\bibfield  {title}
  {\enquote {\bibinfo {title} {{Hidden Markov model tracking of continuous
  gravitational waves from a neutron star with wandering spin}},}\ }\href
  {\doibase 10.1103/PhysRevD.93.123009} {\bibfield  {journal} {\bibinfo
  {journal} {Physical Review D}\ }\textbf {\bibinfo {volume} {93}},\ \bibinfo
  {pages} {123009} (\bibinfo {year} {2016})}\BibitemShut {NoStop}%
\bibitem [{\citenamefont {Sun}\ \emph {et~al.}(2018)\citenamefont {Sun},
  \citenamefont {Melatos}, \citenamefont {Suvorova}, \citenamefont {Moran},\
  and\ \citenamefont {Evans}}]{Sun2018}%
  \BibitemOpen
  \bibfield  {author} {\bibinfo {author} {\bibfnamefont {L.}~\bibnamefont
  {Sun}}, \bibinfo {author} {\bibfnamefont {A.}~\bibnamefont {Melatos}},
  \bibinfo {author} {\bibfnamefont {S.}~\bibnamefont {Suvorova}}, \bibinfo
  {author} {\bibfnamefont {W.}~\bibnamefont {Moran}}, \ and\ \bibinfo {author}
  {\bibfnamefont {R.~J.}\ \bibnamefont {Evans}},\ }\bibfield  {title} {\enquote
  {\bibinfo {title} {Hidden markov model tracking of continuous gravitational
  waves from young supernova remnants},}\ }\href {\doibase
  10.1103/PhysRevD.97.043013} {\bibfield  {journal} {\bibinfo  {journal} {Phys.
  Rev. D}\ }\textbf {\bibinfo {volume} {97}},\ \bibinfo {pages} {043013}
  (\bibinfo {year} {2018})}\BibitemShut {NoStop}%
\bibitem [{\citenamefont {Abbott}\ \emph
  {et~al.}(2017{\natexlab{e}})\citenamefont {Abbott} \emph
  {et~al.}}]{ScoX1ViterbiO1}%
  \BibitemOpen
  \bibfield  {author} {\bibinfo {author} {\bibfnamefont {B.~P.}\ \bibnamefont
  {Abbott}} \emph {et~al.},\ }\bibfield  {title} {\enquote {\bibinfo {title}
  {Search for gravitational waves from scorpius x-1 in the first advanced ligo
  observing run with a hidden markov model},}\ }\href {\doibase
  10.1103/PhysRevD.95.122003} {\bibfield  {journal} {\bibinfo  {journal} {Phys.
  Rev. D}\ }\textbf {\bibinfo {volume} {95}},\ \bibinfo {pages} {122003}
  (\bibinfo {year} {2017}{\natexlab{e}})}\BibitemShut {NoStop}%
\bibitem [{\citenamefont {{Abbott}}\ \emph {et~al.}(2018)\citenamefont
  {{Abbott}} \emph {et~al.}}]{LVC:2018pmr}%
  \BibitemOpen
  \bibfield  {author} {\bibinfo {author} {\bibfnamefont {B.~P.}\ \bibnamefont
  {{Abbott}}} \emph {et~al.},\ }\bibfield  {title} {\enquote {\bibinfo {title}
  {{Search for gravitational waves from a long-lived remnant of the binary
  neutron star merger GW170817}},}\ }\href@noop {} {\  (\bibinfo {year}
  {2018})},\ \Eprint {http://arxiv.org/abs/1810.02581} {arXiv:1810.02581
  [gr-qc]} \BibitemShut {NoStop}%
\bibitem [{\citenamefont {Sun}\ and\ \citenamefont
  {Melatos}(2018)}]{Sun:2018hmm}%
  \BibitemOpen
  \bibfield  {author} {\bibinfo {author} {\bibfnamefont {Ling}\ \bibnamefont
  {Sun}}\ and\ \bibinfo {author} {\bibfnamefont {Andrew}\ \bibnamefont
  {Melatos}},\ }\bibfield  {title} {\enquote {\bibinfo {title} {{Application of
  hidden Markov model tracking to the search for long-duration transient
  gravitational waves from the remnant of the binary neutron star merger
  GW170817}},}\ }\href@noop {} {\  (\bibinfo {year} {2018})},\ \Eprint
  {http://arxiv.org/abs/1810.03577} {arXiv:1810.03577 [astro-ph.IM]}
  \BibitemShut {NoStop}%
\bibitem [{\citenamefont {Jaranowski}\ \emph {et~al.}(1998)\citenamefont
  {Jaranowski}, \citenamefont {Kr{\'{o}}lak},\ and\ \citenamefont
  {Schutz}}]{Jaranowski1998}%
  \BibitemOpen
  \bibfield  {author} {\bibinfo {author} {\bibfnamefont {Piotr}\ \bibnamefont
  {Jaranowski}}, \bibinfo {author} {\bibfnamefont {Andrzej}\ \bibnamefont
  {Kr{\'{o}}lak}}, \ and\ \bibinfo {author} {\bibfnamefont {Bernard~F.}\
  \bibnamefont {Schutz}},\ }\bibfield  {title} {\enquote {\bibinfo {title}
  {{Data analysis of gravitational-wave signals from spinning neutron stars:
  The signal and its detection}},}\ }\href {\doibase
  10.1103/PhysRevD.58.063001} {\bibfield  {journal} {\bibinfo  {journal}
  {Physical Review D}\ }\textbf {\bibinfo {volume} {58}},\ \bibinfo {pages}
  {063001} (\bibinfo {year} {1998})}\BibitemShut {NoStop}%
\bibitem [{\citenamefont {{Dhurandhar}}\ \emph {et~al.}(2008)\citenamefont
  {{Dhurandhar}}, \citenamefont {{Krishnan}}, \citenamefont {{Mukhopadhyay}},\
  and\ \citenamefont {{Whelan}}}]{Dhurandhar2008}%
  \BibitemOpen
  \bibfield  {author} {\bibinfo {author} {\bibfnamefont {S.}~\bibnamefont
  {{Dhurandhar}}}, \bibinfo {author} {\bibfnamefont {B.}~\bibnamefont
  {{Krishnan}}}, \bibinfo {author} {\bibfnamefont {H.}~\bibnamefont
  {{Mukhopadhyay}}}, \ and\ \bibinfo {author} {\bibfnamefont {J.~T.}\
  \bibnamefont {{Whelan}}},\ }\bibfield  {title} {\enquote {\bibinfo {title}
  {{Cross-correlation search for periodic gravitational waves}},}\ }\href@noop
  {} {\bibfield  {journal} {\bibinfo  {journal} {Physical Review D}\ }\textbf
  {\bibinfo {volume} {77}},\ \bibinfo {pages} {082001} (\bibinfo {year}
  {2008})}\BibitemShut {NoStop}%
\bibitem [{\citenamefont {Prix}(2011)}]{F-stat2011}%
  \BibitemOpen
  \bibfield  {author} {\bibinfo {author} {\bibfnamefont {Reinhard}\
  \bibnamefont {Prix}},\ }\href {https://dcc.ligo.org/LIGO-T0900149/public}
  {\emph {\bibinfo {title} {{The $\mathcal{F}$-statistic and its implementation
  in ComputeFstatistic v2}}}},\ \bibinfo {type} {Tech. Rep.}\ \bibinfo {number}
  {LIGO-T0900149}\ (\bibinfo  {institution} {LIGO Laboratory},\ \bibinfo {year}
  {2011})\BibitemShut {NoStop}%
\bibitem [{\citenamefont {Barsotti}\ \emph {et~al.}(2018)\citenamefont
  {Barsotti}, \citenamefont {Fritschel}, \citenamefont {Evans},\ and\
  \citenamefont {Gras}}]{aLIGO_design_sensi}%
  \BibitemOpen
  \bibfield  {author} {\bibinfo {author} {\bibfnamefont {Lisa}\ \bibnamefont
  {Barsotti}}, \bibinfo {author} {\bibfnamefont {Peter}\ \bibnamefont
  {Fritschel}}, \bibinfo {author} {\bibfnamefont {Matthew}\ \bibnamefont
  {Evans}}, \ and\ \bibinfo {author} {\bibfnamefont {Slawomir}\ \bibnamefont
  {Gras}},\ }\href {https://dcc.ligo.org/LIGO-T1800044/public} {\emph {\bibinfo
  {title} {{Updated Advanced LIGO sensitivity design curve}}}},\ \bibinfo
  {type} {Tech. Rep.}\ \bibinfo {number} {LIGO-T1800044}\ (\bibinfo
  {institution} {LIGO Laboratory},\ \bibinfo {year} {2018})\BibitemShut
  {NoStop}%
\bibitem [{\citenamefont {Abbott}\ \emph
  {et~al.}(2016{\natexlab{e}})\citenamefont {Abbott} \emph
  {et~al.}}]{gw150914_pe}%
  \BibitemOpen
  \bibfield  {author} {\bibinfo {author} {\bibfnamefont {B.~P.}\ \bibnamefont
  {Abbott}} \emph {et~al.} (\bibinfo {collaboration} {LIGO Scientific
  Collaboration, Virgo Collaboration}),\ }\bibfield  {title} {\enquote
  {\bibinfo {title} {{Properties of the Binary Black Hole Merger GW150914}},}\
  }\href {\doibase 10.1103/PhysRevLett.116.241102} {\bibfield  {journal}
  {\bibinfo  {journal} {Phys.\ Rev.\ Lett.}\ }\textbf {\bibinfo {volume}
  {116}},\ \bibinfo {pages} {241102} (\bibinfo {year} {2016}{\natexlab{e}})},\
  \Eprint {http://arxiv.org/abs/1602.03840} {arXiv:1602.03840 [gr-qc]}
  \BibitemShut {NoStop}%
\bibitem [{\citenamefont {Abbott}\ \emph
  {et~al.}(2017{\natexlab{f}})\citenamefont {Abbott} \emph
  {et~al.}}]{Evans:2016mbw}%
  \BibitemOpen
  \bibfield  {author} {\bibinfo {author} {\bibfnamefont {B.~P.}\ \bibnamefont
  {Abbott}} \emph {et~al.} (\bibinfo {collaboration} {LIGO Scientific
  Collaboration}),\ }\bibfield  {title} {\enquote {\bibinfo {title} {{Exploring
  the Sensitivity of Next Generation Gravitational Wave Detectors}},}\ }\href
  {\doibase 10.1088/1361-6382/aa51f4} {\bibfield  {journal} {\bibinfo
  {journal} {Class. Quant. Grav.}\ }\textbf {\bibinfo {volume} {34}},\ \bibinfo
  {pages} {044001} (\bibinfo {year} {2017}{\natexlab{f}})},\ \Eprint
  {http://arxiv.org/abs/1607.08697} {arXiv:1607.08697 [astro-ph.IM]}
  \BibitemShut {NoStop}%
\bibitem [{\citenamefont {Abbott}\ \emph
  {et~al.}(2016{\natexlab{f}})\citenamefont {Abbott} \emph
  {et~al.}}]{iswp2016}%
  \BibitemOpen
  \bibfield  {author} {\bibinfo {author} {\bibfnamefont {B.~P.}\ \bibnamefont
  {Abbott}} \emph {et~al.} (\bibinfo {collaboration} {LIGO Scientific
  Collaboration}),\ }\href {https://dcc.ligo.org/LIGO-T1600119/public} {\emph
  {\bibinfo {title} {Instrument Science White Paper}}},\ \bibinfo {type} {Tech.
  Rep.}\ \bibinfo {number} {LIGO-T1600119}\ (\bibinfo  {institution} {LIGO
  Laboratory},\ \bibinfo {year} {2016})\BibitemShut {NoStop}%
\bibitem [{\citenamefont {Team}(2011)}]{et-design}%
  \BibitemOpen
  \bibfield  {author} {\bibinfo {author} {\bibfnamefont {Einstein
  Telescope~Science}\ \bibnamefont {Team}},\ }\href {http://www.et-gw.eu/}
  {\emph {\bibinfo {title} {Einstein gravitational wave Telescope conceptual
  design study}}},\ \bibinfo {type} {Tech. Rep.}\ \bibinfo {number}
  {ET-0106C-10}\ (\bibinfo  {institution} {European Gravitational
  Observatory},\ \bibinfo {year} {2011})\BibitemShut {NoStop}%
\bibitem [{\citenamefont {Sathyaprakash}\ \emph {et~al.}(2012)\citenamefont
  {Sathyaprakash} \emph {et~al.}}]{Sathyaprakash:2012jk}%
  \BibitemOpen
  \bibfield  {author} {\bibinfo {author} {\bibfnamefont {B.}~\bibnamefont
  {Sathyaprakash}} \emph {et~al.},\ }\bibfield  {title} {\enquote {\bibinfo
  {title} {{Scientific Objectives of Einstein Telescope}},}\ }\bibfield
  {booktitle} {\emph {\bibinfo {booktitle} {{Gravitational waves. Numerical
  relativity - data analysis. Proceedings, 9th Edoardo Amaldi Conference,
  Amaldi 9, and meeting, NRDA 2011, Cardiff, UK, July 10-15, 2011}}},\ }\href
  {\doibase 10.1088/0264-9381/29/12/124013, 10.1088/0264-9381/30/7/079501}
  {\bibfield  {journal} {\bibinfo  {journal} {Class. Quant. Grav.}\ }\textbf
  {\bibinfo {volume} {29}},\ \bibinfo {pages} {124013} (\bibinfo {year}
  {2012})},\ \bibinfo {note} {[Erratum: Class. Quant. Grav.30,079501(2013)]},\
  \Eprint {http://arxiv.org/abs/1206.0331} {arXiv:1206.0331 [gr-qc]}
  \BibitemShut {NoStop}%
\bibitem [{\citenamefont {Freise}\ \emph {et~al.}(2009)\citenamefont {Freise},
  \citenamefont {Chelkowski}, \citenamefont {Hild}, \citenamefont {Del~Pozzo},
  \citenamefont {Perreca},\ and\ \citenamefont {Vecchio}}]{Freise:2008dk}%
  \BibitemOpen
  \bibfield  {author} {\bibinfo {author} {\bibfnamefont {A.}~\bibnamefont
  {Freise}}, \bibinfo {author} {\bibfnamefont {S.}~\bibnamefont {Chelkowski}},
  \bibinfo {author} {\bibfnamefont {S.}~\bibnamefont {Hild}}, \bibinfo {author}
  {\bibfnamefont {W.}~\bibnamefont {Del~Pozzo}}, \bibinfo {author}
  {\bibfnamefont {A.}~\bibnamefont {Perreca}}, \ and\ \bibinfo {author}
  {\bibfnamefont {A.}~\bibnamefont {Vecchio}},\ }\bibfield  {title} {\enquote
  {\bibinfo {title} {{Triple Michelson Interferometer for a Third-Generation
  Gravitational Wave Detector}},}\ }\href {\doibase
  10.1088/0264-9381/26/8/085012} {\bibfield  {journal} {\bibinfo  {journal}
  {Class. Quant. Grav.}\ }\textbf {\bibinfo {volume} {26}},\ \bibinfo {pages}
  {085012} (\bibinfo {year} {2009})},\ \Eprint {http://arxiv.org/abs/0804.1036}
  {arXiv:0804.1036 [gr-qc]} \BibitemShut {NoStop}%
\bibitem [{\citenamefont {Hild}\ \emph {et~al.}(2011)\citenamefont {Hild} \emph
  {et~al.}}]{Hild:2010id}%
  \BibitemOpen
  \bibfield  {author} {\bibinfo {author} {\bibfnamefont {S.}~\bibnamefont
  {Hild}} \emph {et~al.},\ }\bibfield  {title} {\enquote {\bibinfo {title}
  {{Sensitivity Studies for Third-Generation Gravitational Wave
  Observatories}},}\ }\href {\doibase 10.1088/0264-9381/28/9/094013} {\bibfield
   {journal} {\bibinfo  {journal} {Class. Quant. Grav.}\ }\textbf {\bibinfo
  {volume} {28}},\ \bibinfo {pages} {094013} (\bibinfo {year} {2011})},\
  \Eprint {http://arxiv.org/abs/1012.0908} {arXiv:1012.0908 [gr-qc]}
  \BibitemShut {NoStop}%
\bibitem [{\citenamefont {Chen}\ \emph {et~al.}(2017)\citenamefont {Chen},
  \citenamefont {Holz}, \citenamefont {Miller}, \citenamefont {Evans},
  \citenamefont {Vitale},\ and\ \citenamefont {Creighton}}]{Chen:2017wpg}%
  \BibitemOpen
  \bibfield  {author} {\bibinfo {author} {\bibfnamefont {Hsin-Yu}\ \bibnamefont
  {Chen}}, \bibinfo {author} {\bibfnamefont {Daniel~E.}\ \bibnamefont {Holz}},
  \bibinfo {author} {\bibfnamefont {John}\ \bibnamefont {Miller}}, \bibinfo
  {author} {\bibfnamefont {Matthew}\ \bibnamefont {Evans}}, \bibinfo {author}
  {\bibfnamefont {Salvatore}\ \bibnamefont {Vitale}}, \ and\ \bibinfo {author}
  {\bibfnamefont {Jolien}\ \bibnamefont {Creighton}},\ }\href@noop {} {\enquote
  {\bibinfo {title} {{Distance measures in gravitational-wave astrophysics and
  cosmology}},}\ } (\bibinfo {year} {2017}),\ \Eprint
  {http://arxiv.org/abs/1709.08079} {arXiv:1709.08079 [astro-ph.CO]}
  \BibitemShut {NoStop}%
\bibitem [{\citenamefont {Ade}\ \emph {et~al.}(2016)\citenamefont {Ade} \emph
  {et~al.}}]{Ade:2015xua}%
  \BibitemOpen
  \bibfield  {author} {\bibinfo {author} {\bibfnamefont {P.~A.~R.}\
  \bibnamefont {Ade}} \emph {et~al.} (\bibinfo {collaboration} {Planck}),\
  }\bibfield  {title} {\enquote {\bibinfo {title} {{Planck 2015 results. XIII.
  Cosmological parameters}},}\ }\href {\doibase 10.1051/0004-6361/201525830}
  {\bibfield  {journal} {\bibinfo  {journal} {Astron. Astrophys.}\ }\textbf
  {\bibinfo {volume} {594}},\ \bibinfo {pages} {A13} (\bibinfo {year}
  {2016})},\ \Eprint {http://arxiv.org/abs/1502.01589} {arXiv:1502.01589
  [astro-ph.CO]} \BibitemShut {NoStop}%
\bibitem [{\citenamefont {Brady}\ \emph {et~al.}(1998)\citenamefont {Brady},
  \citenamefont {Creighton}, \citenamefont {Cutler},\ and\ \citenamefont
  {Schutz}}]{Brady1998}%
  \BibitemOpen
  \bibfield  {author} {\bibinfo {author} {\bibfnamefont {Patrick~R.}\
  \bibnamefont {Brady}}, \bibinfo {author} {\bibfnamefont {Teviet}\
  \bibnamefont {Creighton}}, \bibinfo {author} {\bibfnamefont {Curt}\
  \bibnamefont {Cutler}}, \ and\ \bibinfo {author} {\bibfnamefont {Bernard~F.}\
  \bibnamefont {Schutz}},\ }\bibfield  {title} {\enquote {\bibinfo {title}
  {{Searching for periodic sources with LIGO}},}\ }\href {\doibase
  10.1103/PhysRevD.57.2101} {\bibfield  {journal} {\bibinfo  {journal}
  {Physical Review D}\ }\textbf {\bibinfo {volume} {57}},\ \bibinfo {pages}
  {2101--2116} (\bibinfo {year} {1998})}\BibitemShut {NoStop}%
\bibitem [{\citenamefont {{Brady}}\ and\ \citenamefont
  {{Creighton}}(2000)}]{Brady2000}%
  \BibitemOpen
  \bibfield  {author} {\bibinfo {author} {\bibfnamefont {P.~R.}\ \bibnamefont
  {{Brady}}}\ and\ \bibinfo {author} {\bibfnamefont {T.}~\bibnamefont
  {{Creighton}}},\ }\bibfield  {title} {\enquote {\bibinfo {title} {{Searching
  for periodic sources with LIGO. II. Hierarchical searches}},}\ }\href
  {\doibase 10.1103/PhysRevD.61.082001} {\bibfield  {journal} {\bibinfo
  {journal} {Physical Review D}\ }\textbf {\bibinfo {volume} {61}},\ \bibinfo
  {eid} {082001} (\bibinfo {year} {2000})}\BibitemShut {NoStop}%
\bibitem [{\citenamefont {{Wette}}\ and\ \citenamefont
  {{Prix}}(2013)}]{Wette2013}%
  \BibitemOpen
  \bibfield  {author} {\bibinfo {author} {\bibfnamefont {K.}~\bibnamefont
  {{Wette}}}\ and\ \bibinfo {author} {\bibfnamefont {R.}~\bibnamefont
  {{Prix}}},\ }\bibfield  {title} {\enquote {\bibinfo {title} {{Flat
  parameter-space metric for all-sky searches for gravitational-wave
  pulsars}},}\ }\href {\doibase 10.1103/PhysRevD.88.123005} {\bibfield
  {journal} {\bibinfo  {journal} {\prd}\ }\textbf {\bibinfo {volume} {88}},\
  \bibinfo {eid} {123005} (\bibinfo {year} {2013})}\BibitemShut {NoStop}%
\bibitem [{\citenamefont {{Wette}}(2015)}]{Wette2015}%
  \BibitemOpen
  \bibfield  {author} {\bibinfo {author} {\bibfnamefont {K.}~\bibnamefont
  {{Wette}}},\ }\bibfield  {title} {\enquote {\bibinfo {title}
  {{Parameter-space metric for all-sky semicoherent searches for
  gravitational-wave pulsars}},}\ }\href {\doibase 10.1103/PhysRevD.92.082003}
  {\bibfield  {journal} {\bibinfo  {journal} {\prd}\ }\textbf {\bibinfo
  {volume} {92}},\ \bibinfo {eid} {082003} (\bibinfo {year}
  {2015})}\BibitemShut {NoStop}%
\bibitem [{\citenamefont {Abbott}\ \emph {et~al.}(2018)\citenamefont {Abbott}
  \emph {et~al.}}]{osLRR2018}%
  \BibitemOpen
  \bibfield  {author} {\bibinfo {author} {\bibfnamefont {B.~P.}\ \bibnamefont
  {Abbott}} \emph {et~al.},\ }\bibfield  {title} {\enquote {\bibinfo {title}
  {{Prospects for observing and localizing gravitational-wave transients with
  Advanced LIGO, Advanced Virgo and KAGRA}},}\ }\href {\doibase
  10.1007/s41114-018-0012-9} {\bibfield  {journal} {\bibinfo  {journal} {Living
  Rev. Relativ.}\ }\textbf {\bibinfo {volume} {21}},\ \bibinfo {pages} {3}
  (\bibinfo {year} {2018})}\BibitemShut {NoStop}%
\bibitem [{\citenamefont {Iyer}\ \emph {et~al.}(2011)\citenamefont {Iyer} \emph
  {et~al.}}]{Iyer2011}%
  \BibitemOpen
  \bibfield  {author} {\bibinfo {author} {\bibfnamefont {Bala}\ \bibnamefont
  {Iyer}} \emph {et~al.},\ }\href {https://dcc.ligo.org/LIGO-M1100296/public}
  {\emph {\bibinfo {title} {{LIGO-India, Proposal of the Consortium for Indian
  Initiative in Gravitational-wave Observations (IndIGO)}}}},\ \bibinfo {type}
  {Tech. Rep.}\ \bibinfo {number} {LIGO-M1100296}\ (\bibinfo  {institution}
  {{LIGO Laboratory}},\ \bibinfo {year} {2011})\BibitemShut {NoStop}%
\bibitem [{\citenamefont {Unnikrishnan}(2013)}]{Unnikrishnan:2013qwa}%
  \BibitemOpen
  \bibfield  {author} {\bibinfo {author} {\bibfnamefont {C.~S.}\ \bibnamefont
  {Unnikrishnan}},\ }\bibfield  {title} {\enquote {\bibinfo {title} {{IndIGO
  and LIGO-India: Scope and plans for gravitational wave research and precision
  metrology in India}},}\ }\href {\doibase 10.1142/S0218271813410101}
  {\bibfield  {journal} {\bibinfo  {journal} {Int. J. Mod. Phys.}\ }\textbf
  {\bibinfo {volume} {D22}},\ \bibinfo {pages} {1341010} (\bibinfo {year}
  {2013})}\BibitemShut {NoStop}%
\bibitem [{\citenamefont {Somiya}(2012)}]{Somiya:2011np}%
  \BibitemOpen
  \bibfield  {author} {\bibinfo {author} {\bibfnamefont {Kentaro}\ \bibnamefont
  {Somiya}},\ }\bibfield  {title} {\enquote {\bibinfo {title} {{Detector
  configuration of KAGRA: The Japanese cryogenic gravitational-wave
  detector}},}\ }\bibfield  {booktitle} {\emph {\bibinfo {booktitle}
  {{Gravitational waves. Numerical relativity - data analysis. Proceedings, 9th
  Edoardo Amaldi Conference, Amaldi 9, and meeting, NRDA 2011, Cardiff, UK,
  July 10-15, 2011}}},\ }\href {\doibase 10.1088/0264-9381/29/12/124007}
  {\bibfield  {journal} {\bibinfo  {journal} {Class. Quant. Grav.}\ }\textbf
  {\bibinfo {volume} {29}},\ \bibinfo {pages} {124007} (\bibinfo {year}
  {2012})}\BibitemShut {NoStop}%
\bibitem [{\citenamefont {Aso}\ \emph {et~al.}(2013)\citenamefont {Aso},
  \citenamefont {Michimura}, \citenamefont {Somiya}, \citenamefont {Ando},
  \citenamefont {Miyakawa}, \citenamefont {Sekiguchi}, \citenamefont
  {Tatsumi},\ and\ \citenamefont {Yamamoto}}]{Aso:2013eba}%
  \BibitemOpen
  \bibfield  {author} {\bibinfo {author} {\bibfnamefont {Yoichi}\ \bibnamefont
  {Aso}}, \bibinfo {author} {\bibfnamefont {Yuta}\ \bibnamefont {Michimura}},
  \bibinfo {author} {\bibfnamefont {Kentaro}\ \bibnamefont {Somiya}}, \bibinfo
  {author} {\bibfnamefont {Masaki}\ \bibnamefont {Ando}}, \bibinfo {author}
  {\bibfnamefont {Osamu}\ \bibnamefont {Miyakawa}}, \bibinfo {author}
  {\bibfnamefont {Takanori}\ \bibnamefont {Sekiguchi}}, \bibinfo {author}
  {\bibfnamefont {Daisuke}\ \bibnamefont {Tatsumi}}, \ and\ \bibinfo {author}
  {\bibfnamefont {Hiroaki}\ \bibnamefont {Yamamoto}},\ }\bibfield  {title}
  {\enquote {\bibinfo {title} {{Interferometer design of the KAGRA
  gravitational wave detector}},}\ }\href {\doibase 10.1103/PhysRevD.88.043007}
  {\bibfield  {journal} {\bibinfo  {journal} {Phys. Rev.}\ }\textbf {\bibinfo
  {volume} {D88}},\ \bibinfo {pages} {043007} (\bibinfo {year}
  {2013})}\BibitemShut {NoStop}%
\bibitem [{\citenamefont {Mroue}\ \emph {et~al.}(2013)\citenamefont {Mroue}
  \emph {et~al.}}]{Mroue:2013xna}%
  \BibitemOpen
  \bibfield  {author} {\bibinfo {author} {\bibfnamefont {Abdul~H.}\
  \bibnamefont {Mroue}} \emph {et~al.},\ }\bibfield  {title} {\enquote
  {\bibinfo {title} {{Catalog of 174 Binary Black Hole Simulations for
  Gravitational Wave Astronomy}},}\ }\href {\doibase
  10.1103/PhysRevLett.111.241104} {\bibfield  {journal} {\bibinfo  {journal}
  {Phys. Rev. Lett.}\ }\textbf {\bibinfo {volume} {111}},\ \bibinfo {pages}
  {241104} (\bibinfo {year} {2013})}\BibitemShut {NoStop}%
\bibitem [{\citenamefont {Lovelace}\ \emph {et~al.}(2015)\citenamefont
  {Lovelace} \emph {et~al.}}]{Lovelace:2014twa}%
  \BibitemOpen
  \bibfield  {author} {\bibinfo {author} {\bibfnamefont {Geoffrey}\
  \bibnamefont {Lovelace}} \emph {et~al.},\ }\bibfield  {title} {\enquote
  {\bibinfo {title} {{Nearly extremal apparent horizons in simulations of
  merging black holes}},}\ }\href {\doibase 10.1088/0264-9381/32/6/065007}
  {\bibfield  {journal} {\bibinfo  {journal} {Class. Quant. Grav.}\ }\textbf
  {\bibinfo {volume} {32}},\ \bibinfo {pages} {065007} (\bibinfo {year}
  {2015})}\BibitemShut {NoStop}%
\bibitem [{\citenamefont {Scheel}\ \emph {et~al.}(2015)\citenamefont {Scheel},
  \citenamefont {Giesler}, \citenamefont {Hemberger}, \citenamefont {Lovelace},
  \citenamefont {Kuper}, \citenamefont {Boyle}, \citenamefont {Szilagyi},\ and\
  \citenamefont {Kidder}}]{Scheel:2014ina}%
  \BibitemOpen
  \bibfield  {author} {\bibinfo {author} {\bibfnamefont {Mark~A.}\ \bibnamefont
  {Scheel}}, \bibinfo {author} {\bibfnamefont {Matthew}\ \bibnamefont
  {Giesler}}, \bibinfo {author} {\bibfnamefont {Daniel~A.}\ \bibnamefont
  {Hemberger}}, \bibinfo {author} {\bibfnamefont {Geoffrey}\ \bibnamefont
  {Lovelace}}, \bibinfo {author} {\bibfnamefont {Kevin}\ \bibnamefont {Kuper}},
  \bibinfo {author} {\bibfnamefont {Michael}\ \bibnamefont {Boyle}}, \bibinfo
  {author} {\bibfnamefont {B.}~\bibnamefont {Szilagyi}}, \ and\ \bibinfo
  {author} {\bibfnamefont {Lawrence~E.}\ \bibnamefont {Kidder}},\ }\bibfield
  {title} {\enquote {\bibinfo {title} {{Improved methods for simulating nearly
  extremal binary black holes}},}\ }\href {\doibase
  10.1088/0264-9381/32/10/105009} {\bibfield  {journal} {\bibinfo  {journal}
  {Class. Quant. Grav.}\ }\textbf {\bibinfo {volume} {32}},\ \bibinfo {pages}
  {105009} (\bibinfo {year} {2015})}\BibitemShut {NoStop}%
\bibitem [{\citenamefont {Chu}\ \emph {et~al.}(2016)\citenamefont {Chu},
  \citenamefont {Fong}, \citenamefont {Kumar}, \citenamefont {Pfeiffer},
  \citenamefont {Boyle}, \citenamefont {Hemberger}, \citenamefont {Kidder},
  \citenamefont {Scheel},\ and\ \citenamefont {Szilagyi}}]{Chu:2015kft}%
  \BibitemOpen
  \bibfield  {author} {\bibinfo {author} {\bibfnamefont {Tony}\ \bibnamefont
  {Chu}}, \bibinfo {author} {\bibfnamefont {Heather}\ \bibnamefont {Fong}},
  \bibinfo {author} {\bibfnamefont {Prayush}\ \bibnamefont {Kumar}}, \bibinfo
  {author} {\bibfnamefont {Harald~P.}\ \bibnamefont {Pfeiffer}}, \bibinfo
  {author} {\bibfnamefont {Michael}\ \bibnamefont {Boyle}}, \bibinfo {author}
  {\bibfnamefont {Daniel~A.}\ \bibnamefont {Hemberger}}, \bibinfo {author}
  {\bibfnamefont {Lawrence~E.}\ \bibnamefont {Kidder}}, \bibinfo {author}
  {\bibfnamefont {Mark~A.}\ \bibnamefont {Scheel}}, \ and\ \bibinfo {author}
  {\bibfnamefont {Bela}\ \bibnamefont {Szilagyi}},\ }\bibfield  {title}
  {\enquote {\bibinfo {title} {{On the accuracy and precision of numerical
  waveforms: Effect of waveform extraction methodology}},}\ }\href {\doibase
  10.1088/0264-9381/33/16/165001} {\bibfield  {journal} {\bibinfo  {journal}
  {Class. Quant. Grav.}\ }\textbf {\bibinfo {volume} {33}},\ \bibinfo {pages}
  {165001} (\bibinfo {year} {2016})}\BibitemShut {NoStop}%
\bibitem [{\citenamefont {Gaebel}\ and\ \citenamefont
  {Veitch}(2017)}]{Gaebel:2017zys}%
  \BibitemOpen
  \bibfield  {author} {\bibinfo {author} {\bibfnamefont {Sebastian~M.}\
  \bibnamefont {Gaebel}}\ and\ \bibinfo {author} {\bibfnamefont {John}\
  \bibnamefont {Veitch}},\ }\bibfield  {title} {\enquote {\bibinfo {title}
  {{How would GW150914 look with future gravitational wave detector
  networks?}}}\ }\href {\doibase 10.1088/1361-6382/aa82d9} {\bibfield
  {journal} {\bibinfo  {journal} {Class. Quant. Grav.}\ }\textbf {\bibinfo
  {volume} {34}},\ \bibinfo {pages} {174003} (\bibinfo {year}
  {2017})}\BibitemShut {NoStop}%
\bibitem [{\citenamefont {Vitale}\ and\ \citenamefont
  {Evans}(2017)}]{Vitale:2016icu}%
  \BibitemOpen
  \bibfield  {author} {\bibinfo {author} {\bibfnamefont {Salvatore}\
  \bibnamefont {Vitale}}\ and\ \bibinfo {author} {\bibfnamefont {Matthew}\
  \bibnamefont {Evans}},\ }\bibfield  {title} {\enquote {\bibinfo {title}
  {{Parameter estimation for binary black holes with networks of third
  generation gravitational-wave detectors}},}\ }\href {\doibase
  10.1103/PhysRevD.95.064052} {\bibfield  {journal} {\bibinfo  {journal} {Phys.
  Rev.}\ }\textbf {\bibinfo {volume} {D95}},\ \bibinfo {pages} {064052}
  (\bibinfo {year} {2017})}\BibitemShut {NoStop}%
\bibitem [{\citenamefont {Remillard}\ and\ \citenamefont
  {McClintock}(2006)}]{Remillard:2006fc}%
  \BibitemOpen
  \bibfield  {author} {\bibinfo {author} {\bibfnamefont {Ronald~A.}\
  \bibnamefont {Remillard}}\ and\ \bibinfo {author} {\bibfnamefont
  {Jeffrey~E.}\ \bibnamefont {McClintock}},\ }\bibfield  {title} {\enquote
  {\bibinfo {title} {{X-ray Properties of Black-Hole Binaries}},}\ }\href
  {\doibase 10.1146/annurev.astro.44.051905.092532} {\bibfield  {journal}
  {\bibinfo  {journal} {Ann. Rev. Astron. Astrophys.}\ }\textbf {\bibinfo
  {volume} {44}},\ \bibinfo {pages} {49--92} (\bibinfo {year}
  {2006})}\BibitemShut {NoStop}%
\bibitem [{\citenamefont {{Middleton}}(2016)}]{Middleton2016}%
  \BibitemOpen
  \bibfield  {author} {\bibinfo {author} {\bibfnamefont {M.}~\bibnamefont
  {{Middleton}}},\ }\bibfield  {title} {\enquote {\bibinfo {title} {{Black Hole
  Spin: Theory and Observation}},}\ }in\ \href {\doibase
  10.1007/978-3-662-52859-4_3} {\emph {\bibinfo {booktitle} {Astrophysics of
  Black Holes: From Fundamental Aspects to Latest Developments}}},\ \bibinfo
  {series} {Astrophysics and Space Science Library}, Vol.\ \bibinfo {volume}
  {440},\ \bibinfo {editor} {edited by\ \bibinfo {editor} {\bibfnamefont
  {C.}~\bibnamefont {{Bambi}}}}\ (\bibinfo {year} {2016})\ p.~\bibinfo {pages}
  {99}\BibitemShut {NoStop}%
\bibitem [{\citenamefont {Miller}\ and\ \citenamefont
  {Miller}(2014)}]{Miller:2014aaa}%
  \BibitemOpen
  \bibfield  {author} {\bibinfo {author} {\bibfnamefont {M.~Coleman}\
  \bibnamefont {Miller}}\ and\ \bibinfo {author} {\bibfnamefont {Jon~M.}\
  \bibnamefont {Miller}},\ }\bibfield  {title} {\enquote {\bibinfo {title}
  {{The Masses and Spins of Neutron Stars and Stellar-Mass Black Holes}},}\
  }\href {\doibase 10.1016/j.physrep.2014.09.003} {\bibfield  {journal}
  {\bibinfo  {journal} {Phys. Rept.}\ }\textbf {\bibinfo {volume} {548}},\
  \bibinfo {pages} {1--34} (\bibinfo {year} {2014})}\BibitemShut {NoStop}%
\bibitem [{\citenamefont {Suvorova}\ \emph {et~al.}(2017)\citenamefont
  {Suvorova}, \citenamefont {Clearwater}, \citenamefont {Melatos},
  \citenamefont {Sun}, \citenamefont {Moran},\ and\ \citenamefont
  {Evans}}]{Suvorova2017}%
  \BibitemOpen
  \bibfield  {author} {\bibinfo {author} {\bibfnamefont {S.}~\bibnamefont
  {Suvorova}}, \bibinfo {author} {\bibfnamefont {P.}~\bibnamefont
  {Clearwater}}, \bibinfo {author} {\bibfnamefont {A.}~\bibnamefont {Melatos}},
  \bibinfo {author} {\bibfnamefont {L.}~\bibnamefont {Sun}}, \bibinfo {author}
  {\bibfnamefont {W.}~\bibnamefont {Moran}}, \ and\ \bibinfo {author}
  {\bibfnamefont {R.~J.}\ \bibnamefont {Evans}},\ }\bibfield  {title} {\enquote
  {\bibinfo {title} {{Hidden Markov model tracking of continuous gravitational
  waves from a binary neutron star with wandering spin. II. Binary orbital
  phase tracking}},}\ }\href {\doibase 10.1103/PhysRevD.96.102006} {\bibfield
  {journal} {\bibinfo  {journal} {Phys. Rev.}\ }\textbf {\bibinfo {volume}
  {D96}},\ \bibinfo {pages} {102006} (\bibinfo {year} {2017})},\ \Eprint
  {http://arxiv.org/abs/1710.07092} {arXiv:1710.07092 [astro-ph.IM]}
  \BibitemShut {NoStop}%
\bibitem [{\citenamefont {Orosz}\ \emph {et~al.}(2011)\citenamefont {Orosz},
  \citenamefont {McClintock}, \citenamefont {Aufdenberg}, \citenamefont
  {Remillard}, \citenamefont {Reid}, \citenamefont {Narayan},\ and\
  \citenamefont {Gou}}]{Orosz2011}%
  \BibitemOpen
  \bibfield  {author} {\bibinfo {author} {\bibfnamefont {Jerome~A.}\
  \bibnamefont {Orosz}}, \bibinfo {author} {\bibfnamefont {Jeffrey~E.}\
  \bibnamefont {McClintock}}, \bibinfo {author} {\bibfnamefont {Jason~P.}\
  \bibnamefont {Aufdenberg}}, \bibinfo {author} {\bibfnamefont {Ronald~A.}\
  \bibnamefont {Remillard}}, \bibinfo {author} {\bibfnamefont {Mark~J.}\
  \bibnamefont {Reid}}, \bibinfo {author} {\bibfnamefont {Ramesh}\ \bibnamefont
  {Narayan}}, \ and\ \bibinfo {author} {\bibfnamefont {Lijun}\ \bibnamefont
  {Gou}},\ }\bibfield  {title} {\enquote {\bibinfo {title} {{The Mass of the
  Black Hole in Cygnus X-1}},}\ }\href {\doibase 10.1088/0004-637X/742/2/84}
  {\bibfield  {journal} {\bibinfo  {journal} {Astrophys. J.}\ }\textbf
  {\bibinfo {volume} {742}},\ \bibinfo {pages} {84} (\bibinfo {year} {2011})},\
  \Eprint {http://arxiv.org/abs/1106.3689} {arXiv:1106.3689 [astro-ph.HE]}
  \BibitemShut {NoStop}%
\bibitem [{\citenamefont {{Reid}}\ \emph {et~al.}(2011)\citenamefont {{Reid}},
  \citenamefont {{McClintock}}, \citenamefont {{Narayan}}, \citenamefont
  {{Gou}}, \citenamefont {{Remillard}},\ and\ \citenamefont
  {{Orosz}}}]{Reid2011}%
  \BibitemOpen
  \bibfield  {author} {\bibinfo {author} {\bibfnamefont {M.~J.}\ \bibnamefont
  {{Reid}}}, \bibinfo {author} {\bibfnamefont {J.~E.}\ \bibnamefont
  {{McClintock}}}, \bibinfo {author} {\bibfnamefont {R.}~\bibnamefont
  {{Narayan}}}, \bibinfo {author} {\bibfnamefont {L.}~\bibnamefont {{Gou}}},
  \bibinfo {author} {\bibfnamefont {R.~A.}\ \bibnamefont {{Remillard}}}, \ and\
  \bibinfo {author} {\bibfnamefont {J.~A.}\ \bibnamefont {{Orosz}}},\
  }\bibfield  {title} {\enquote {\bibinfo {title} {{The Trigonometric Parallax
  of Cygnus X-1}},}\ }\href {\doibase 10.1088/0004-637X/742/2/83} {\bibfield
  {journal} {\bibinfo  {journal} {\apj}\ }\textbf {\bibinfo {volume} {742}},\
  \bibinfo {eid} {83} (\bibinfo {year} {2011})},\ \Eprint
  {http://arxiv.org/abs/1106.3688} {arXiv:1106.3688 [astro-ph.HE]} \BibitemShut
  {NoStop}%
\bibitem [{\citenamefont {{Iorio}}(2008)}]{Iorio2008}%
  \BibitemOpen
  \bibfield  {author} {\bibinfo {author} {\bibfnamefont {L.}~\bibnamefont
  {{Iorio}}},\ }\bibfield  {title} {\enquote {\bibinfo {title} {{On the orbital
  and physical parameters of the HDE 226868/Cygnus X-1 binary system}},}\
  }\href {\doibase 10.1007/s10509-008-9839-y} {\bibfield  {journal} {\bibinfo
  {journal} {Astrophysics and Space Science}\ }\textbf {\bibinfo {volume}
  {315}},\ \bibinfo {pages} {335--340} (\bibinfo {year} {2008})}\BibitemShut
  {NoStop}%
\end{thebibliography}%

\end{document}